\documentclass[preprint]{aastex}
\usepackage{graphicx}
\usepackage{amsmath}

\shorttitle{Variable Stars with ATLAS}
\shortauthors{Heinze et al.}

\begin{document}

\title{A First Catalog of Variable Stars Measured by the Asteroid Terrestrial-impact Last Alert System (ATLAS)}

\author{A. N. Heinze\altaffilmark{1}, J. L. Tonry\altaffilmark{1}, L. Denneau\altaffilmark{1}, H. Flewelling\altaffilmark{1}, B. Stalder\altaffilmark{2}, A. Rest\altaffilmark{3,4} K. W. Smith\altaffilmark{5}, S. J. Smartt\altaffilmark{5}, and H. Weiland\altaffilmark{1}}

\altaffiltext{1}{Institute for Astronomy, University of Hawaii, 2680 Woodlawn, Honolulu, HI, 96822, USA; aheinze@hawaii.edu}
\altaffiltext{2}{LSST, 950 N. Cherry Ave, Tucson, AZ 85719, USA}
\altaffiltext{3}{Space Telescope Science Insititute, 3700 San Martin Drive, Baltimore, MD 21218, USA}
\altaffiltext{4}{Department of Physics and Astronomy, Johns Hopkins University, Baltimore, MD 21218, USA}
\altaffiltext{5}{Astrophysics Research Centre, School of Mathematics and Physics, Queen's University Belfast, Belfast, BT7 1NN, UK}

\begin{abstract}
The Asteroid Terrestrial-impact Last Alert System (ATLAS) carries out its primary planetary defense mission by surveying about 13000 deg$^2$ at least four times per night. The resulting data set is useful for the discovery of variable stars to a magnitude limit fainter than $r \sim 18$, with amplitudes down to 0.02 mag for bright objects. Here we present a Data Release One catalog of variable stars based on analyzing the lightcurves of 142 million stars that were measured at least 100 times in the first two years of ATLAS operations. Using a Lomb-Scargle periodogram and other variability metrics, we identify 4.7 million candidate variables. Through Space Telescope Science Institute, we publicly release lightcurves for all of them, together with a vector of 169 classification features for each star. We do this at the level of unconfirmed candidate variables in order to provide the community with a large set of homogeneously analyzed photometry and avoid pre-judging which types of objects others may find most interesting. We use machine learning to classify the candidates into fifteen different broad categories based on lightcurve morphology. About 10\% (427,000 stars) pass extensive tests designed to screen out spurious variability detections: we label these as `probable' variables. Of these, 214,000 receive specific classifications as eclipsing binaries, pulsating, Mira-type, or sinusoidal variables: these are the `classified' variables. New discoveries among the probable variables number 315,000, while 141,000 of the classified variables are new, including about 10,400 pulsating variables, 2,060 Mira stars, and 74,700 eclipsing binaries.
\end{abstract}


\section{Introduction}

\subsection{Variable Stars and Wide-Field Surveys} \label{sec:reviewvar}

Variable stars have profound and wide-ranging value for astrophysics. Pulsating variables, especially Cepheids, are a central link in the cosmic distance ladder that is foundational to our understanding of cosmology. Detached eclipsing binaries offer some of the best opportunities to get precise masses and radii of distant stars. Contact binaries present us with a rich variety of interesting phenomena, and some of them represent intermediate stages in systems evolving toward novae, stellar mergers, X-ray binaries, and (possibly) Type Ia supernovae. Flare stars and spotted rotators give insight into stellar magnetic fields across the HR diagram. Pulsating red giants (especially the huge-amplitude Mira stars), and a vast diversity of exotic types of variables probe interesting astrophysics and stellar evolution scenarios.

Going back at least to the early results of the Optical Gravitational Lens Experiment \citep[OGLE,][]{OGLE}, sky surveys using wide-field CCD imagers have greatly increased the number of known variable stars. Even though many of the surveys do not have variable stars as their primary objective, the data they produce is revolutionizing the field of variable star research. This trend will only accelerate in the future as Gaia \citep{Gaia}, the Zwicky Transient Facility \citep{ZTF}, and LSST publish their first time-series photometry while ongoing surveys continue releasing interesting results. 

A full review of variable star results from wide-field surveys is beyond the scope of the present work, but we briefly note a few publications and statistics for context. Surveys that have produced data used for variable star discovery and analysis include the gravitational microlensing surveys MACHO \citep{MACHO} and OGLE \citep{OGLE}; supernova/transient surveys including the All-Sky Automated Survey \citep[ASAS,][]{ASAS}, the All-Sky Automated Survey for Supernovae \citep[ASAS-SN,][]{Shappee2014}, the Palomar Transient Factory \citep[PTF,][]{PTF}, and the China-based Tsinghua University-NAOC Transient Survey \citep[TNTS,][]{TNTS}; Pan-STARRS1 \citep{PS1A,Flewelling2016,PS1B,PS1C,PS1D,PS1E}; the Vista Variables in the Via Lactea \citep[VVV,][]{VVV}; the Robotic Optical Transient Search Experiment (ROTSE-I), which was built to look for optical counterparts of gamma-ray bursts \citep{ROTSE}; and asteroid surveys including the Lowell Observatory Near-Earth Object Search \citep[LONEOS,][]{LONEOS}, the Lincoln Near-Earth Asteroid Research \citep[LINEAR,][]{LINEAR} and the Catalina Sky Survey \citep{CSS}.

Data from these surveys have been used in many publications analyzing and presenting catalogs of variable stars. We list only a few examples here, reserving the OGLE surveys for a separate paragraph. \citet{Alcock1998,Alcock2000} used data from the MACHO survey to find RR Lyrae and $\delta$ Scuti stars in the Galactic bulge, while the same authors also published numerous papers on MACHO variables in the Magellanic Clouds. \citet{Pojmanski2002,Pojmanski2003,Pojmanski2004,Pojmanski2005a}; and \citet{Pojmanski2005b} used the ASAS survey to identify a total of 46,756 variable stars at declinations south of +28$^{\circ}$. These include 9,581 eclipsing binaries, 4,921 pulsating stars, and 2,758 Mira variables. Using data from ROTSE-I, \citet{Kinemuchi2006} discovered 1197 RR Lyrae stars and analyzed their metalicity using metalicity-dependence in their pulse waveforms. \citet{Miceli2008} discovered and analyzed 838 RR Lyrae stars in the Galactic halo, using data from the LONEOS asteroid survey. Using the LINEAR survey data, \citet{Palaversa2013} discovered and classified 7000 variable stars, while \citet{Sesar2013} analyzed a partly-overlapping set of 5000 RR Lyrae stars in the same data. About 60,000 variable stars were discovered in asteroid search data from the Catalina Sky Survey by \citet{Drake2013a,Drake2014a}.  \citet{Drake2013b}, \citet{Hernitscheck2016}, and \citet{Cohen2017} explored star-streams in the outer halo of the Milky Way using RR Lyrae candidates identified in data from the Catalina Sky Survey, Pan-STARRS1, and the PTF, respectively. Using the same class of variables stars to probe the inner rather than the outer Milky Way, \citet{VVVRR} measured the distance to the Galactic center by analyzing 4194 RR Lyrae stars from the VVV survey. \citet{Yao2015} have released a meticulously analyzed list of 1237 variables stars from the TNTS. \citet{ASASSN} have released a catalog of 66,533 variable stars discovered in data from ASAS-SN. We mention in passing that a host of interesting variable star results have also been obtained using photometry from the Kepler mission (e.g. \citet{Benko2010,Banyai2013}; and many others), but we will not discuss them herein because most of the dramatic Kepler discoveries have come from probing a regime of small-amplitude, high-precision photometry that is inaccessible from the ground and hence of limited relevance to the ground-based ATLAS survey results that are the subject of this paper.

The Optical Gravitational Lens Experiment \citep[OGLE,][]{OGLE} surveys deserve a separate discussion becuse they have produced the largest homogeneous catalogs of variable stars thus far (by an order of magnitude). The OGLE surveys of the Galactic bulge have revealed about 700,000 new variables among 400 million stars analyzed \citep{Soszynski2011a,Soszynski2011b,Soszynski2013,Soszynski2014,Mroz2015,Soszynski2015,Soszynski2016,Soszynski2017}, while several hundred thousand more variable stars have been found at more southerly declinations in the Magellanic Clouds. Besides these huge numbers of stars, the OGLE catalogs significantly exceed most of the others described here in temporal span and numbers of photometric points per star. This wealth of data has enabled many important results. These include \citet{Soszynski2015}, who present the shortest-period known main-sequence eclipsing binary, together with a fascinating astrophysical discussion of the existence (and rarity) of eclipsing binaries with periods shorter than the well-known 0.22 day cutoff; and \citet{Soszynski2017}, who discover and classify Cepheids toward the galactic center using Fourier phase coefficients; as well as others too numerous to list.

\subsection{The ATLAS Survey}
The Asteroid Terrestrial-impact Last Alert System \citep[ATLAS;][]{Tonry2018} is designed to detect small (10--140m) asteroids on their `final plunge' toward impact with Earth. Because such asteroids can come from any direction and go from undetectable to impact in less than a week, ATLAS scans the whole accessible sky every few days. To achieve this, we use fully robotic 0.5m f/2 Wright Schmidt telescopes with 10560$\times$10560 pixel STA1600 CCDs yielding a 5.4$\times$5.4 degree field of view with 1.86 arcsec pixels. The first ATLAS telescope commenced operations in mid-2015 on the summit of Haleakal\=a on the Hawaiian island of Maui, and the second was installed in January/February 2017 at Maunaloa Observatory on the big island of Hawaii. On a typical night, each ATLAS telescope takes four 30-second exposures of 200--250 target fields covering approximately one fourth of the accessible sky. Together, the two telescopes cover half the accessible sky each night. The four observations of a given target field on a given night are typically obtained over a period of somewhat less than one hour.

The wide-field, high-cadence observations ATLAS makes to discover near-Earth asteroids are also well-suited to the discovery and characterization of variable stars down to a magnitude limit fainter than $r=18$. We present herein the first catalog of variable stars measured by ATLAS, including characterization of known variable stars and the discovery of about 300,000 new variables. 

This initial data release is based on the first two years of operation of the Haleakal\=a telescope only, and covers observations taken up through the end of June, 2017. This date marked the end of a series of changes, which included the switch to dual-telescope operations; upgraded optics for both telescopes; re-collimation of the telescopes to take advantage of the new optics; and changes in our observing cadence, processing pipeline, and calibration data. The optical upgrades changed the full width at half maximum (FWHM) of the typical point spread function (PSF) delivered by the  Haleakal\=a telescope from 7 arcsec to 4 arcsec. The conclusion of these significant changes made it natural to consider the data before the end of June 2017 as a closed chapter, and accordingly we re-analyzed all of it with optimized and homogeneous methodology. This is the data set we analyze herein to produce ATLAS variable star Data Release One (ATLAS DR1; see Table \ref{tab:numbers}). The more recent data are expected to be even better photometrically, but the ATLAS DR1 data set enables the discovery and/or characterization of several hundred thousand variable stars. We anticipate generating additional data releases (ATLAS DR2, DR3, etc) approximately once a year, which will include homogenously processed data from both telescopes, with adjustments to the calibration and analysis to take advantage of optical improvements.

The ATLAS telescope on Haleakal\=a observes with two customized, wide filters designed to optimize detection of faint objects while still providing some color information. The `cyan' filter ($c$, covering 420--650 nm) is used during the two weeks surrounding new moon; and the `orange' filter ($o$, 560--820 nm) is used in lunar bright time. As described in \citet{Tonry2018}, the $o$ and $c$ filters are well-defined photometric bands with known color transformations linking them to the Pan-STARRS $g$, $r$, and $i$ bands \citep{PS1E}.

During the period covered by ATLAS DR1, the ATLAS Haleakal\=a telescope cycled through four bands of declination (`Dec bands'), observing one band each night. Cumulatively, the Dec bands extended from Dec $-30^{\circ}$ to $+60^{\circ}$ in their narrowest configuration. Within the scheduled Dec band for a given night, the telescope took four to six 30-second exposures of each of typically 200 fields covering the accessible range in right ascension (RA). The accessible range in RA was defined by an altitude limit of 20$^{\circ}$, which enabled dark-sky observations (Sun more than 18$^{\circ}$ below the horizon) at solar elongations as small as 45$^{\circ}$ at the beginning and end of the night. Thus, observations for Dec bands north of the equator could span as much as 270$^{\circ}$ in RA on a single night. Pointings near the Moon were avoided by modeling the sky background and skipping areas where the predicted degradation of the signal-to-noise ratio (SNR) amounted to more than 1 magnitude. This resulted in a lunar avoidance radius of about $30^{\circ}$ at full Moon, decreasing to about $10^{\circ}$ for the crescent phases.  The exact thickness of each Dec band was adjusted night-by-night depending on the phase of the Moon: a bright Moon would render a large area of some Dec bands unobservable, and hence the Dec range would be widened in order to obtain enough viable pointings to fill the night.

The exposures of each given field were all taken within a period of typically 0.5--1.5 hours, with small ($\sim0.05^{\circ}$) dithers between them. The exact cadence varied from night-to-night due to the details of automated schedule optimization and also to deliberate experiments we made to find the survey parameters that would produce the greatest efficiency for discovering near-Earth asteroids. Such variations are  preferable to a strictly regular cadence for the detection of variable stars, since the latter would produce unnecessary period-aliasing (beyond the diurnal aliasing that is unavoidable for ground-based observations from a single longitude). To mitigate any systematic effects dependent on field position, a random offset of amplitude $\sim1^{\circ}$ was selected and homogeneously imposed on all the pointings from each night, to ensure that over a long period there would be a large diversity of pointings in each Dec band. During the period covered by DR1, various trial adjustments were made to the extent of the Dec bands (in both RA and Dec); to the number of observations of each field per night; and to the dithering strategy. These resulted in some observations being conducted north and south of the $-30^{\circ}$ to $+60^{\circ}$ Dec range, but they were not sufficiently numerous to discover many variable stars. Using observations from both telescopes, variable star measurements in ATLAS DR2 will cover the entire sky north of Dec $-50^{\circ}$. ATLAS DR2 will contain 70\% more stars and at least three times more photometric measurements than the current data release.

\subsection{ATLAS Variable Stars}
The ATLAS DR1 catalog we present herein makes a major contribution even in the context of the great expansion of known variable stars described in \S \ref{sec:reviewvar}. It is based on analyzing the photometric time series (lightcurves) of 142 million stars, which we refer to herein as the `ATLAS lightcurve set', and of which we identify 4.7 million as candidate variables. The on-sky distributions of both the lightcurve set and the candidate variables are shown in Figure \ref{fig:allstars}. All of the photometry for each of these candidate varaiables is being publicly released: the largest catalog of candidate variables yet. With 430,000 confirmed variables (of which 300,000 are new), ATLAS DR1 is also the largest homogeneous catalog of {\em confirmed} variables apart from OGLE, and the largest to span the sky (since the OGLE variables are confined to relatively small areas targeting the Galactic bulge and the Magellanic Clouds). By using two filters ($c$ and $o$) with well-defined photometric properties \citep{Tonry2018}, ATLAS obtains quantitative color information for every star. We provide AB magnitudes in the $c$ and $o$ bands that are free of any known systematic bias, together with realistic uncertainties for every measurement\footnote{To obtain these realistic uncertainties, the original sources of noise (read noise, photon shot noise, and dark current) are propagated through our complex reduction pipeline to produce an accurate pixel-by-pixel variance map of every ATLAS image. Our photometric code (an enhanced version of \texttt{DoPHOT} \citep{dophot2}; see also \S \ref{sec:data}) performs a mathematically rigorous fit to the PSF of each star, calculating the uncertainties on each fit parameter using individual-pixel variances from the variance map.}. Table \ref{tab:numbers} gives the numbers of stars, images, and photometric measurements used for various stages of our analysis, and assigns names to various subsets of stars that we will use frequently below.

\begin{deluxetable}{lll}
\tabletypesize{\small}
\tablewidth{0pt}
\tablecaption{Scale of ATLAS DR1 by the Numbers \label{tab:numbers}}
\tablehead{ \colhead{Name} & \colhead{Quantity} & \colhead{Description} }
\startdata
Input data & 284,000 images & Raw data of our analysis\\ \\
All photometry & $\sim 60$ billion measurements & Total photometric data \\ \\
Lightcurve photometry & 30 billion measurements & Photometric data that contributed to \\
                      &                         &lightcurves analyzed herein \\ \\
Object-matching catalog & 302 million stars & Pan-STARRS based input catalog \\
                        &                   & used to assign ATLAS photometric \\
                        &                   & measurements to specific stars \\\\
Lightcurve set\tablenotemark{a} & 142 million stars & Subset of the object-matching catalog \\
               &                   & consisting of stars for which ATLAS acquired \\
               &                   & at least 100 photometric data points \\\\
Candidate variables\tablenotemark{b} & 4.7 million stars & Objects from the lightcurve set for \\
                    &                   & which ATLAS data showed evidence of \\
                    &                   & variability indicating more detailed \\
                    &                   & analysis \\\\
Probable variables & 427,000 stars & Candidate variables indicated by detailed \\
                   &               & analysis as probably real (any category \\
                   &               & other than `dubious'; see \S \ref{sec:classdesc}) \\ \\
Classified variables & 214,000 stars & Probable variables that received specific \\
                     &               & classifications (excludes generic IRR, LPV, \\
                     &               & NSINE and STOCH classes; see \S \ref{sec:classdesc})\\
\enddata
\tablenotetext{a}{Note that each group of stars described in this table is a subset of the one immediately above it.}
\tablenotetext{b}{All photometry of the candidate variables has been publicly released through STScI; see \S \ref{sec:atquery}.}
\end{deluxetable}

Besides the photometry, we are publicly releasing an extensive set of 169 variability features for each of the candidate variables, which we hope other researchers will find useful for developing the rich scientific potential of the new catalog. The payoff for developing effective data mining techniques to extract astrophysical discoveries from this and other current variable star catalogs will only increase in the future. New, larger catalogs will continue to be released by Gaia \citep{Gaia}; the Zwicky Transient Facility \citep{ZTF}; expanded versions of ongoing surveys including OGLE, ATLAS, and the VVV \citep{VVV}; and ultimately the Large Synoptic Survey Telescope. The potential for major discoveries from these data is enormous and spans almost all of astronomy, from star formation and planetary habitability to supernovae and cosmology.

\section{The Data: Images and Detections} \label{sec:data}

A customized, fully automated pipeline processes every image from an ATLAS telescope, outputting a flatfielded, calibrated image with both astrometric and photometric solutions. For asteroid detection, we subtract from each of these images a template extracted from the low-noise static sky image, or `wallpaper' we have built up by stacking tens of thousands of ATLAS images taken under excellent conditions and covering the accessible sky (to a coverage depth of a few tens of images per filter at most locations). We perform this subtraction using a customized version of the `HOTPANTS' program \citep{hotpants}, which is based on the methodology developed by \citet{Alard1998} and \citet{Alard2000} to match the PSF of two images by convolving one of them with a kernel that is a linear combination of functions in a basis set composed of radial Gaussians multiplied by polynomials. We adopt a policy of matching the PSF of the template image (extracted from the wallpaper) to that of the science image, rather than modifying the latter in any way prior to subtraction. This is similar to the approach of \citet{Alcock1999}, which is often referred to as Difference Image Analysis (DIA). It requires that the template be at least as sharp as the science image, an outcome we achieve by making the wallpaper out of the sharpest available ATLAS images, and applying additional sharpening via Richardson-Lucy deconvolution \citep{RLdeconR,RLdeconL} as necessary. The subtraction of constant sources by differencing each image relative to the wallpaper is essential to the sensitive discovery of asteroids with a low false-positive rate. However, the variable star results we present herein are based primarily on photometry of the images prior to image differencing, because deviations around the mean flux in the wallpaper are less useful for variability analyses than the total flux of a star.
%we currently achieve more precise photometric calibration for the unsubtracted images. 
We use the differenced images as a final check to confirm the nature of stars with only tentative variability detections based on the unsubtracted photometry.

The analysis we present herein is based on 284,000 images, which span the sky from Dec -30$^{\circ}$ to +60$^{\circ}$, with some additional coverage north and south of these limits. Within this range, most areas of the sky are covered by more than 200 images. 

We perform photometry of the unsubtracted images using the {\tt DoPHOT} code. \linebreak{\tt DoPHOT} \citep{dophot1} measures a star's position and flux by adopting a point spread function (PSF) model, and iteratively finding, fitting, and subtracting each star from the image.  The PSF model and aperture magnitudes are derived from the brightest stars as the program iterates.  We use a Fortran-90 version of {\tt DoPHOT} \citep{dophot2} that has a number of enhancements including floating point input and, most importantly, the ability to perform accurate fits when the FWHM and shape of the PSF vary from one part of the image to another. 

{\tt DoPHOT} fits each star with a PSF whose functional form is based on an elliptical Gaussian but altered to better match real stellar images, which have broader wings than a strict Gaussian \citep{dophot1}. Three parameters define the FWHM and shape of the PSF: major axis, minor axis, and position angle\footnote{The formalism of \citet{dophot1} uses the functionally equivalent $\sigma_{xy}$ rather than position angle, and defines user-adjustable parameters $\beta_4$ and $\beta_6$, which control the wings of the PSF and are fixed to 1 and $\frac{1}{2}$ for ATLAS data.}. The enhanced {\tt DoPHOT} of \citet{dophot2} allows the three PSF shape parameters to vary smoothly across the image by fitting each of them as a polynomial function of the x,y position in the image. Thus, {\tt DoPHOT} will accurately fit the stars in images that are sharp in one area and blurry in another, or even that exhibit optical aberration causing an elongated PSF that rotates from one part of the image to another.

In successive iterations, {\tt DoPHOT} measures and subtracts fainter and fainter stars until no significant sources remain in the image. This permits better photometry of faint stars by first removing bright neighbors. In a given iteration, {\tt DoPHOT} first finds the best-fit PSF for each stellar image, where the three shape paramers are allowed to vary freely from star to star. It uses the results to produce the polynomial fits (referred to above) that model the variation of the PSF across the image, and then re-fits each star with the PSF shape constrained to match the shape given by the model evaluated at that star's location. Finally, {\tt DoPHOT} subtracts all of the measured stars and proceeds to a new iteration in which it fits a new cohort of fainter stars that can be accurately measured now that their brighter neighbors have been subtracted away.

Our particular {\tt DoPHOT} code is further developed from that of \citet{dophot2}, to enhance performance and correct a few minor `bugs' that only become manifest when the code is used on extremely large images. The corrections predominately relate to the robustness of the spatially varying PSF fit. They include a change to double-precision model fitting (at single precision, the fit to spatial variations of the PSF shape could fail when attempting to process millions of stars across the 10560$\times$10560 pixel ATLAS images) and a change to calculating the sky backgrounds with a median not over all pixels in each region of the image (which was very slow) but only over an optimally-sized subsample. Enhancements include multi-threading, performing all calculations (not just the spatial PSF fit) in double precision, and the input of an external variance image (described above) to enable mathematically rigorous propagation of photometric uncertainties for images produced by our complex pipeline. The problems we have corrected would not be considered as actual bugs in the code of \citet{dophot2}, since we have seen them to cause incorrect results only when this code is applied to CCD images from a monolithic chip larger than any in astronomical use at the time it was written. We have not found bugs of any kind in the original {\tt DoPHOT} code of \citet{dophot1}.

For stars of sufficient brightness (e.g., detection SNR $\gtrsim 50$), {\tt DoPHOT} calculates two different fluxes: an `aperture' magnitude which is the sum of the flux within a large aperture (e.g. 30 arcsec), and a `fit' magnitude which is the integral of the PSF fit. The fit magnitude is expected to be less noisy than the aperture magnitude, but it is more vulnerable to systematic effects because the  3-parameter {\tt DoPHOT} PSF (even with the parameters varying smoothly across the image) is not expected to capture the full complexity of the PSF in a real astronomical image, especially from a wide-field system such as ATLAS. Hence, we perform further processing of the {\tt DoPHOT} output to capture the best characteristics of both the fit magnitudes and the aperture magnitudes. We model the spatial variation of the difference between the aperture and fit magnitudes across the image, and correct all of the fit magnitudes according to this  `ap minus fit' model. Hence, we obtain low-noise instrumental magnitudes for all stars, referenced to the large-aperture fluxes to minimize systematic effects. Since fit magnitudes exist for even the faintest stars measured by {\tt DoPHOT}, these corrected instrumental magnitudes are obtained for all measured stars, not just those bright enough to have aperture magnitudes.

We perform additional optimizations of our photometry even beyond the `ap minus fit' correction described above, in order to remove remaining photometric variations from a variety of sources (e.g. imperfect flatfield and uneven atmospheric transparency). To do this, we first calculate the offset between the measured magnitude of each star (above a flux threshold to ensure low-noise measurements) and its expected magnitude from our object-matching catalog (based on Pan-STARRS1 DR1 \citep{Flewelling2016}; see \S \ref{sec:obmatch} and \S \ref{pscat}), using known transformations we have derived between the Pan-STARRS $gri$ photometry and the wider ATLAS filters \citep[][see also Equation \ref{eq:phottrans}]{Tonry2018}. We then perform a bicubic fit on $8\times8$ cells to model the variation in observed minus expected magnitude over the image, and we correct the measured magnitudes based on this fit. Since we have thousands of bright stars per image, we are able to make the fit robust against outliers due to stellar variability and other effects.

\begin{equation} \label{eq:phottrans}
c \sim 0.49 g + 0.51 r \; \; \; \; \; \; \; \; o \sim 0.55 r + 0.45 i
\end{equation}

In this way, we obtain tens of thousands (the median number is 110,000) of highly precise photometric measurements per image. The mean number of stars measured per image is more than twice as large as the median because of extremely dense starfields near the Galactic plane. Although we use {\tt DoPHOT} with a sensitive, $3\sigma$ threshold in order to detect the faintest measurable objects, a majority of these measurements are still expected to correspond to real stars. Under good conditions (i.e. uncrowded fields observed under clear, moonless skies), {\tt DoPHOT} measures objects significantly fainter than 19th magnitude, and the median uncertainty at magnitude 18.0 is about 0.095 mag in $c$ and slightly better than 0.15 mag in $o$. The total number of photometric measurements in our analysis may be conservatively estimated by multiplying the approximate mean of 220,000 per image times 284,000 images: more than 60 billion individual measurements. Note well that all of the above statistics apply to the DR1 data analyzed herein. We already have about twice this much data on disk (to be relased in DR2), and the sharper PSF of the new images enables significantly more precise photometry.

\section{Photometric Analysis} \label{sec:photan}

\subsection{The Object-Matching Catalog} \label{sec:obmatch}

As described in the previous section, we have obtained about 60 billion precise photometric measurements of stars and other objects detected in ATLAS images. To use this data to find variable stars, we must first assign the measurements to specific objects. We elect to do this using an external object-matching catalog, constructed from survey data with a higher resolution and (where possible) a fainter limiting magnitude than ATLAS. The advantages of using a higher-resolution external catalog include more precise positions for every star, and fewer instances of multiple blended stars being incorrectly analyzed as a single object. The disadvantage is that we may miss objects that have only recently become visible. Thus, we would not expect novae or supernovae to appear in our current analysis, and we might also miss some extremely long-period, high amplitude variables that have been coming out of a deep minimum in the last three years. An ATLAS catalog of transients, focused on supernovae, is currently in preparation (Smith et al., in prep).

We construct our object-matching catalog primarily from the Pan-STARRS1 DR1 catalog \citep{Flewelling2016}, which covers the sky north of Dec -30$^{\circ}$. The resolution of Pan-STARRS images ($\sim$ 1 arcsec), and hence their astrometric accuracy, are much better than ATLAS images, which in the current data set have a typical PSF width of 7 arcsec. Pan-STARRS also goes at least three magnitudes deeper than ATLAS in the $g$, $r$, and $i$ bands. To construct a subset of the Pan-STARRS1 DR1 catalog suitable for matching to ATLAS photometric detections, we require that each star be brighter than magnitude 19 in at least one of the $g$, $r$, $i$, or $z$ bands. To obtain the best list of PS1 objects, we require that the objects exist in the PS1 stack catalogs, and we use various flags to select the best position when objects are duplicated (see \S \ref{sec:appendixB} for more details and a sample query).

To include objects south of Dec -30$^{\circ}$, and bright stars that saturate in Pan-STARRS images, we augment our object-matching catalog using the TYCHO \citep{TYCHO} and APASS \citep{APASS} catalogs. These have magnitude limits considerably brighter than our intrinsic limit of $\sim18$th mag, so we monitor only bright stars south of Dec -30$^{\circ}$. In total, the object-matching catalog we use herein contains about 302 million stars. For use in ATLAS DR2, we are currently constructing an updated object-matching catalog (combining data from Gaia, Pan-STARRS, and several other surveys) that will have a uniform limiting magnitude of 19 over the whole sky \citep{refcat}.

\subsection{The Photometric Data} \label{sec:photdat}
We associate our individual photometric detections to particular stars by cross-matching the RA and Dec output by {\tt DoPHOT} with objects in our object-matching catalog, using a radius of $0.0003^{\circ}$, or slightly more than 1 arcsec. This matching radius is smaller than our 7 arcsec FWHM, but considerably larger than our astrometric precision, except for the faintest stars. Using a small matching radius is important to minimize spurious matches in crowded fields. In cases of stars resolved in the object-matching catalog but blended together in the ATLAS images, the small radius will often prevent matching: a desirable outcome since the photometry of unresolved blends would be inaccurate, unstable with respect to changes in the FHWM, and unsuitable for variable star analysis. Measurements of very faint isolated stars near our detection limit will occasionally fail to match due to random astrometric error, but this is an acceptable loss.

\begin{figure}
\includegraphics[scale=0.75]{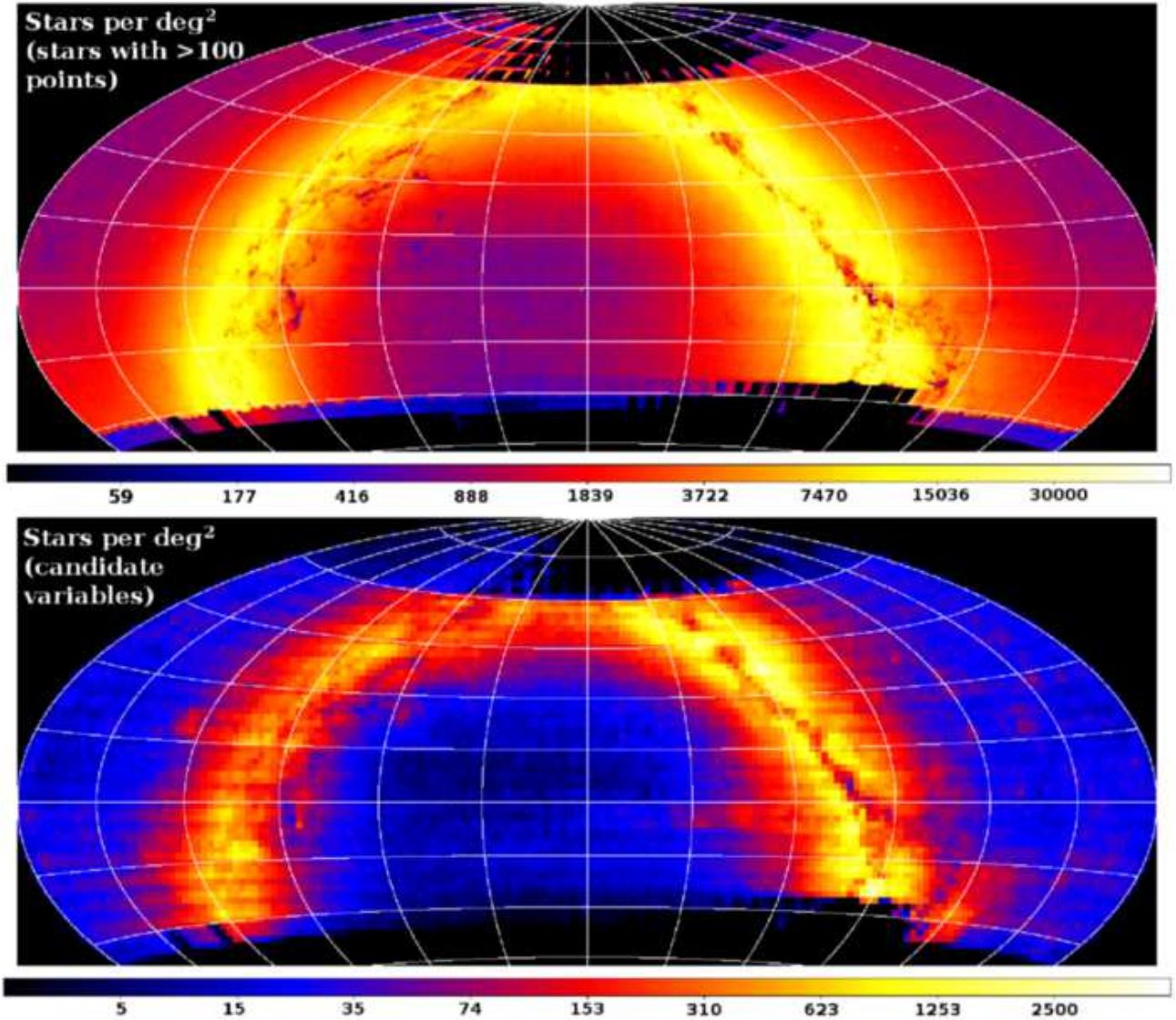}
\caption{\textbf{Top:} Density of well-measured ATLAS stars (the lightcurve set) over the whole sky, in units of stars/deg$^2$. These stars extend down to about $c$ magnitude 19 or $o$ magnitude 18.5, with brighter effective limits in crowded regions. \textbf{Bottom:} Density of candidate variable stars, in the same units. Grid lines are spaced at 30 degree intervals in RA and 15 degree intervals in Dec, with 0,0 in the center of the plot. Except for a narrow, southerly band near the Galactic center, there are no significant gaps in coverage between Dec -30$^{\circ}$ and +60$^{\circ}$. Uneven observations outside this Dec range enabled the discovery of some additional variables, but at much lower completeness.
\label{fig:allstars}}
\end{figure}

\begin{figure}
\includegraphics[scale=0.57]{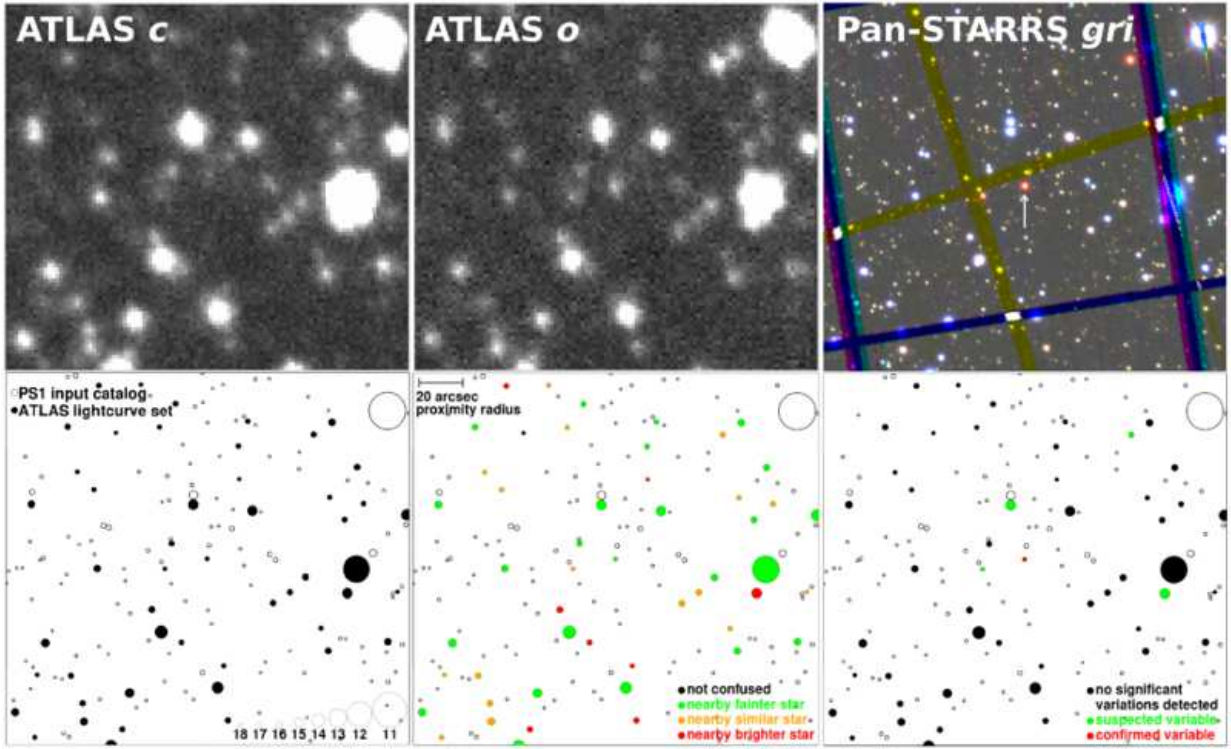}
\caption{A dense Galactic-plane starfield centered on ATO J296.1011+19.9265, a new Mira variable. Panels are 3 arcmin square. \textbf{Top left, center:} ATLAS $c$ and $o$-band single images (pixel scale 1.86 arcsec). \textbf{Top right:} color image made from single $g$, $r$, and $i$-band images from Pan-STARRS1 (pixel scale 0.25 arcsec). \textbf{Bottom left:} Stars in our Pan-STARRS based object-matching catalog (symbol size gives $r$ mag). Solid symbols identify stars measured at least 100 times by ATLAS and hence included in our lightcurve set. Stars could fail this criterion by being too faint, too confused, or too bright (saturating in ATLAS images). \textbf{Bottom center:} Same chart with stars in the lightcurve set color-coded with ATLAS confusion flags. In a dense field such as this, almost all stars are potentially confused, raising the bar for identification as confirmed variables. \textbf{Bottom right:} Same chart with stars in the lightcurve set color-coded as suspected (green) and confirmed (red) variables. ATO J296.1011+19.9265 itself is the only example of the latter in this field.
\label{fig:imexamp}}
\end{figure}

\begin{figure}
\includegraphics[scale=0.8]{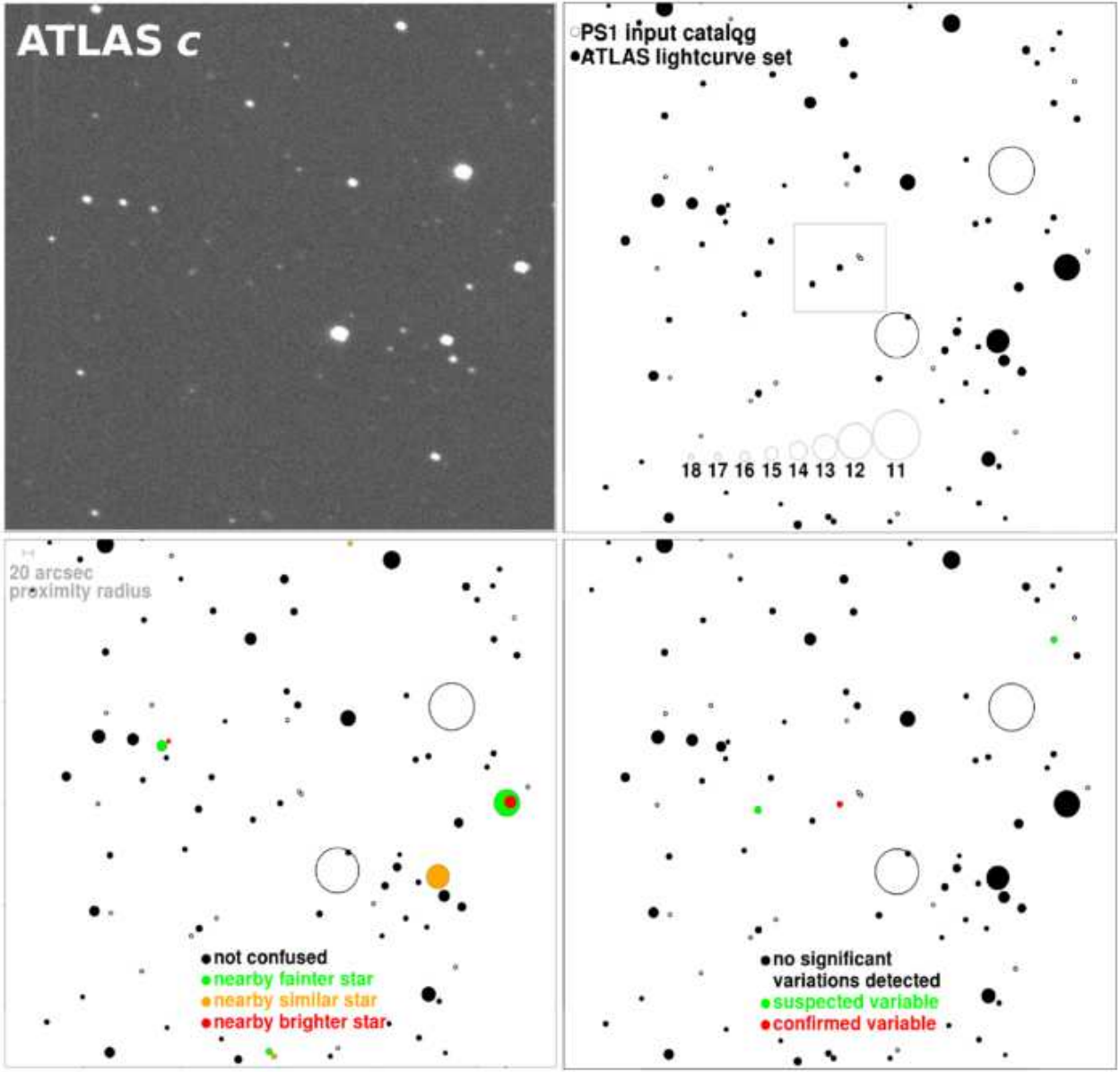}
\caption{Similar to Figure \ref{fig:imexamp}, but showing a much wider (18 arcmin square) view of an uncrowded starfield far from the Galactic plane, centered on known RRab variable CSS\_J133208.4+213245. \textbf{Top left:} ATLAS $c$-band single image. \textbf{Top right:} Stars in our Pan-STARRS based object-matching catalog. Solid symbols identify those in the ATLAS lightcurve set. The gray square at center shows the angular size of Figure \ref{fig:imexamp}, emphasizing the difference in star density. \textbf{Bottom left:} Same chart  with stars in the lightcurve set color-coded with ATLAS confusion flags. In contrast to the dense field of Figure \ref{fig:imexamp}, most stars here are not confused. The bright orange-coded star at lower right is flagged as confused because it is an equal-brightness double, resolved by Pan-STARRS but not ATLAS. \textbf{Bottom right:} Same chart with stars in the lightcurve set color-coded as suspected (green) and confirmed (red) variables. The known RR Lyrae star at center is the only example of the latter in this field.
\label{fig:imexamp2}}
\end{figure}

\begin{figure}
\plottwo{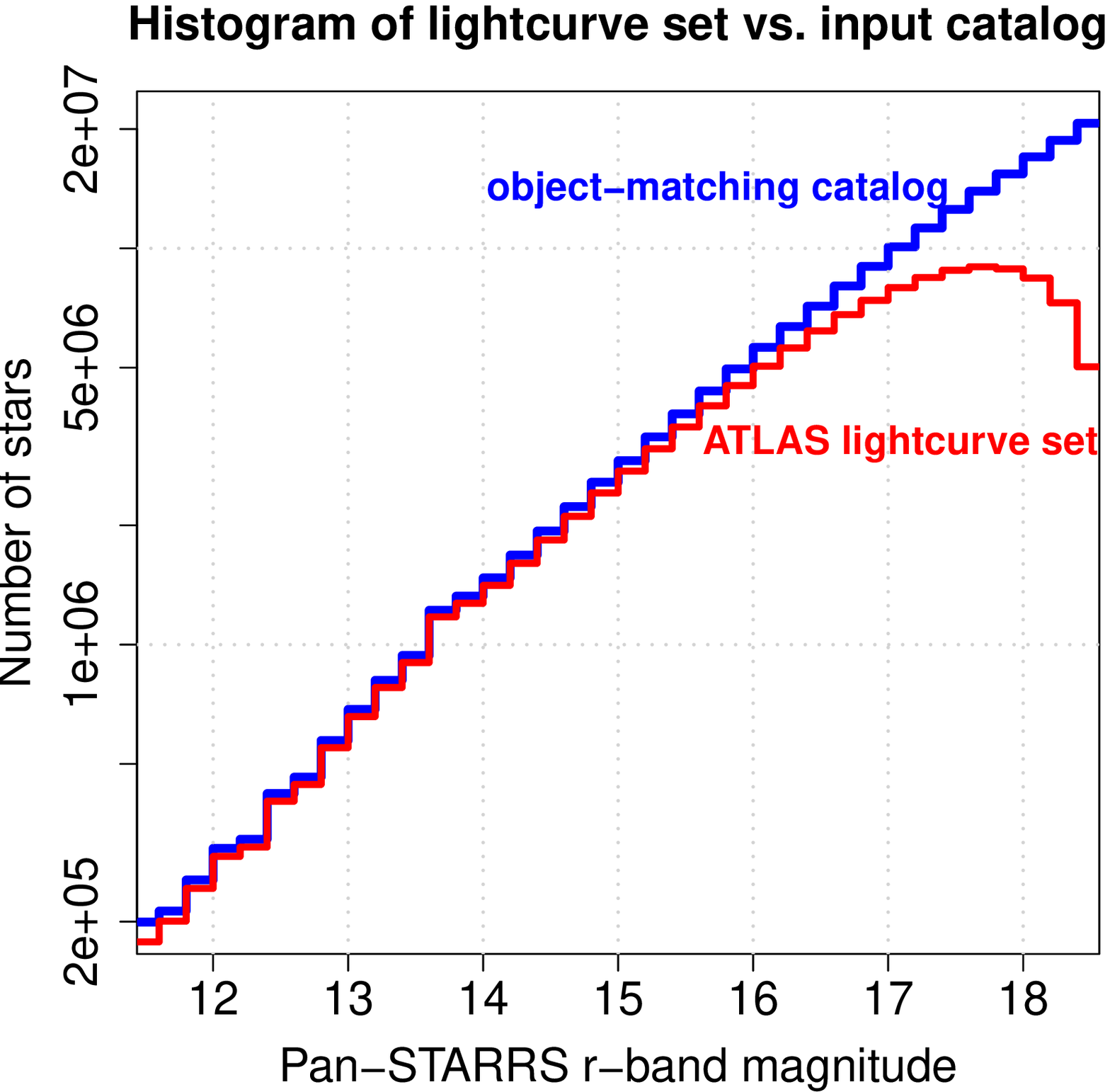}{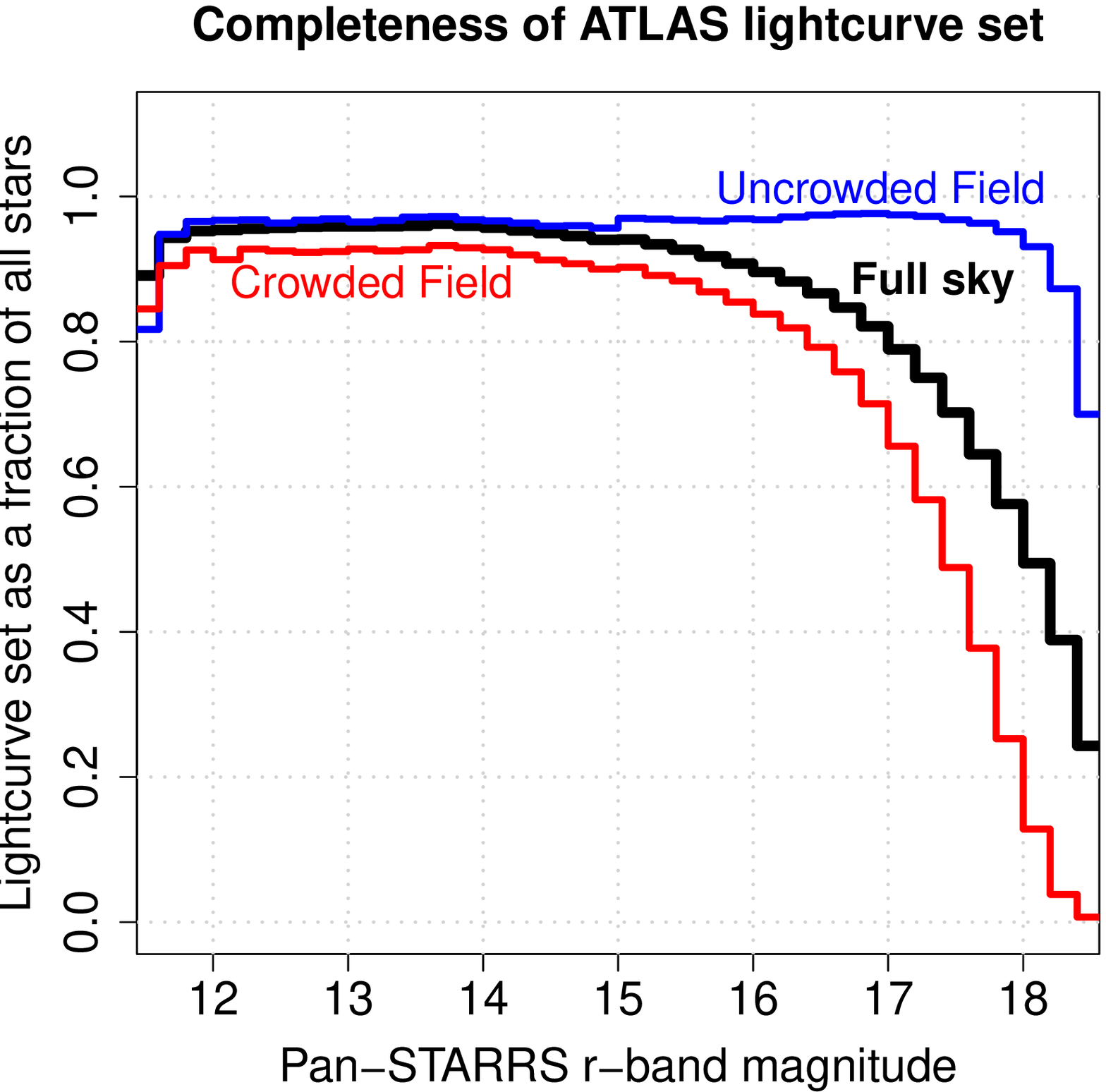}
\caption{\textbf{Left:} Magnitude histograms for our Pan-STARRS based object-matching catalog and for the lightcurve set. \textbf{Right:} Fraction of all stars in the object-matching catalog that were measured at least 100 times by ATLAS and hence included in the lightcurve set. As expected, faint stars were less likely to make it into the lightcurve set in a crowded field, because blending prevented ATLAS from obtaining good measurements of them.
\label{fig:meashist}}
\end{figure}

To avoid expending effort on stars with insufficient data for useful characterization, we confine our current analysis to stars for which ATLAS has at least 100 photometric measurements. Since most areas of the sky have been covered more than 200 times, this is not extremely restrictive, but stars that are so faint (or so confused with nearby neighbors) that they are detected and matched with less than 50\% probability will not be included in the current catalog. We find that ATLAS has more than 100 measurements for 142 million out of the 302 million stars in the object-matching catalog. As stated above, we refer to this subset of 142 million stars as the lightcurve set. The stars in the object-matching catalog that did not make it into the lightcurve set must, by construction, have been photometrically measured by ATLAS fewer than 100 times during the period covered by DR1. This could be because they are outside the Dec range of good coverage; fainter than 18th mag in the ATLAS bands; or located in crowded fields where they form unresolved blends with other objects. Figures \ref{fig:imexamp} and \ref{fig:imexamp2} provide example images and star charts of crowded and uncrowded fields, respectively, showing which stars in the object-matching catalog made it into the lightcurve set in each case. Figure \ref{fig:allstars} shows the distribution of the ATLAS lightcurve set on the sky, while Figure \ref{fig:meashist} shows the magnitude-dependent completeness of the lightcurve set (as a fraction of the object-matching catalog) for uncrowded fields, crowded fields, and averaged over the sky. The lightcurve set has well over 90\% completeness from $r$ mag 12 to 18 in uncrowded fields, while severe crowding (e.g. Figure \ref{fig:imexamp}) brings the completeness below 90\% at $r=15.5$ mag and 50\% at $r=17.5$ mag.

The median number of measurements per analyzed star is 208, while the mean is 213.3. The total number of photometric measurements we analyze herein is therefore 213 $\times$ 142 million, or about 30 billion measurements. Since about twice this many measurements were obtained, roughly half of them must have corresponded to stars too faint or confused to accumulate 100 measurements, or to transients/artifacts. In DR2 we plan to extend our analysis to some of these hard-to-measure stars, likely using our difference images (which play only a minor role herein) to overcome the confusion limit in crowded fields, as has been done so effectively by the OGLE project \citep[e.g.][]{Alard1998}.

\subsection{Selecting Candidate Variables}
We begin our variability analysis with photometric time series (i.e. lightcurves) for all stars in the ATLAS lightcurve set. Each lightcurve comprises at least 100 photometric data points. Following \citet{heatherAAS} and \citet{Drake2013a,Drake2014a}, we calculate the Lomb-Scargle periodogram \citep{Lomb1976,Scargle1982} of the lightcurve for every star, and use the output false alarm probability (FAP) for each star as our initial screening for variability. The Lomb-Scargle periodogram is more computationally intensive than traditional means of identifying variables (e.g. the Stetson indices), but it is much more sensitive to low-amplitude periodic variables, and the analysis is entirely tractable with modern facilities. A Lomb-Scargle periodogram can also sensitively detect variability that is not strictly periodic, as long as it has some type of coherent behavior with time. 

This initial Lomb-Scargle periodogram is carried out by a customized program called {\tt lombscar}. This program is based on the code of \citet{lombscar}, but is enhanced to do all calculation in double precision and to accept a vector of photometric uncertainties and perform a weighted analysis. The nominal processing carried out by {\tt lombscar} is to read the time (applying a light travel-time correction to translate the times into Heliocentric Julian Days), magnitude, mag error, and filter for each measurement of a given star, and then perform two iterations of fitting.  It initializes with a fit to the light curve that consists of a constant brightness equal to the median magnitude (in each filter). All magnitude uncertainties are softened by addition of 0.03 mag in quadrature. The benefit of this softening is to reduce the impact of rare points with large systematic errors, while the cost is a reduction in the statistical power of good points with very low photometric uncertainties. The softening parameter of 0.03 mag was chosen as small enough not to hamper the period search, but large enough to significantly reduce the effect of the occasional systematic error.

For each iteration, {\tt lombscar} does the following:

\begin{itemize}
 \item Prunes ``bad'' photometric data points, which either:
   \begin{itemize}
     \item have a photometric uncertainty that is bigger than the larger of 0.3 mag or 2 times the upper quartile of the photometric uncertainties of data points for this star, or
     \item have a residual with respect to the current fit which is greater than 0.8 mag (first iteration) or 0.4 mag (second iteration), or
     \item have a residual with respect to the current fit which is greater than 30 sigma (first iteration) or 15 sigma (second iteration).
     \end{itemize}
 \item Performs a quadratic polynomial fit to the light curve minus the current Fourier fit, including separate constant terms for each filter, but all filters sharing the same time behavior.
 \item Calculates a Lomb-Scargle periodogram of the light curve after subtraction of the polynomial fit, using HIFAC=100 and OFAC=4 (parameters explained below). 
 \item Re-scales the frequency axis of the Lomb-Scargle periodogram, doubling all periods and halving all frequencies, in order to fit eclipsing binaries correctly.
 \item Does a Fourier fit of the data for every frequency $f_p$ in the periodogram that has a probability at least 85\% as large as the highest probability. Also does a Fourier fit at $2f_p$ and $3f_p$, thereby including the base frequency output by the Lomb-Scargle analysis, since it is equal to $2f_p$ for the highest peak. All of these Fourier fits use a frequency sampling 5$\times$ finer than that of the periodogram, and report a $\chi^2/N$ that rejects the worst 10\% of points.  The very best $\chi^2/N$ from all of these fits is deemed to indicate the correct period for this iteration.
 \item At the conclusion of the iteration, computes Fourier fits at aliased frequencies of +/-0.5 day$^{-1}$ and +/-1.0 day$^{-1}$.
\end{itemize}

The sampling factors `HIFAC' and `OFAC' \citep[see, e.g., the discussion in][]{nrc} with which a Lomb-Scargle periodogram is run are important for determining the range of variables to which it is sensitive. The `oversampling' factor OFAC determines how fine the sampling is in frequency space, such that the maximum phase error is approximately 1/OFAC. By viewing plots of the periodogram, one can easily evaluate whether OFAC is large enough to capture all the structure, and adjust if necessary. The HIFAC parameter determines the maximum detectable frequency. For a data set with total temporal span of $T$ and number of data points $N$, this maximum frequency is $\mathrm{HIFAC} N/2T$ \citep{nrc}. HIFAC may therefore be interpreted as the factor by which the highest frequency probed exceeds the Nyquist limit that would apply {\em if the measurements were equally spaced in time}. In the case of unevenly spaced data such as ours, the Lomb-Scargle periodogram can accurately measure frequencies many times higher than the equally-spaced Nyquist limit \citep{nrc}. Our data have a median temporal span of about 620 days, while the median value of $N$ is 208. The maximum frequency with HIFAC=100 is therefore typically 16.8 cycles/day, corresponding to a period of 0.06 days or 1.4 hours. Eclipsing binaries and pulsating objects such as RR Lyrae stars and many $\delta$ Scuti stars have periods longer than this; however, some $\delta$ Scuti objects, subdwarf B stars, and pulsating white dwarfs have periods too short for detection in our current analysis. These objects are rare and often have amplitudes too small ($<<0.02$ mag) for ATLAS to detect anyway. Since the runtime of the Lomb-Scargle periodogram increases linearly with HIFAC, probing down to a period of, e.g, 0.5 hours would almost triple the computational cost. For the present, we have elected not to make such a large investment to obtain a small increase in variable discoveries. We will probe shorter periods, at least for a subset of the brightest stars, in DR2.

The outlier-clipping applied by {\tt lombscar}, as well as its subtraction of the best-fit 2nd-order polynomial from the original time series, are intended to remove bad points and systematic trends and hence to increase the detectability of variable stars with periods shorter than the temporal span of our data. They can, however, decrease our sensitivity to long-period variables and very high-amplitude variables. Since most of our data do not appear to suffer from significant long term systematics, we have the potential to be very sensitive to long-period variables, and we have taken steps to recover this sensitivity as detailed below in \S \ref{sec:selection}.

The most significant variables are those with the smallest FAP values output by the Lomb-Scargle analysis, and these probabilities range down to extremely small values (e.g. $<10^{-60}$). Hence, we adopt $-$log10(FAP) as our primary measure of the strength of a variability detection. For convenience, we will refer to $-$log10(FAP) as PPFAP, meaning `Power of the Periodogram False Alarm Probability'. Besides PPFAP, we record 31 additional statistics output by {\tt lombscar}. These include the number of points (total and post-clipping); the median magnitudes; the coefficients of the initial polynomial; the period identified by the Lomb-Scargle analysis; the coefficients and reduced $\chi^2$ value of the Fourier fit; and others described in \S \ref{sec:appendixA} below.

\subsubsection{Variable Features}
We augment these 32 statistics from {\tt lombscar} by calculating, for each star in the lightcurve set, a set of 22 additional statistics intended to be sensitive to non-periodic as well as periodic variability, using a program called {\tt varfeat}. Calculated features include the 5th, 10th, 25th, 75th, 90th, and 95th percentile magnitudes; a statistic we call Hday that probes the median nightly $\chi^2$ value to identify significant variability on a timescale shorter than a night; and a statistic we call Hlong that probes night-to-night variability relative to the intra-night scatter. The {\tt varfeat} analysis also includes many statistics described in \citet{Sokolovsky2017}: the weighted standard deviation; inter-quartile range; $\chi^2/N$ for a constant-brightness model; robust median statistic; normalized excess variance; normalized peak-to-peak amplitude; inverse von Neumann ratio; Welch-Stetson I; Stetson J; and Stetson K. These are described in more detail in \S \ref{sec:appendixA}.

\subsubsection{Final Selection of Candidates} \label{sec:selection}
We wish to select a subset of the ATLAS lightcurve stars for more intensive variability analysis that would be computationally intractable as applied to the full lightcurve set. We do this in three stages.

First, we select the stars that appear to be strongly variable based on the initial analysis with {\tt lombscar}. For these, we adopt a threshold of PPFAP$=$10.0, corresponding to a formal false alarm probability of $10^{-10}$.  We also add all stars from AAVSO Variable Star Index \citep[VSX;][downloaded as of November 2017]{VSX} for which we have at least 100 measurements. The number of stars in the union of strong {\tt lombscar} variables with known VSX stars is 1.1 million, or 0.77\% of the lightcurve set. VSX stars that would not have been independently included make only a small ($\sim 7$\%) contribution to the total of 1.1 million candidates identified at this stage.

%1093664 = 1.1 million

Next, to avoid excluding stars with low-amplitude variability, or objects whose variability was suppressed by the outlier-clipping or polynomial subtraction applied by {\tt lombscar}, we select stars with weaker Lomb-Scargle variability detections, having PPFAP between 5.0 and 10.0. This adds 2.4 million stars (1.67\% of the lightcurve set) to our list of candidate variables. 

%2370116 = 2.4 million

Finally, to catch any additional variables that may have been missed by {\tt lombscar}, we use the {\tt varfeat} analysis to select a set of potentially interesting stars that all have PPFAP less than 5.0. To determine which {\tt varfeat} outputs are most useful for selecting candidate variables, we make use of the fact that all of the {\tt varfeat} statistics are expected to be capable of detecting periodic as well as unperiodic variability. Thus, we can examine their degree of correlation with Lomb-Scargle PPFAP to identify those that are most sensitive to generic variability. We do this by calculating the 90th percentile envelope of PPFAP as a function of each of the {\tt varfeat} statistics. The most useful statistics are those for which the envelope reaches the highest values while still in a regime populated with a significant number of stars. We find that the best ones are $\chi^2/N$; the Robust Median Statistic; the Inverse von Neumann ratio; the Welch-Stetson I and Stetson J indices \citep[all described in][]{Sokolovsky2017}; the two that we invented to probe inter- and intra-night variability (Hday and Hlong, see \S \ref{sec:appendixA}); and the inter-quartile range \citep{Sokolovsky2017}. For all of these except the inter-quartile range, there is a value for which the 90th percentile envelope of PPFAP rises above 20.0, corresponding to a nominal false alarm probability of $10^{-20}$. We choose thresholds for each statistic that correspond to envelope values between 10 and 20. These thresholds are 2.5 for the Robust Median Statistic; 1.4 for the Inverse von Neumann ratio; 8.0 for Welch-Stetson I; 6.0 for Stetson J; 7 for Hday, and 20 for Hlong. We combined all these criteria with a logical OR, and thus identified 1.3 million potentially interesting stars (0.90\% of the lightcurve set) with PPFAP values of less than 5.0 in the initial screening with {\tt lombscar}.

%1270435 = 1.3 million, 

The total number of candidate variables identified by these three selections is 4.7 million, or 3.34\% of the lightcurve set (1.6\% of the object-matching catalog). The bottom panel of Figure \ref{fig:allstars} shows the distribution of these candidate variables on the sky.

%4734215 = 4.7 million

\subsection{Fourier Fitting}

We characterize each of our candidate variables with a program called {\tt fourierperiod}, which performs a sophisticated Fourier analysis aimed at resolving any period aliases and probing the lightcurve morphology in detail. We begin this analysis with another Lomb-Scargle periodogram, which differs from the initial one in three ways. First, there is no pre-subtraction of a polynomial fit. Second, OFAC=20 is used rather than OFAC=4, ensuring finer sampling of the periods. Third, the outlier-clipping is less aggressive. We reject all points with nominal uncertainties greater than 0.2 mag, corresponding to detections with less than 5$\sigma$ significance. We calculate a Lomb-Scargle periodogram without any additional clipping. However, since surviving outliers can sometimes distort a truly periodic signal and greatly reduce the value of PPFAP, we also perform three iterations of 3$\sigma$-clipping relative to a constant model, and then recalculate the periodogram of the clipped data. Whichever data set (unclipped or clipped) produces the strongest variability detection (highest value of PPFAP) is retained for further analysis. Note that the value of $\sigma$ used in the sigma-clipping is a simpleminded RMS scatter around the median in each filter, and hence will be elevated by the star's own variability. This makes the clipping very conservative and ensures that, for example, no points from a pure sinusoid would be rejected regardless of its amplitude.

At each period $P$, {\tt fourierperiod} subtracts the median magnitude in each filter and then fits the data with a truncated Fourier series of the form:

\begin{equation} \label{eq:fourmod}
\mathrm{mag}(t) = C_0 + \sum_{m=1}^n a_m \sin \left( m \cdot \frac{2 \pi t}{P} \right)+ b_m \cos \left( m \cdot \frac{2 \pi t}{P} \right).
\end{equation}

Where $C_0$ is a constant term, allowed to be different for each filter. We scan through a finely-sampled range of values for the master period $P$, selecting the optimal order $n$ of the Fourier fit as described below.

The analysis defaults to the assumption that every star is a long-period variable. The reasons are, first, that long-period variability can be aliased to short periods in the Lomb-Scargle analysis (so in general it isn't safe to assume that a high-frequency periodogram peak means a short period), and second, a search for long-period variability is computationally cheap because only a relatively small number of periods must be probed. Therefore we begin by probing periods from 5 to 1500 days. At each period $P$, we calculate a sampling step $\Delta P$ based on a maximum phase error $\phi_{err}$:

\begin{equation} \label{eq:phaseerr}
\Delta P = 2 \phi_{err} P^2/T
\end{equation}

Where $T$ is the temporal span of the data, as before. We set $\phi_{err}$ to 0.025: thus, whatever the actual period of the star, we will fit some period $P$ such that no point is incorrectly phased by more than 0.025 cycles. Note that this is approximately equivalent to OFAC=40 in a Lomb-Scargle analysis. The $P^2$ dependence of the period sampling interval illustrates why probing long periods is cheap.

We begin by fitting a pure sinusoid ($n=1$ in Equation \ref{eq:fourmod}) at every period $P$ from 5 to 1500 days, with the spacing between successive values of $P$ dictated by Equation \ref{eq:phaseerr}. We identify the period producing the best fit based on the $\chi^2$ value, and then evaluate the remaining signal by taking the Lomb-Scargle periodogram of the residuals. If PPFAP for the residuals is greater than 4.0, we add another Fourier term and scan all the periods again. Since we have two Fourier terms now, the lightcurve could be more complex and a phasing error correspondingly more serious: thus, we reduce $\phi_{err}$ by a factor of two relative to its initial value of 0.025. If the residuals from the two-term Fourier fit still have PPFAP greater than 4.0, we add a third term and reduce $\phi_{err}$ to 1/3 its initial value. We proceed until we reach a maximum number $n$ of Fourier terms. For the long-period analysis, we use a maximum of 4 Fourier terms. Since long-period variables (e.g. Mira stars) often have very different amplitudes in our different ATLAS filters, the Fourier coefficients $a_m$ and $b_m$ are allowed to be different for each filter, although the master period $P$ has to be the same.

If the periodogram of the residuals still shows PPFAP$>$4.0 after the subtraction of a 4-term Fourier fit, we conclude that the long period analysis did not find a satisfactory fit and we proceed to the short-term analysis. Here, in the interest of computational tractability, we do not probe every possible period in a wide range. Instead, we probe a set of narrow ranges based on the initial Lomb-Scargle period, intended to include all plausible values for the true period. As in the long-period fit, we start with a pure sinusoid and add additional terms, but the maximum is now $n=6$, and the criterion for a good fit is more strict: residual PPFAP$<$2.0 rather than $<$4.0. Also, since short-period variables usually don't have huge differences in amplitude and lightcurve shape between the cyan and orange filters, the Fourier coefficients $a_m$ and $b_m$ are required to be the same for both filters, although each filter still gets its own constant term $C_0$. Where the amplitude and/or the shape of the lightcurve is somewhat different in the $c$ vs. the $o$ band, the fit finds an approximate average lightcurve and no serious error results.

For a fit with $n$ Fourier terms, we probe base periods $P_f$ that are 1, 2, 3...$n$ times longer than the Lomb-Scargle output period $P_0$. For each base period, we probe the aliases of the Earth's sidereal day, so the full set of trial periods $P_{f,j}$ that we probe is given by:

\begin{equation} \label{eq:alias}
P_{f,j} = \frac{t_{sid}}{t_{sid}/(f P_0) + j}
\end{equation}

Where $t_{sid} = 0.99726957$ days is the sidereal rotation period; the alias index $j$ is allowed to take on values of -3, -2, -1, -0.5, 0, 0.5, 1, 2, and 3; and $f$ is an integer ranging from 1 to the number $n$ of Fourier terms being used in the fit. If the right hand side of Equation \ref{eq:alias} turns out to be negative, we simply take its absolute value. We note that such `negative aliases' are mathematically legitimate, and that they have the initially bewildering effect of time-reversing the folded lightcurve. For example, a pulsating star with a nominal period of 2.45433 days could be exhibiting the $j=-2$ alias of a true period of 0.625769 days, even though the left hand side of Equation \ref{eq:alias} becomes negative if we plug in $f=1$, $P_0=2.45433$ days, and $j=-2$. In this case, the lightcurve folded at the nominal period of 2.45433 days will show a slow brightening and then a rapid fading rather than the classic `sawtooth' lightcurve with its rapid rise and slow fall. Re-folding the data with the correct period of 0.625769 days will correct the time-reversal and recover the familiar sawtooth in its normal orientation.

We note that Equation \ref{eq:alias} probes both aliases and multiples of the initial Lomb-Scargle period, as it should since eclipsing binaries, multi-mode pulsators, and other objects often have true periods that are a multiple of the period corresponding to the dominant frequency that will be identified by Lomb-Scargle. Specifically, Equation \ref{eq:alias} probes \textit{aliases} of \textit{multiples} of the nominal period: it does the period multiplication first and then calculates the aliases. The reverse procedure, probing \textit{multiples} of \textit{aliased} periods, is almost certainly more realistic in terms of the actual aliasing that occurs in a Lomb-Scargle analysis. This would produce:

\begin{equation} \label{eq:mult}
P_{j,f} = f \frac{t_{sid}}{t_{sid}/P_0 + j}
\end{equation}

The sets of periods produced by Equations \ref{eq:alias} and \ref{eq:mult} are not entirely identical, and we have used Equation \ref{eq:alias} herein only because we did not realise its sub-optimal characteristics until the computation was substantially complete. The errors incurred thereby are not likely to be significant: all but the rarest types of period ambiguity would be covered by both equations --- especially since we include half-integer aliases in our application of Equation \ref{eq:alias}. We will use Equation \ref{eq:mult} for DR2.

Around each value of $P_{f,j}$ given by  Equation \ref{eq:alias}, we search a narrow range in period that corresponds to $\pm2$ cycles over the whole temporal span $T$ (except in the un-aliased case $j=0$, when we search a wider range corresponding to $\pm6$ cycles). In each case, the period sampling is given by Equation \ref{eq:phaseerr}, and the maximum phase error $\phi_{err}$ is set to 0.025 divided by the number $n$ of Fourier terms being fit.

When the best-fit period (based on the minimum $\chi^2$ criterion) has been identified for a given number $n$ of Fourier terms, the periodogram FAP of the residuals from this optimal fit is calculated. If PPFAP$<$2 for the residuals, the fit is considered to have captured all the variability and fitting stops. Otherwise, another Fourier term is added and the period search begins again, unless the maximum number $n=6$ of Fourier terms has already been reached.

Note that the Fourier fitting rapidly becomes more computationally expensive as additional terms are added in the short-period fit. In the last iteration, with 6 Fourier terms, six different values of $f$ are explored; for each of which we probe the usual 9 different values of the alias $j$, making 54 different period ranges in all. The ranges also are required to be more finely sampled, since $\phi_{err}$ has been reduced by a factor of 6 relative to its initial value of 0.025. 

The respective FAP thresholds and maximum numbers of Fourier terms for the long and short period fits are sensitive and important parameters, and we arrived at the current values to optimize results after considerable experimentation. The maximum number $n=6$ of Fourier terms that can be used in the short period fit is an optimum because it usually produces very good fits to eclipsing binaries and pulsating stars, but yet is low enough that the computation does not become intractable. For the long-period fit, we found that allowing more than four Fourier terms could sometimes enable a formally acceptable long-period fit even to a strong and obvious short-period object --- e.g. an RR Lyrae star vulnerable to aliasing because of having a period near 0.5 sidereal days. Such cases are extremely problematic because then the Fourier code does not even attempt the short-period fit that would yield the correct solution. On the other hand, giving the long-period fit an insufficient number of Fourier terms (or a too-tight threshold in terms of the acceptable FAP) results in much time being wasted in futile attempts to obtain short-period fits to long-period variables.

We note that here (and throughout the current paper) we focus on the time domain rather than the frequency domain. Our intent with the Fourier series is to find a periodic function that fits the data, not to analyze the frequency content of the signal. The terms of the Fourier series have fixed frequencies $1/P$, $2/P$, $3/P$, etc, dictated by the master period $P$ that is being explored. Thus, we are {\em not} performing a CLEAN algorithm-like subtraction of successive best-fit sine waves at arbitrary frequencies until the residuals are consistent with random noise. The latter type of analysis is required, e.g., for detailed characterization of stars that pulsate with multiple periods --- while our aim at present is simply a very generalized characterization of variability that will identify stars worthy of further study. We suspect the ATLAS data would support sophisticated frequency analyses of many stars, and as we are making our photometric data public, we hope the current paper will serve to guide other researchers toward promising objects of study.

Our Fourier analysis code calculates and saves 92 different statistics, which are described in detail in \S \ref{sec:appendixA}. These include the period and PPFAP of the initial periodogram; the numbers of points used for the final analysis; the original RMS scatter of the data from the mean (overall and in each filter); the master period adopted in the long period fit; the residual RMS and $\chi^2$ for this fit; the number of Fourier terms used; the minimum and maximum fitted brightness (confined to times where the fit is constrained by the data); the constant terms in the final Fourier fits; the sine and cosine coefficients for each Fourier term in each filter; the residual PPFAP after subtracting each successive Fourier fit; the Fourier index of the term that has the most power; analogous quantities for the short-period fit, if applicable, including the specifications on the aliasing and period-multiplication of the final adopted period relative to the initial Lomb Scargle output; and two statistics measuring the degree of invariance of the short-period Fourier fit under time-reversal and 180-degree phase-shifting, respectively.

Our (rather arbitrary) definition of short-period is $P<5.0$ days, and applies to the highest-frequency Fourier term. Thus, the shortest master period that counts as `long' in our analysis is 5 days for a pure sinusoid and 10, 15, and 20 days for fits with 2, 3, and 4 fourier terms respectively. If the long-period analysis finds a satisfactory fit (which will necessarily have a period at least as long as these respective values), no short-period fit will be attempted. If the best long-period fit is {\em not} satisfactory, a short period fit can (and will) be performed even if the period found by the initial periodogram is long. This is true because any possible input period will have aliases shorter than 5 days for some value of the alias parameter $j$.

The limit of 5.0 days for the highest-frequency Fourier term applies to the short periods as well, so that the \textit{longest} master period that counts as short is 5.0 days for a pure sinusoid but can be as long as 30 days if 6 Fourier terms are used in the fit. Thus there is some potential overlap in the regimes probed by the long- and short-period fits. Note, however, that the short-period fit is performed only if the long-period fit did not find an acceptable solution, defined as a fit with residual PPFAP$<4.0$.

\subsection{Statistics from Difference Imaging}
In order to detect asteroids, all ATLAS images are `differenced' by the subtraction of a static sky template produced from earlier ATLAS data. Both the original and difference images are saved, and our variable star analysis thus far is based on the former. However, the difference images could be very useful in identifying variable stars, especially doubtful cases. 

Hence, we wrote a program to calculate 19 potentially relevant statistics from the difference images for each candidate variable star (15 of which turn out to be sufficiently useful for variable identification that we release them publicly and list them in \S \ref{sec:appendixA}). These are {\em not} based on re-accessing the pixels of the difference image (e.g. by doing forced photometry at the locations of suspected variable stars). Rather, they are based on existing detection catalogs (`ddc files') automatically produced from the each difference image for purposes of asteroid detection. Besides basic astrometry and photometry, the ddc files present a concise yet sophisticated list of analytics for each detection, all aimed at distinguishing between various types of real objects and spurious detections. These analytics are critical to ATLAS' primary mission of asteroid discovery, and hence are hightly evolved and optimized. Many of them are produced by an image analysis program called {\tt vartest} that supports ATLAS asteroid discovery by automatically performing a pixel-based analysis to classify detected objects in the difference images and rule out false positives. For each detection in a difference image, {\tt vartest} assigns the probability that it is a noise fluctuation (Pno); a cosmic ray (Pcr); an electronic artifact (Pbn, Pxt); a star-subtraction residual (Psc); a bona fide asteroid or transient (Ptr) --- or a variable star (Pvr). To identify possible variable stars, {\tt vartest} uses astrometric consistency between the original and difference images; unusual levels of residual flux; and a bias away from zero in the statistics of nearby pixels (which should have mean zero if the detection is a subtraction residual from a non-varying star). All of these are synthesized into a single value, (Pvr), which is an integer ranging from 999 (certainly a variable star) to 0 (certainly something else).

The 19 statistics we calculate from the ddc files include the number of times there was any detection corresponding to the star's position; the median magnitude and SNR of such detections; the median $\chi^2/N$ of the PSF-fit; and several more statistics based on the {\tt vartest} probabilities. The most useful of the calculated statistics turn out to be the number of detections, and the median and rank 2 values of Pvr from {\tt vartest}. We identify thresholds on these statistics that are able to select a set of stars with median PPFAP$>$10 in the {\tt lombscar} analysis. The significance of this is that the ddc statistics are entirely independent of the {\tt lombscar} results and hence can provide an independent confirmation of variablity. The required thresholds on the ddc statistics are hard to meet: most stars, variable and not, don't pass the test. Of randomly selected stars regardless of variability, only 0.09\% meet the criteria. We had to adopt such strict thresholds to meet the requirement of median PPFAP$>$10, in order to reasonably claim that a star only tentatively identified as variable can, if it passes, be declared variable with some confidence. For stars meeting these demanding criteria, we assign a value ddcSTAT=1, indicating that the statistics from the difference images provide strong evidence of genuine variability independent of other considerations. All other stars are assigned ddcSTAT=0.

\subsection{Stellar Proximity Statistics} \label{sec:proxSTAT}

Due to its hierarchical approach --- detecting and subtracting away the brightest objects, prior to attempting to measure fainter ones --- {\tt DoPhot} is able to extract good photometry even from dense starfields where some of the stellar images overlap and are confused. Where the PSF changes over time, however, the total number of stars detected in a confused field may change: on the blurrier images, some stars will blend together and be measured as one that were identified as distinct objects in sharper frames. This change in the number of detected stars can also affect the photometry.

\begin{figure}
\plottwo{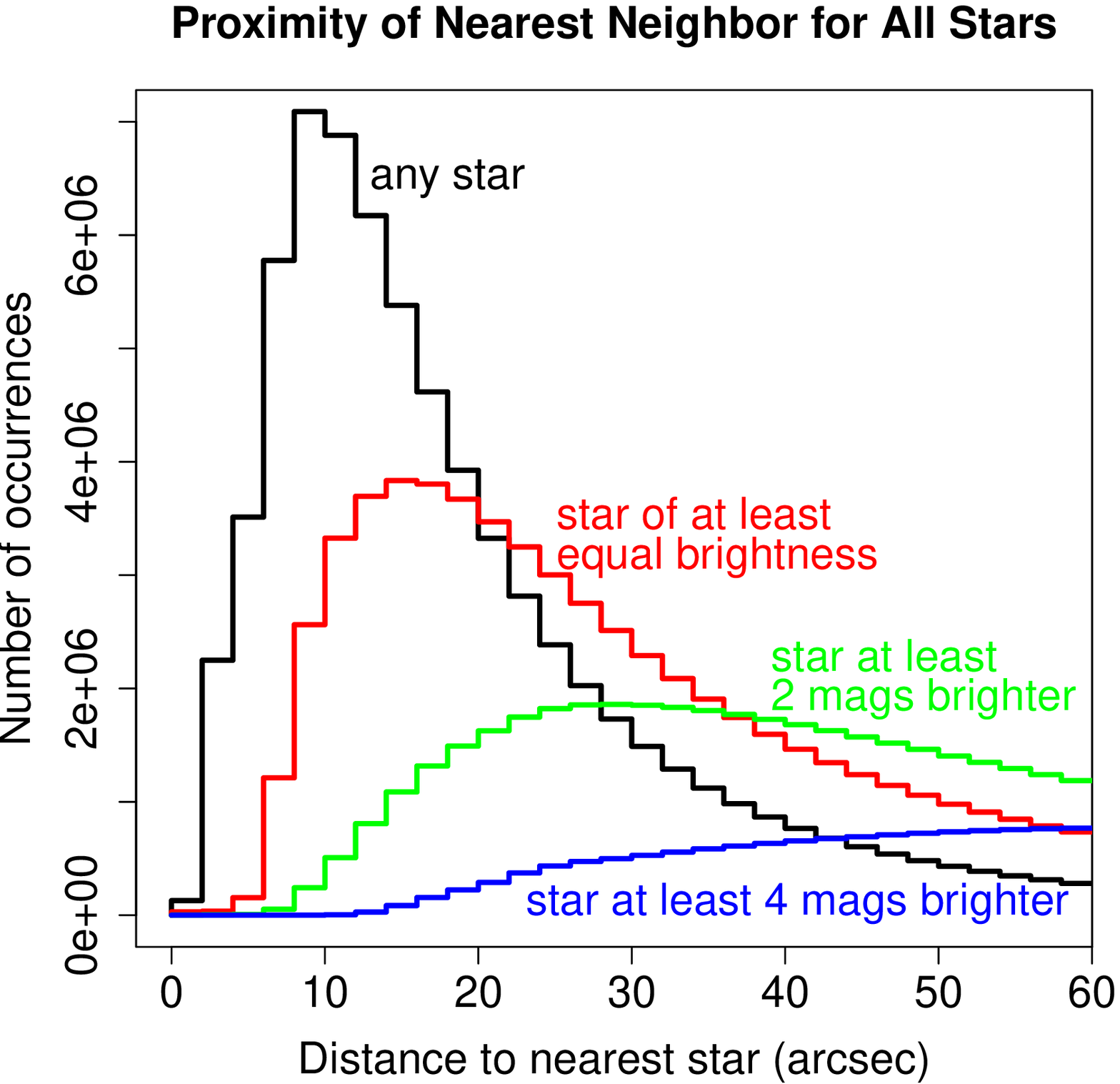}{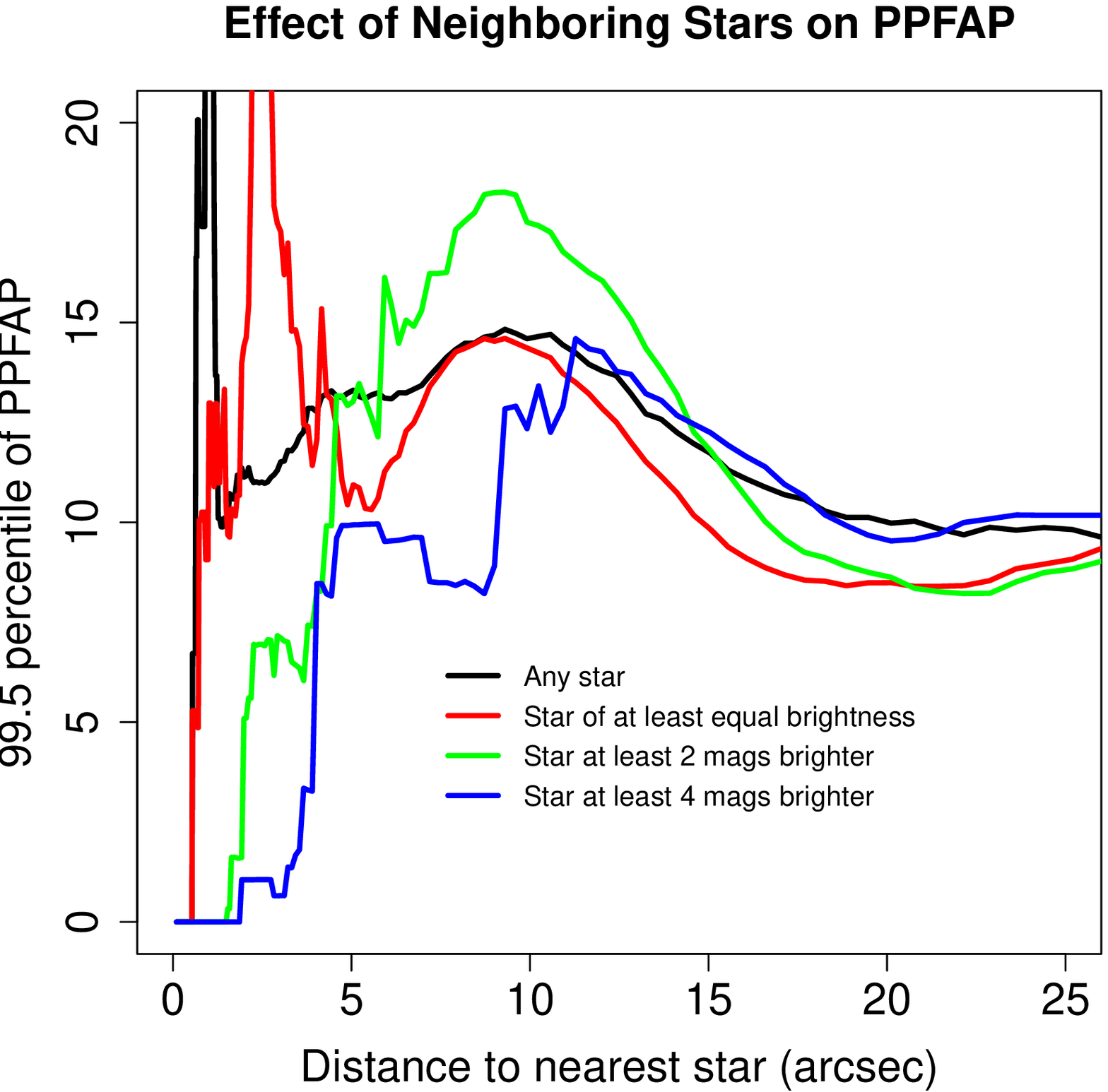}
\caption{\textbf{Left:} Absolute (un-normalized) histograms of angular distances from each star in the ATLAS lightcurve set to its neighbors in the object-matching catalog, which has higher resolution and is far more complete in crowded fields. A majority (64.7\%) of stars in the lightcurve set have a neighbor within 20 arcsec, and for a substantial minority (31.7\%) the neighbor is least equally bright. \textbf{Right:} Effect of neighbor proximity on apparent variability as measured by the PPFAP from our Lomb-Scargle analysis. Spurious variability in stars with near neighbors is expected to be caused by blending or incorrect/inconsistent assignment of ATLAS photometric measurements to stars in the object-matching catalog. We used this plot to determine thresholds for the binary statistic proxSTAT, which indicates potential spurious variability, as described in \S \ref{sec:proxSTAT}.
\label{fig:proxSTAT}}
\end{figure}

To probe the effect of confusion on our photometry, we used our object-matching catalog, described in \S \ref{sec:photdat}. For each star in the lightcurve set, we calculated the distance to the nearest star in the object-matching catalog (dist); the distance to the nearest star of at least equal brightness (dist0); the distance to the nearest star at least two magnitudes brighter (dist2); and the distance to the nearest star at least four magnitudes brigher (dist4). We then plotted the 99.5\% upper envelope of the PPFAP in a sliding box as a function of these distances (Figure \ref{fig:proxSTAT}). The PPFAP envelope, near PPFAP=10.0 for isolated stars, rises at distances smaller than 20 arcsec. We choose to regard as potentially affected any stars with dist $<1.5$ arcsec or dist0 $<5.0$ arcsec regardless of PPFAP; dist or dist0$<$20 arcsec and PPFAP$<$15.0; and dist2$<$20 arcsec with PPFAP$<$20.0. Since PPFAP=10.0 is our nominal boundary between strong and weak variability candidates for isolated stars, our objective here is to set conservative, but approximately equivalent, thresholds for stars that may be affected by blending from neighbors. We find no evidence that dist4 provides a meaningful constraint not already captured by dist, dist0, and dist2.

We find that 64.7\% of stars in the lightcurve set have a neighbor in the object-matching catalog within 20 arcsec. Hence, the variability for all these stars is potentially spurious unless PPFAP$>$15.0. Meanwhile, 7.88\% of the stars have a neighbor 2 mag brighter within 20 arcsec: their variability might be spurious up to PPFAP$=20.0$. Only 0.16\% of stars have a neighbor within 1.5 arcsec or a neighbor of equal brightness within 5.0 arcsec. The photometry of these last stars will certainly be affected by blending, and their variability is suspect regardless of the value of PPFAP.

To all stars with variability that is potentially suspect based on the criteria above, we assign proxSTAT=0, indicating that proximity statistics call their variability into question. Isolated stars or stars with values of PPFAP above the respective thresholds get proxSTAT=1, indicating their variability status is secure, at least as far as proximity effects are concerned.

\clearpage

\section{Classification of Variable Stars} \label{sec:classify}

In \S \ref{sec:photan} we have described how we analyzed our lightcurves using {\tt lombscar}; the calculation of additional statistics with {\tt varfeat}; detailed Fourier analysis using {\tt fourierperiod}; the calculation of statistics from the difference images; and finally the stellar proximity analysis to probe the extent to which confusion creates spurious variability. Of these analyses, {\tt lombscar}, {\tt varfeat}, and the proximity analysis are applied to all stars in the lightcurve set; while the Fourier fit and the difference statistics are calculated only for candidate variables.

For the candidate variables, on which all five analyses were performed, we calculate and save 169 different features, including the binary proxSTAT and ddcSTAT values described above. For a description of these features, see \S \ref{sec:appendixA}. For the candidate variables, all of these statistics are publicly available through STScI\footnote{http://mastweb.stsci.edu/ps1casjobs/}, in addition to the lightcurves.

Based on visual examination of a few tens of thousands of lightcurves, we identified 13 broad categories into which all stars could be classified, and developed a training set for input into machine learning algorithms, which we used to classify the remainder of the candidate variables. The 13 categories are CBF (close eclipsing binary, full period correctly identified by {\tt fourierperiod}; CBH (close eclipsing binary, period found by {\tt fourierperiod} is half the true orbital period); DBF and DBH (detached eclipsing binaries with either the full or half period identified; PULSE (pulsating variables of any kind for which the period found by {\tt fourierperiod} corresponds to a single pulse); MPULSE (pulsating variables for which the period corresponds to multiple pulses: hence, likely multi-mode pulsators); SINE (pure sine wave); NSINE (pure sine wave was fit, but the data are noisy and/or residuals indicate non-sinusoidal variations); MSINE (modulated sine wave; period corresponds to multiple cycles: analogous to MPULSE); MIRA (Mira-type long-period, high-amplitude variables); LPV (generic hard-to-classify variable without much power at frequencies corresponding to periods less than 5 days); IRR (generic hard-to-classify variable with significant power at high frequencies); and `dubious' (probably not a real variable). These categories were chosen based on extensive visual examination as capturing most of the morphological types of lightcurves present in our data.

We performed machine training and classification using the Google TensorFlow machine learning library on a standard Linux  platform with a single GPU card.  39,100 hand-classified variable stars were selected for the TensorFlow training set. Seventy  features were selected for training from the full set of 169 variable star features output by the five analyses described above. We employed the TensorFlow {\tt DNNClassifier} model, a simple deep neural network, with three hidden layers of 400, 800, and 400 nodes respectively in each layer. This architecture was selected after iterating with models with different numbers of hidden layers and nodes as the simplest model capable of attaining high training accuracy.

The seventy features used for machine learning are described in \S \ref{sec:appendixA}. They include the PPFAP from the Lomb-Scargle periodogram run by {\tt fourierperiod}; the filter-specific raw RMS scatter; the master period, min and max brightness, residual RMS, and Fourier coefficients from both the long and (if applicable) short-period fits performed by {\tt fourierperiod}; and the two parameters that describe the invariance of the lightcurve under 180-degree phase shift and under time-reversal centered on the deepest minimum. They also include several statistics output by {\tt varfeat}: the median magnitudes and 5th, 10th, 25th, 75th, 90th, and 95th percentile magnitudes; Hday, Hlong, $\chi^2/N$, the robust median statistic,  Inverse von Neumann ratio, Welch-Stetson I, and Stetson J statistics. 

In the extended trial-and-error process of finding a satisfactory methodology for the machine classification, one important breakthrough was achieved when we converted the Fourier terms from sine and cosine coefficients to amplitude and phase. Feeding the machine phase and amplitude information produced markedly more accurate classifications. We defined the amplitude and phase coefficents so that the mth Fourier term, previously given as in Equation \ref{eq:fourmod} by:

\begin{equation} \label{eq:onetermsincos}
f_m(t) = a_m \sin \left( m \cdot \frac{2 \pi t}{P} \right)+ b_m \cos \left( m \cdot \frac{2 \pi t}{P} \right).
\end{equation}

is instead expressed as:

\begin{equation} \label{eq:onetermamp}
f_m(t) = d_m \cos \left( m \left( \frac{2 \pi t}{P} - \phi_m \right) \right)
\end{equation}

We choose this particular formulation because it has the property that the minimum brightness (maximum magnitude) for a given Fourier term will occur whenever the argument of the cosine is zero, and if $\phi_m$ is the same for all values of $m$, the minimum brightness will occur at the same time for all Fourier terms. Of course, different values of $\phi_m$ are equivalent if separated by $2 \pi k/m$ for any integer $k$. We regularize the interpretation of $\phi_m$ for terms with $m>1$ by choosing $k$ so that $\phi_m$ will be as close as possible to, but greater than, $\phi_1$. Combined with the definition in Equation \ref{eq:onetermamp}, this also has the implication that the phase offset between $\phi_m$ and $\phi_1$ cannot be greater than $2 \pi/m$. %Given the sine and cosine coefficients $a_m$ and $b_m$ from Equation \ref{eq:onetermsincos}, the amplitude and phase $d_m$ and $\phi_m$ in Equation \ref{eq:onetermamp} are given by:

%\begin{equation} \label{eq:phaseconv1}
%d_m = \sqrt{a_m^2 + b_m^2} \\
%\end{equation}

%\begin{equation} \label{eq:phaseconv1}
%\phi_m = (\theta_m + 2 \pi k)/m
%\end{equation}

%Where for $m>0$, $k$ is an integer chosen to minimize the difference $\phi_m - \phi_0$, and $\theta_m = \pi/2 - \mathrm{atan}(b_m/a_m)$ if $a_m<0$ or $\theta_m = 3\pi/2 - \mathrm{atan}(b_m/a_m)$ if $a_m>0$. We include these transformation equations to enable the reader to easily duplicate our conversion to amplitude and phase if desired: it is the cosine and sine amplitudes $a_m$ and $b_m$ that are given by our table at STScI.

Another breakthrough was the training of the machine classifier in two stages. In stage 1, we pool the LPV, IRR, and `dubious' classifications into a single classification called HARD. This step allows the classifier to train on the most distinct classes of variable stars, achieving an accuracy of 94.1\%. For stage 2, we train a second classifier using the same training set to separate HARD variable stars into LPV, IRR and `dubious' classes, with an accuracy of 96.8\%. Training the {\tt DNNClassifier} model typically takes up to 10 minutes on our single-GPU system, and classifying all 4.7 million candidate stars using the trained model took about 10 minutes.

The probabilities output by the machine classifier for each of the 13 classes of variables, as well as a generic `HARD' probability, are provided for each star along with the vector of 169 features already mentioned. Including the proxSTAT and ddtSTAT values, we thus provide a total of 185 statistics for each candidate variable. All of these are publicly available from STScI, in addition to the lightcurves.

After the final round of classification with machine learning, we use parameters output by {\tt fourierperiod} to identify subsets of most of the categories that are atypical and hence potentially misclassified. We investigate these by hand and re-classify them where appropriate --- an exceedingly interesting exercise since some of them are very unusual objects (see \S \ref{sec:mystery}). Among the high-amplitude stars classified by machine learning as MIRA (and a few other classes) are a handful of objects with Mira-like amplitudes but colors not red enough for actual Mira variables. In addition to their relatively blue colors, they often show less smooth lightcurves than real Mira stars. We have invented a new class for these objects: SHAV, for `slow high-amplitude variable'. They include known AGN, variables of the R Coronae Borealis type (which, though red, are not as red as Miras), and other exotic objects.

Lastly, we make use of the proxSTAT and ddcSTAT values to adjust classification as follows. We assume that stars classified as any type of eclipsing binary (CBF, CBH, DBF, and DBH), pulsator (PULSE and MPULSE), coherent sinusoid (SINE and MSINE), or Mira variable have lightcurves with specific characteristics that are unlikely to be spuriously produced by blending. Hence, we do not adjust classifications for stars in any of these types due to proxSTAT=0. On the other hand, the generic categories IRR and LPV, as well as the lower significance variables in the NSINE category, could be contaminated by spurious variables due to blending. Hence, we reclassify all IRR, LPV, and NSINE variables that have proxSTAT=0 as `dubious' {\em unless} they also had ddcSTAT=1, in which case their classifications were left unchanged. This exception makes sense because blended stars subtract just as cleanly as unblended ones in our difference images, so ddcSTAT=1 rules out a blend as the cause of the original variability detection. Given this fact, we should also reclassify all the `dubious' stars with ddcSTAT=1 as something else. Reclassifying them as IRR could be a reasonable choice, but we elected to invent a new classification to reflect the unique analytical history of these stars. Since the machine learning did {\em not} classify them as IRR, it seems reasonable that they might be even farther from coherent periodicity than most stars in the IRR category. In order to communicate this, we have elected to call them `STOCH', for stochastic. Thus, our final classification includes fifteen categoires: the thirteen listed above plus SHAV and STOCH. These are given in Table \ref{tab:classfirst}, and described in more detail in \S \ref{sec:classdesc}.

\begin{deluxetable}{ll}
\tabletypesize{\small}
\tablewidth{0pt}
\tablecaption{ATLAS Variable Classes \label{tab:classfirst}}
\tablehead{ \colhead{Class} & \colhead{Description} }
\startdata
CBF & Close binary, full period \\
CBH & Close binary, half period \\
DBF & Distant binary, full period \\
DBH & Distant binary, half period \\
dubious & Star might not be a real variable\\
IRR & Irregular: catch-all for difficult short-period cases\\
LPV & Long period variable: catch-all for difficult cases\\
MIRA & High-amplitude, long-period red variable\\
MPULSE & Modulated Pulse: likely multi-modal pulsator\\
MSINE & Modulated Sine: multiple cycles of sine-wave were fit\\
NSINE & Noisy Sine: pure sine was fit, but residuals are large or non-random\\
PULSE & Pulsating variable \\
SHAV & Slow High-Amplitude Variable, too blue or irregular for Mira\\
SINE & Pure sine was fit with small residuals \\
STOCH & Stochastic: certainly variable, yet more incoherent even than IRR \\
\enddata
\end{deluxetable}

With its broad classes derived from visual investigation of lightcurve morphologies in our particular data set, our classification scheme is quite different from the schemes adopted by most previous works on variables detected in sky surveys \citep[e.g.][]{Drake2013a,Drake2013b,Drake2014a,ASASSN}, which have generally adopted pre-existing, astrophysically-based classification schemes with larger total numbers of categories. Both approaches have their merits, and the ATLAS data we present herein would certainly support more categories of classification. We adopt the broad, morphological categories partly with the objective of handing the machine classifier an easier problem and hence obtaining more reliable results from it --- an important consideration since the huge number of stars we have classified precludes checking more than a small fraction of them by hand. Our broad categories may also lend themselves to the detection of unusual objects or new classes of variables: each broad category provides a helpful context of objects that are in some way similar, while at the same time containing considerable substructure on which the classifier has not yet passed any judgement. In effect, this can allow the stars to tell us how they want to be classified --- a topic we explore further in \S \ref{sec:substructure}. One disadvantage of our current scheme, which we intend to correct for DR2, is that it has no separate class for spotted rotators and other periodic variables that are not eclipsing binaries, pulsators, Mira stars, or sinusoids such as ellipsoidal variables. Many spotted rotators have likely been classified as IRR or LPV, even though they may show quite regular periodicity, and a few may also have been misclassifed as eclipsing binaries or pulsating stars.

Table \ref{tab:class} gives the total number of stars finally classified in each category, as well as the number that were re-screened by hand (if any) and the number that turned out to be new. The new objects are identified by excluding every star recorded in the VSX or GCVS catalogs (downloaded on March 15, 2018); the catalog from the ASAS-SN survey presented by \citet{ASASSN}; the catalogs from the Catalina Sky Survey presented by \citet{Drake2013a,Drake2013b,Drake2014a}; and the OGLE catalogs published since 2010 and covering variable stars north of Dec -50 \citep{Soszynski2011a,Soszynski2011b,Soszynski2013,Soszynski2014,Mroz2015,Soszynski2015,Soszynski2016,Soszynski2017}. We note that most of the Catalina Sky Survey variables had already been incorporated into the VSX at the time of our download, and that due to the mostly southerly coverage of the OGLE surveys, there was very little overlap between our data set and the OGLE variables.

\begin{deluxetable}{lccccc}
\tabletypesize{\small}
\tablewidth{0pt}
\tablecaption{Statistics of Variable Classes \label{tab:class}}
\tablehead{ \colhead{Class} & \colhead{Total} & \colhead{Re-screened} & \colhead{New} & \colhead{Percent New} & \colhead{Percent ddcSTAT=1}}
\startdata
CBF & 44165 & 810 & 25901 & 58.65 & 30.98 \\
CBH & 36582 & 789 & 26196 & 71.61 & 16.65 \\
DBF & 11338 & 458 & 8487 & 74.85 & 10.55 \\
DBH & 17672 & 1392 & 14121 & 79.91 & 9.05 \\
dubious & 4307019 & 0 & 4218985 & 97.96 & 0.00 \\
IRR & 82960 & 0 & 72137 & 86.95 & 9.87 \\
LPV & 50909 & 0 & 29968 & 58.87 & 38.03 \\
MIRA & 7626 & 627 & 2063 & 27.05 & 55.23 \\
MPULSE & 5514 & 873 & 2357 & 42.75 & 33.71 \\
MSINE & 36285 & 229 & 30702 & 84.61 & 2.74 \\
NSINE & 64726 & 0 & 58777 & 90.81 & 1.03 \\
PULSE & 25162 & 5749 & 8031 & 31.92 & 40.33 \\
SHAV & 17 & 17 & 2 & 11.76 & 58.82 \\
SINE & 29404 & 0 & 23422 & 79.66 & 3.14 \\
STOCH & 14834 & 0 & 13076 & 88.15 & 100.00 \\
All EBIN & 109757 & 3449 & 74705 & 68.06 & 20.56 \\
All pulsators & 30676 & 6622 & 10388 & 33.86 & 39.14 \\
\enddata
\end{deluxetable}

\begin{figure}
\includegraphics{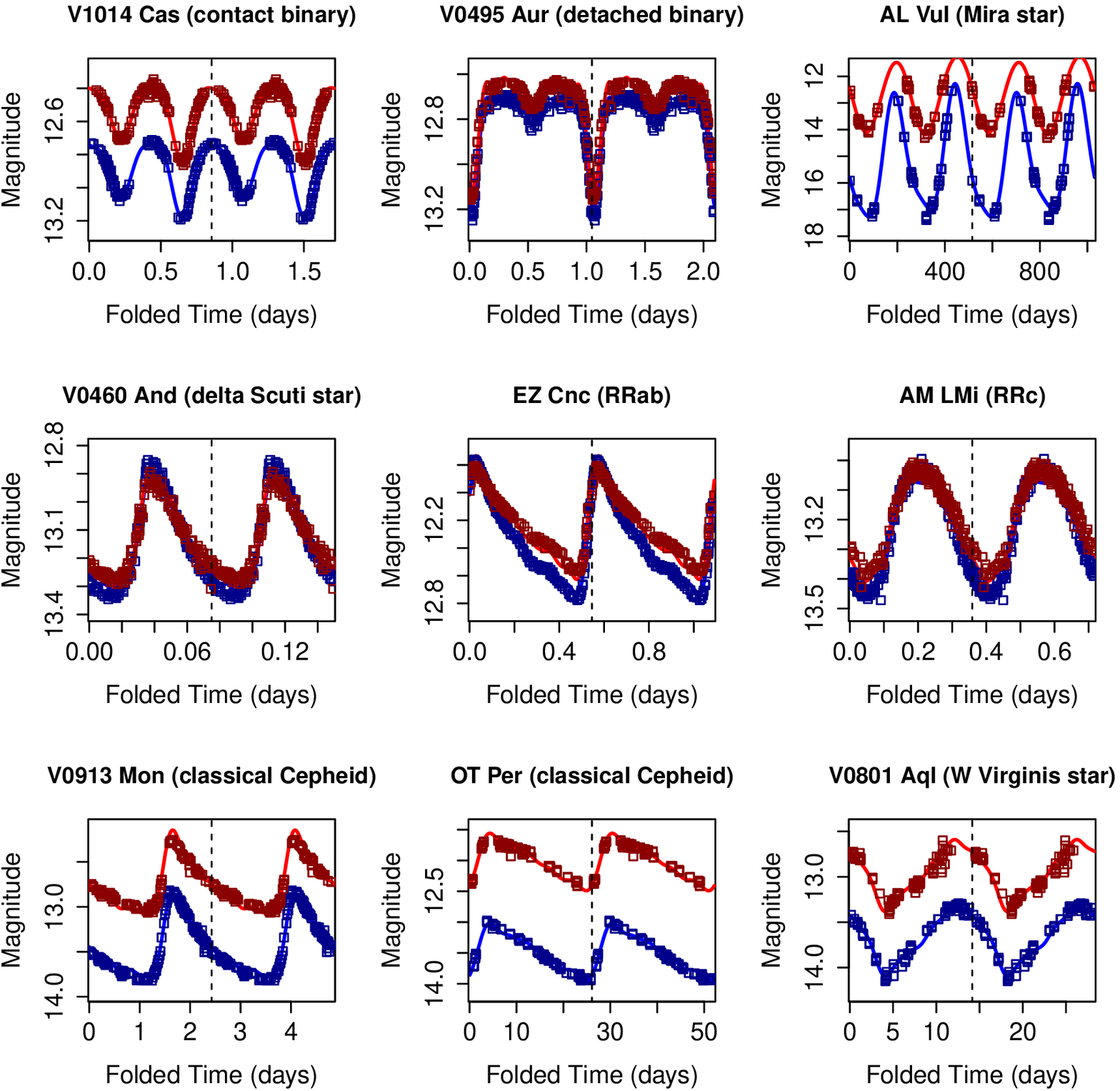}
\caption{Characteristic examples of bright known variables of well-known and astrophysically significant types measured and correctly classified by ATLAS. Note that a W Virginis star is the same as a type II Cepheid. Our $c$ band data and the corresponding Fourier fit are shown in blue, with the $o$ band photometry and fit in red. The data are phase-folded, and two periods are shown for each object, divided by a vertical dotted line whose intersection with the x axis gives the period. 
\label{fig:knownvar}}
\end{figure}

In Figure \ref{fig:knownvar}, we show characteristic examples of ATLAS lightcurves of bright known variables in 9 different important classes: contact binary, detached binary, Mira, $\delta$ Scuti, RRab, RRc, short-period classical Cepheid, long-period classical Cepheid, and W Virginis star (i.e., type II Cepheid). ATLAS classifies all of these correctly: the contact binary as CBF, the detached binary as DBF, the Mira star as MIRA, and all the rest as PULSE.

\subsection{Categories of Variables: Examples and Discussion} \label{sec:classdesc}

\begin{figure}
\includegraphics[scale=0.85]{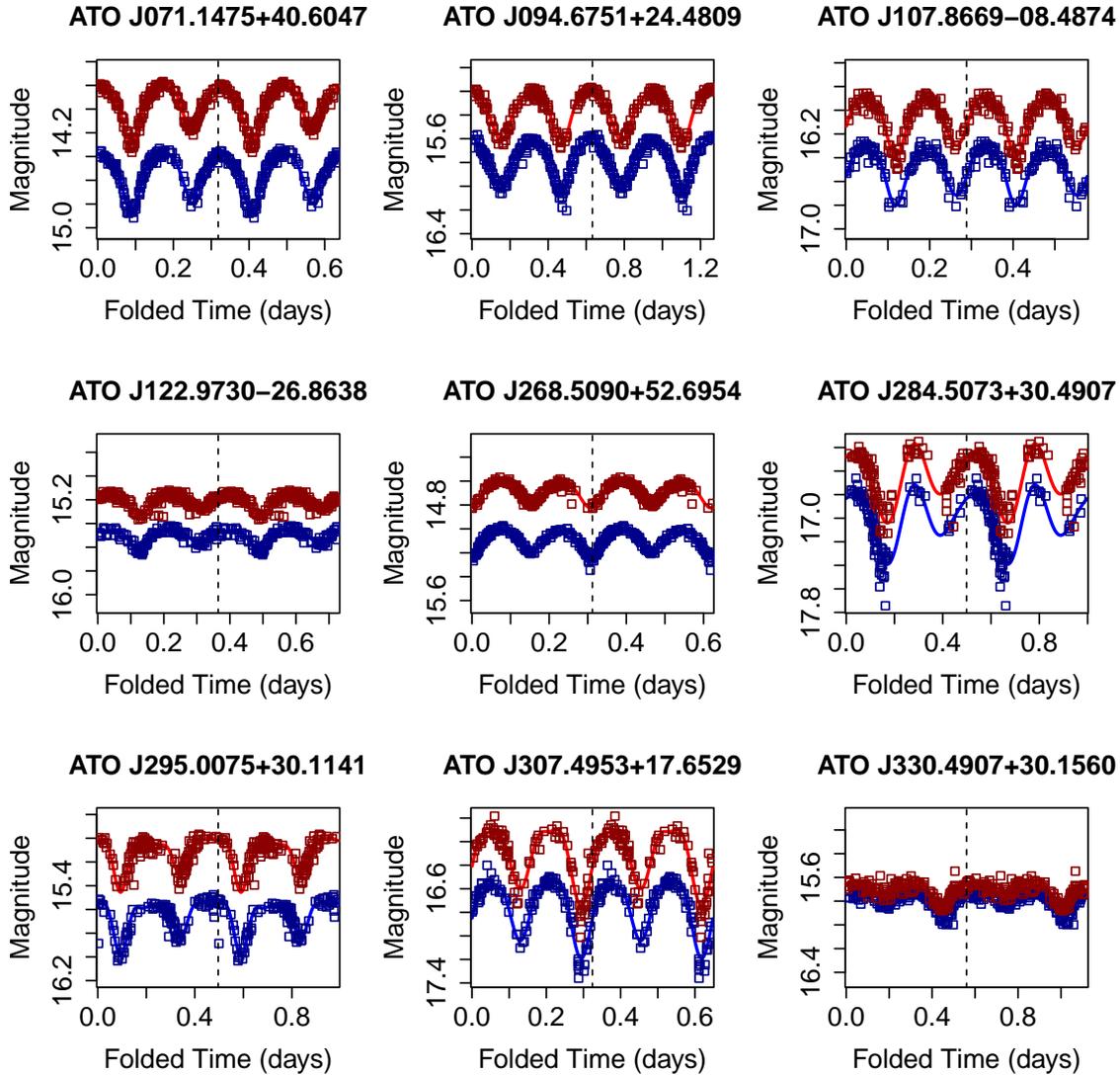} 
\caption{\textbf{CBF:} Examples of our CBF class, which consists of close eclipsing binaries for which our analysis using {\tt fourierperiod} correctly identified the full orbital period. Our $c$ band data and the corresponding Fourier fit are shown in blue, with the $o$ band photometry and fit in red. The data are phase-folded, and two periods are shown for each object, divided by a vertical dotted line whose intersection with the x axis gives the period. A consistent magnitude scale has been used for all panels, so the diversity of amplitudes can be seen at a glance. Since measured magnitudes are shown without re-scaling, the offset between the $c$ and $o$ band data indicates the star's color. The objects plotted here are randomly selected from our newly discovered variables that received the CBF classification. 
\label{fig:CBF}}
\end{figure}

\begin{figure}
\includegraphics[scale=0.85]{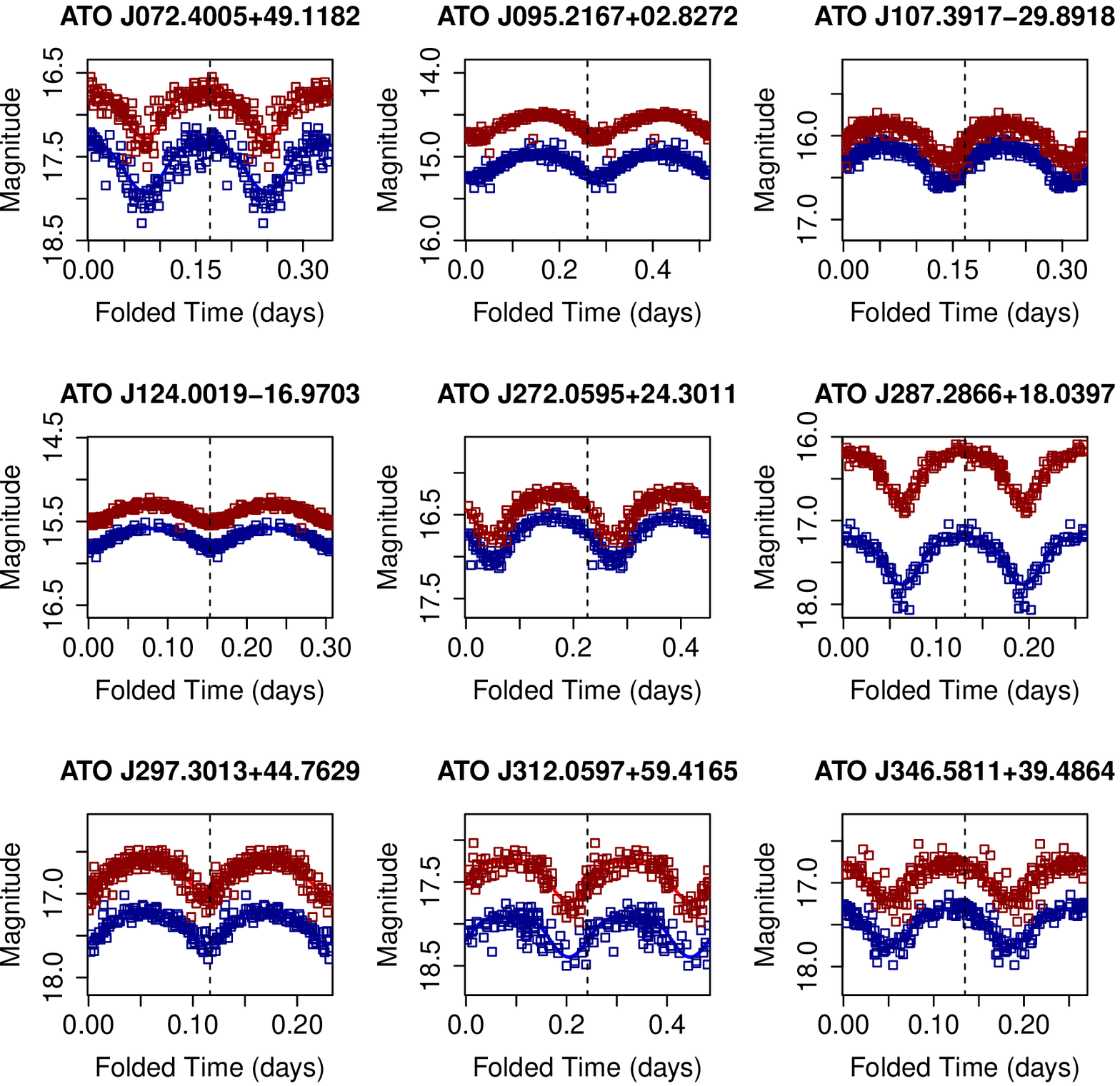}
\caption{\textbf{CBH:} Examples of our CBH class, which consists of close eclipsing binaries for which {\tt fourierperiod} selected a period equal to half the true orbital period. The plots are constructed exactly as in Figure \ref{fig:CBF}, so the offset between the $c$ and $o$ band data indicates a star's color. The objects are randomly selected from our newly discovered variables classified as CBH. 
\label{fig:CBH}}
\end{figure}

\begin{figure}
\includegraphics[scale=0.85]{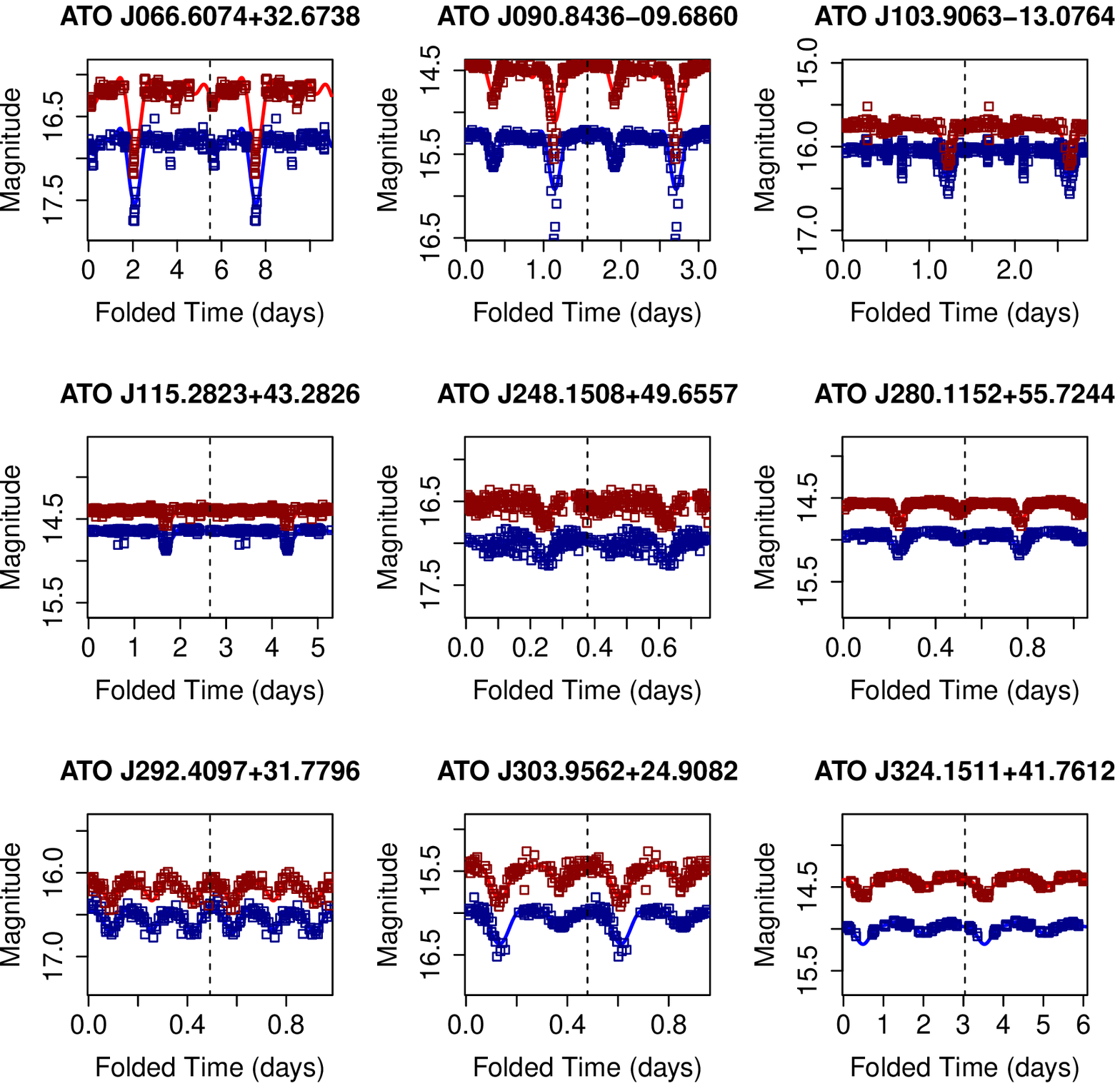}
\caption{\textbf{DBF:} Examples of our DBF class, which consists of detached eclipsing binaries for which the {\tt fourierperiod} appears to have correctly identified the full orbital period. The plots are constructed exactly as in Figure \ref{fig:CBF}, so the offset between the $c$ and $o$ band data indicates a star's color. The objects plotted here are randomly selected from our newly discovered variables classified as DBF. ATO J292.4097+31.7796 is evidently misclassified: it should have been a CBF, and the nominal period is 1.5 times the true value. 
\label{fig:DBF}}
\end{figure}

\begin{figure}
\includegraphics[scale=0.85]{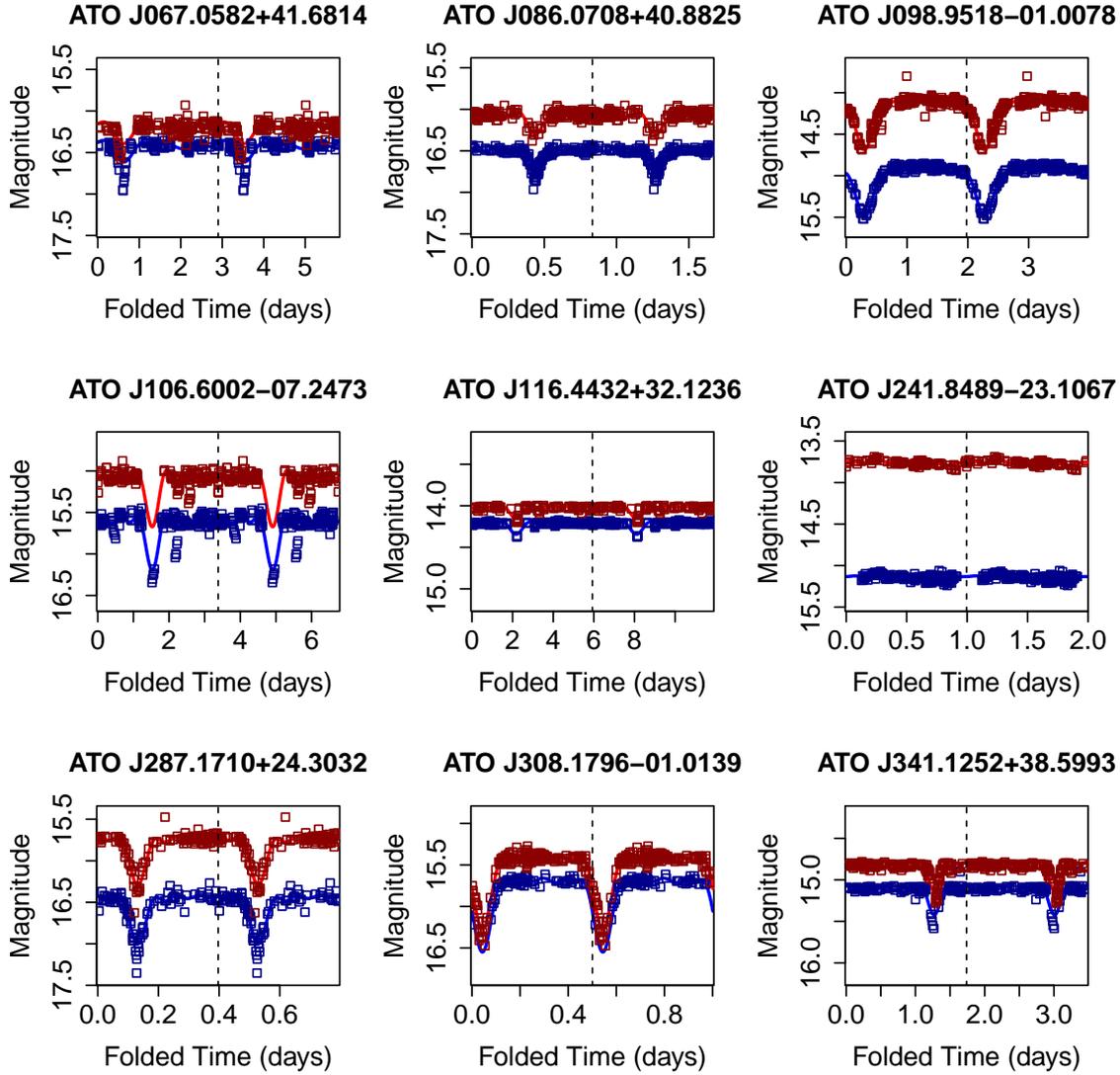}
\caption{\textbf{DBH:} Examples of our DBH class, which consists of detached eclipsing binaries for which the period found by {\tt fourierperiod} appears to be half the true orbital period. The plots are constructed exactly as in Figure \ref{fig:CBF}, so the offset between the $c$ and $o$ band data indicates a star's color. The objects plotted here are randomly selected from our newly discovered variables classified as DBH. ATO J241.8489-323.106 is evidently misclassified: it should be IRR or `dubious'. 
\label{fig:DBH}}
\end{figure}

\begin{figure}
\includegraphics[scale=0.85]{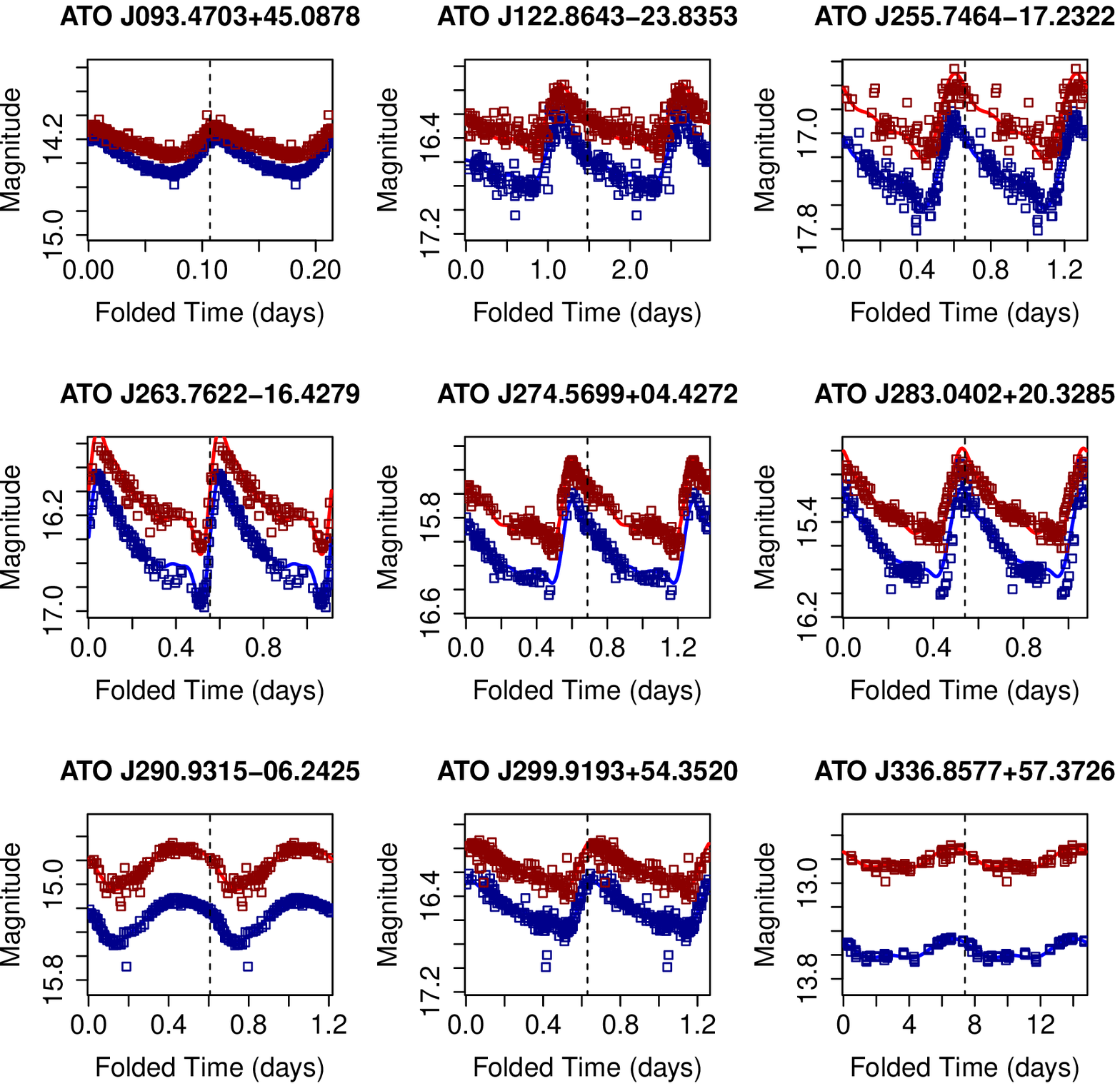}
\caption{\textbf{PULSE:} Examples of our PULSE class, which consists of stars exhibiting non-sinusoidal variability due to pulsations. The plots are constructed exactly as in Figure \ref{fig:CBF}, so the offset between the $c$ and $o$ band data indicates a star's color. The objects plotted here are randomly selected from our newly discovered variables classified as PULSE. ATO J093.4703+45.0978 is a $\delta$ Scuti star; ATO J122.8643-23.8353 may be a short-period Cepheid; ATO J336.8577+57.3726 belongs to a subclass of variable stars that may be a discovery of ATLAS (see \S \ref{sec:UCBH}); and the remainder are RR Lyrae variables (all RRab except for ATO J290.9315-06.2425). 
\label{fig:PULSE}}
\end{figure}

\begin{figure}
\includegraphics[scale=0.85]{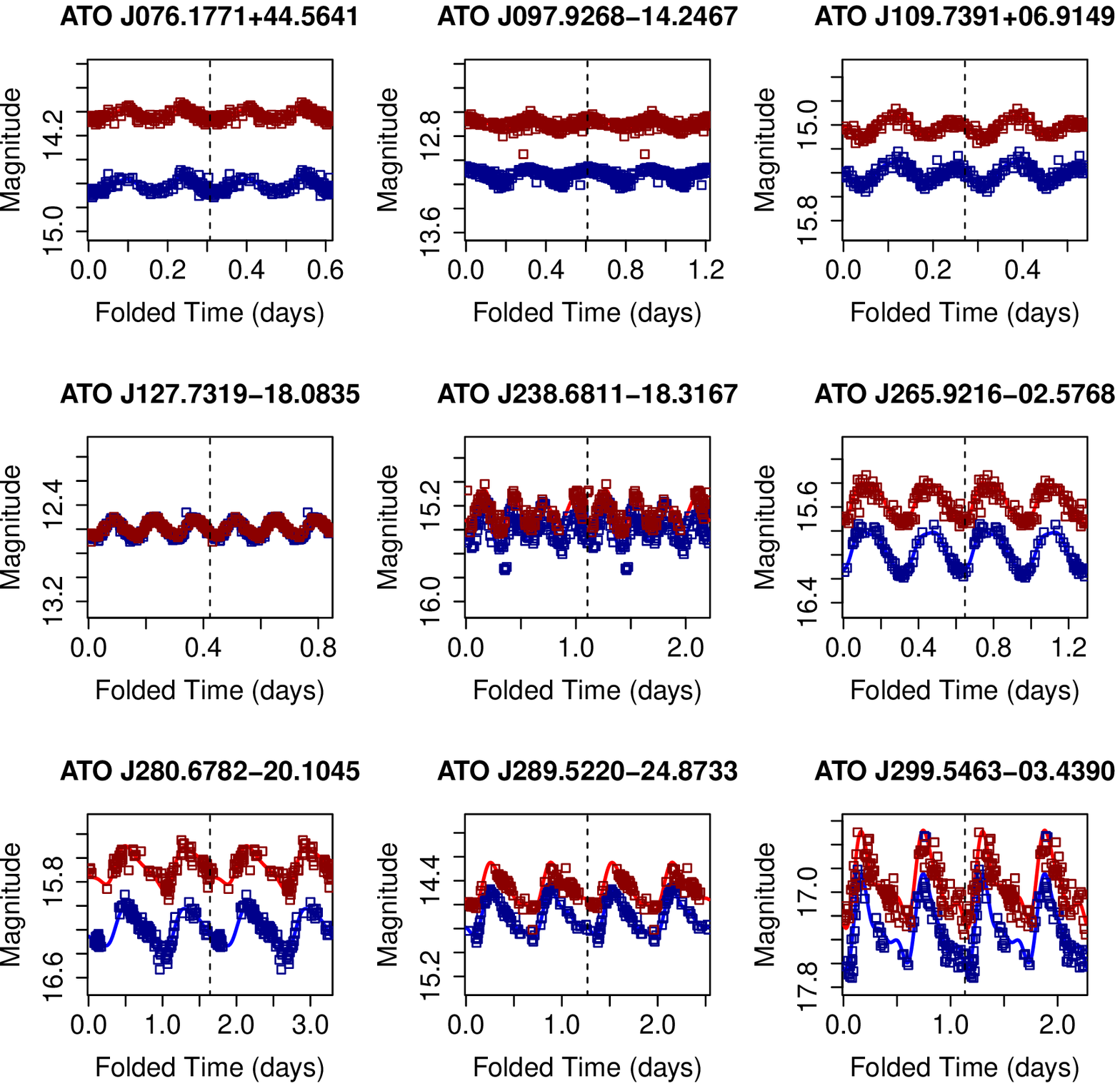}
\caption{\textbf{MPULSE:} Examples of our MPULSE class, which consists of stars having waveforms like the PULSE class, but for which {\tt fourierperiod} has fitted multiple pulses per nominal period. The plots are constructed exactly as in Figure \ref{fig:CBF}, so the offset between the $c$ and $o$ band data indicates a star's color. The fitting by {\tt fourierperiod} of more than one pulse indicates a modulation in pulse height, likely due to the presenence of multiple modes. The objects plotted here are randomly selected from our newly discovered variables classified as MPULSE. All of them appear to be either $\delta$ Scuti or RR Lyrae stars, though the MPULSE category as a whole also includes longer-period variables likely to be Cepheids. 
\label{fig:MPULSE}}
\end{figure}

\begin{figure}
\includegraphics[scale=0.85]{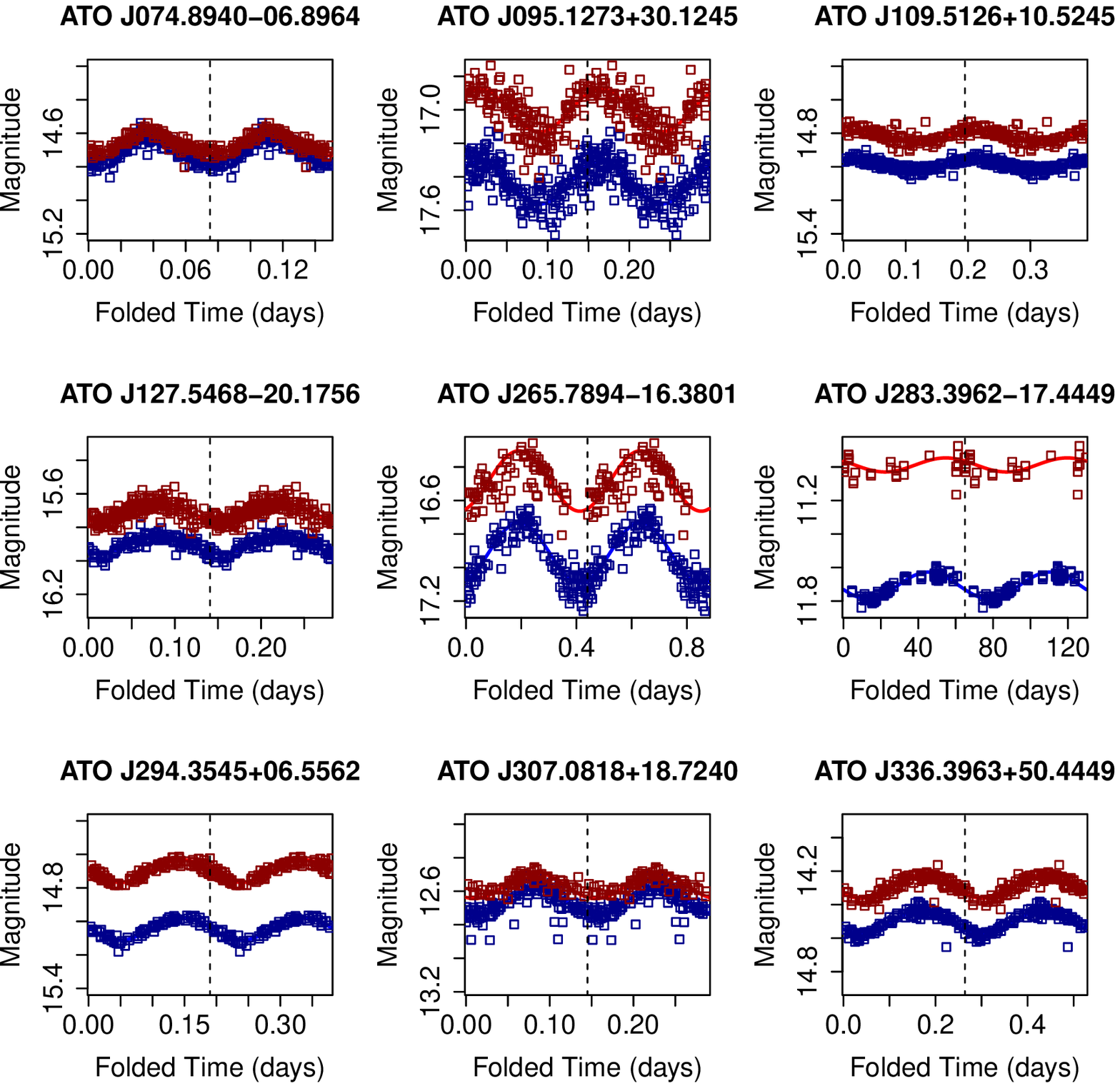}
\caption{\textbf{SINE:} Examples of our SINE class, which consists of stars fitted with a pure sine wave. The plots are constructed exactly as in Figure \ref{fig:CBF}, so the offset between the $c$ and $o$ band data indicates a star's color. The objects plotted here are randomly selected from our newly discovered variables classified as SINE. Many ellipsoidal variables likely fall into this class, together with some low-amplitude contact binaries, pulsating stars, and spotted rotators. ATO J074.8940-06.8964 is certainly a $\delta$ Scuti star --- a type of pulsator prone to being given a SINE classification because their low amplitudes and very short periods inhibit {\tt fourierperiod} from finding a more complex, non-sinusoidal fit.
\label{fig:SINE}}
\end{figure}

\begin{figure}
\includegraphics[scale=0.85]{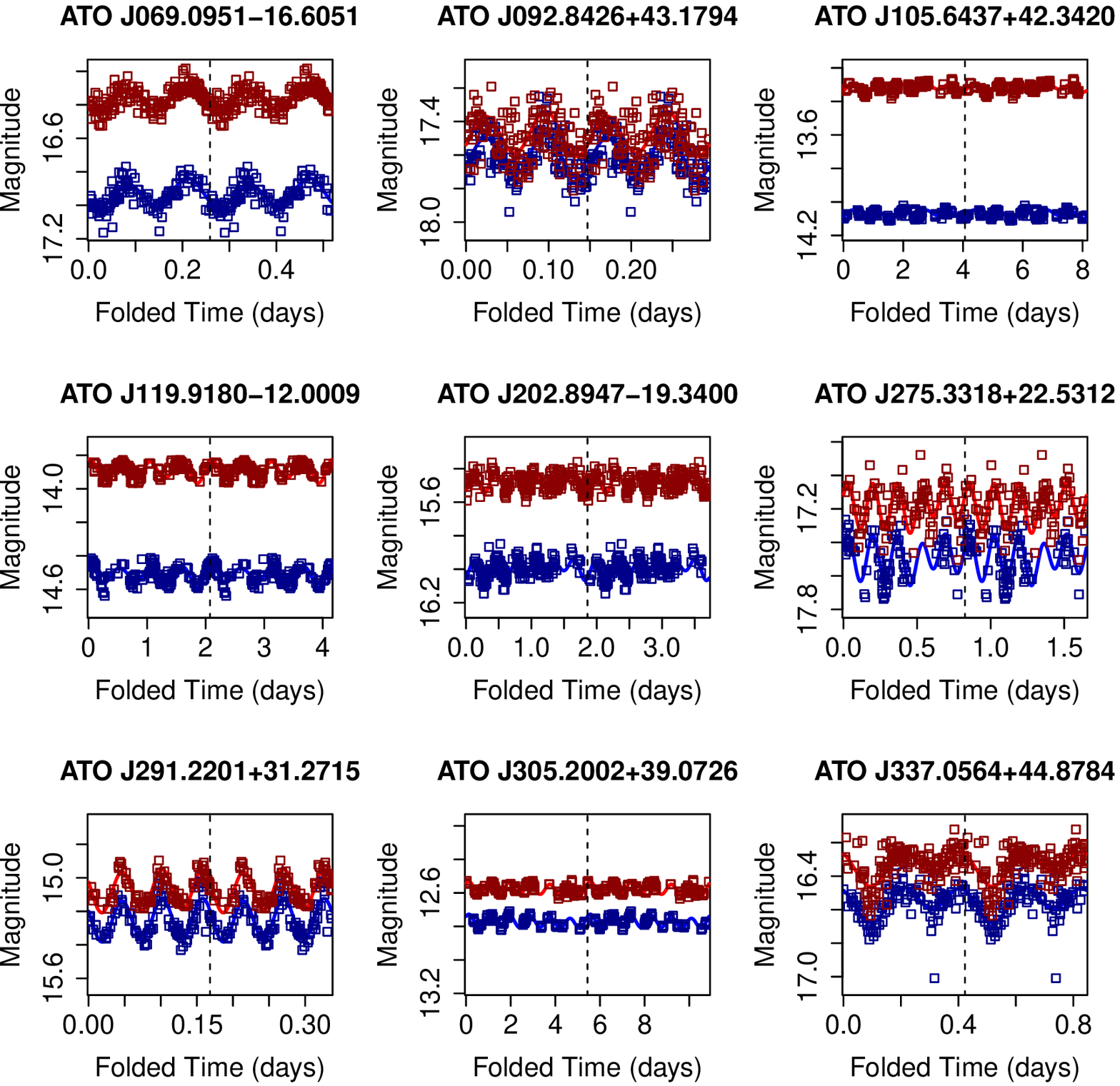}
\caption{\textbf{MSINE:}  Examples of our MSINE class, which consists of stars fitted with a modulated sine wave having multiple cycles for each nominal period. The plots are constructed exactly as in Figure \ref{fig:CBF}, so the offset between the $c$ and $o$ band data indicates a star's color. The objects plotted here are randomly selected from our newly discovered variables classified as MSINE. They differ from MPULSE in that the individual cycles have a sinusoidal rather than sawtooth appearance. This class likely includes spotted rotators and ellipsoidal variables, as well as some multi-mode pulsators with very symmetrical pulses. 
\label{fig:MSINE}}
\end{figure}

\begin{figure}
\includegraphics[scale=0.85]{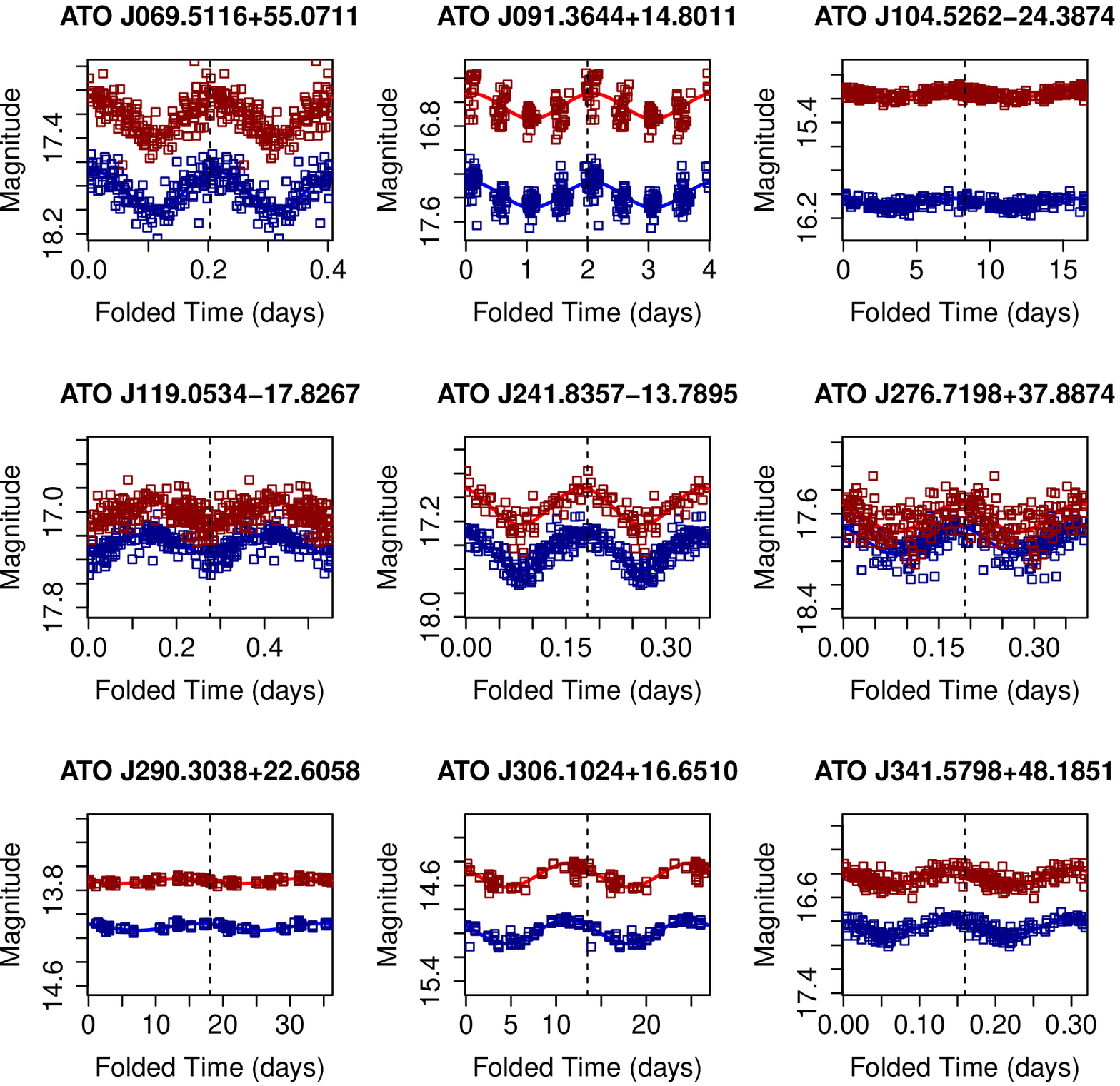}
\caption{\textbf{NSINE:} Examples of our NSINE class, which consists stars fitted with a noisy sine (`NSINE') or a sine wave with obvious non-random residuals. The plots are constructed exactly as in Figure \ref{fig:CBF}, so the offset between the $c$ and $o$ band data indicates a star's color. The objects plotted here are randomly selected from our newly discovered variables classified as NSINE. This class includes spotted rotators and ellipsoidal variables, as well as some contact binaries and pulsators that were too faint and noisy to classify definitively. 
\label{fig:NSINE}}
\end{figure}

\begin{figure}
\includegraphics[scale=0.85]{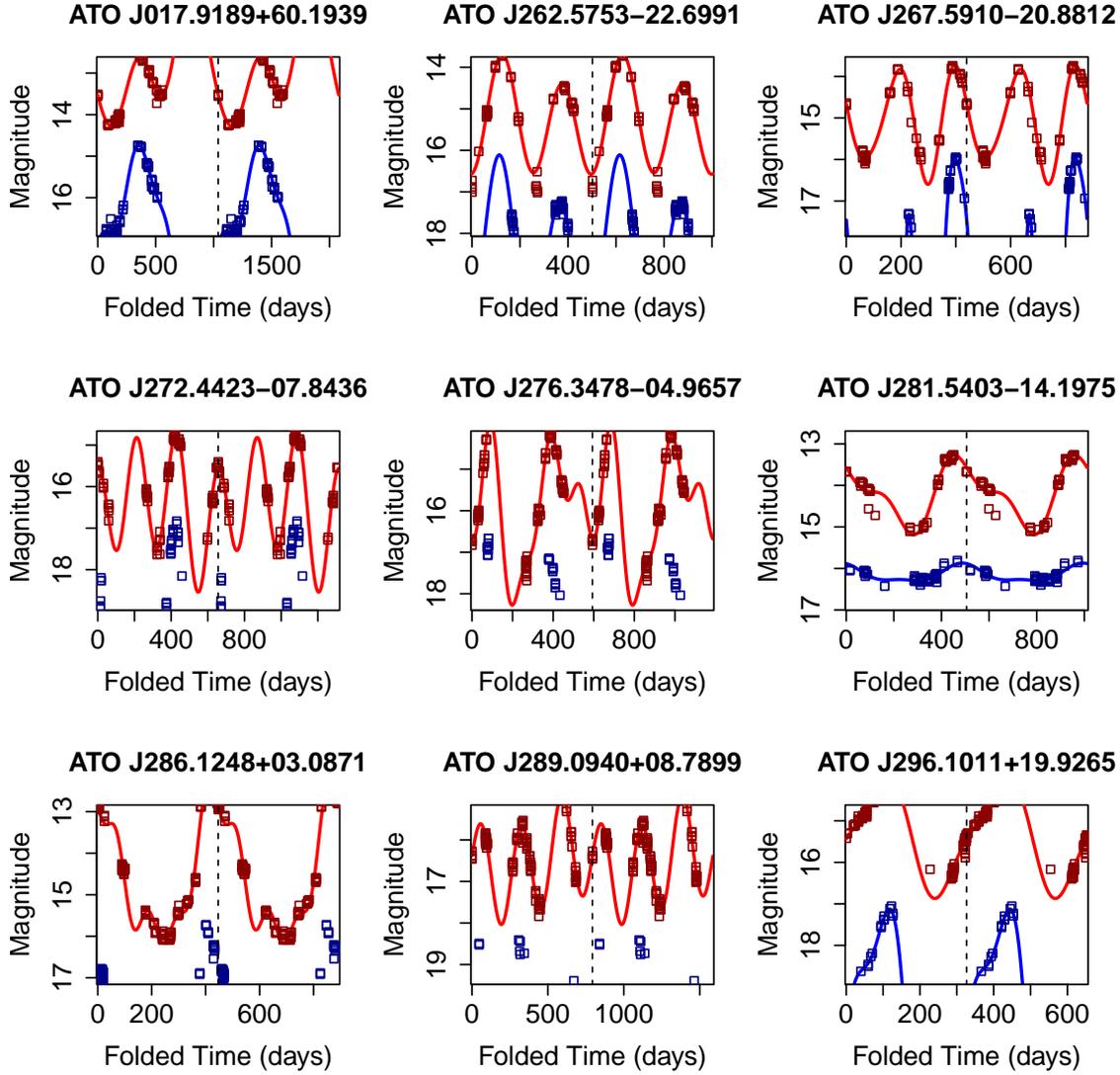}
\caption{\textbf{MIRA:} Examples of our MIRA class, which consists of stars with long period, high amplitude variability like that of the pulsating red giant Mira. The plots are constructed exactly as in Figure \ref{fig:CBF}, so the offset between the $c$ and $o$ band data indicates a star's color. A paucity of $c$ band data, due to the stars' extremely red colors, caused the $c$ band Fourier fit to run away in some cases. These runaway fits do not compromise our analysis, but are not plotted because they are distracting and meaningless. The objects plotted here are randomly selected from our newly discovered variables classified as MIRA. ATO J281.5403-14.1975 has Mira-like $o$ band amplitude, but its much smaller $c$ band amplitude suggests it is blended with a luminous blue companion or else is a different type of variable. 
\label{fig:MIRA}}
\end{figure}

\begin{figure}
\includegraphics[scale=0.85]{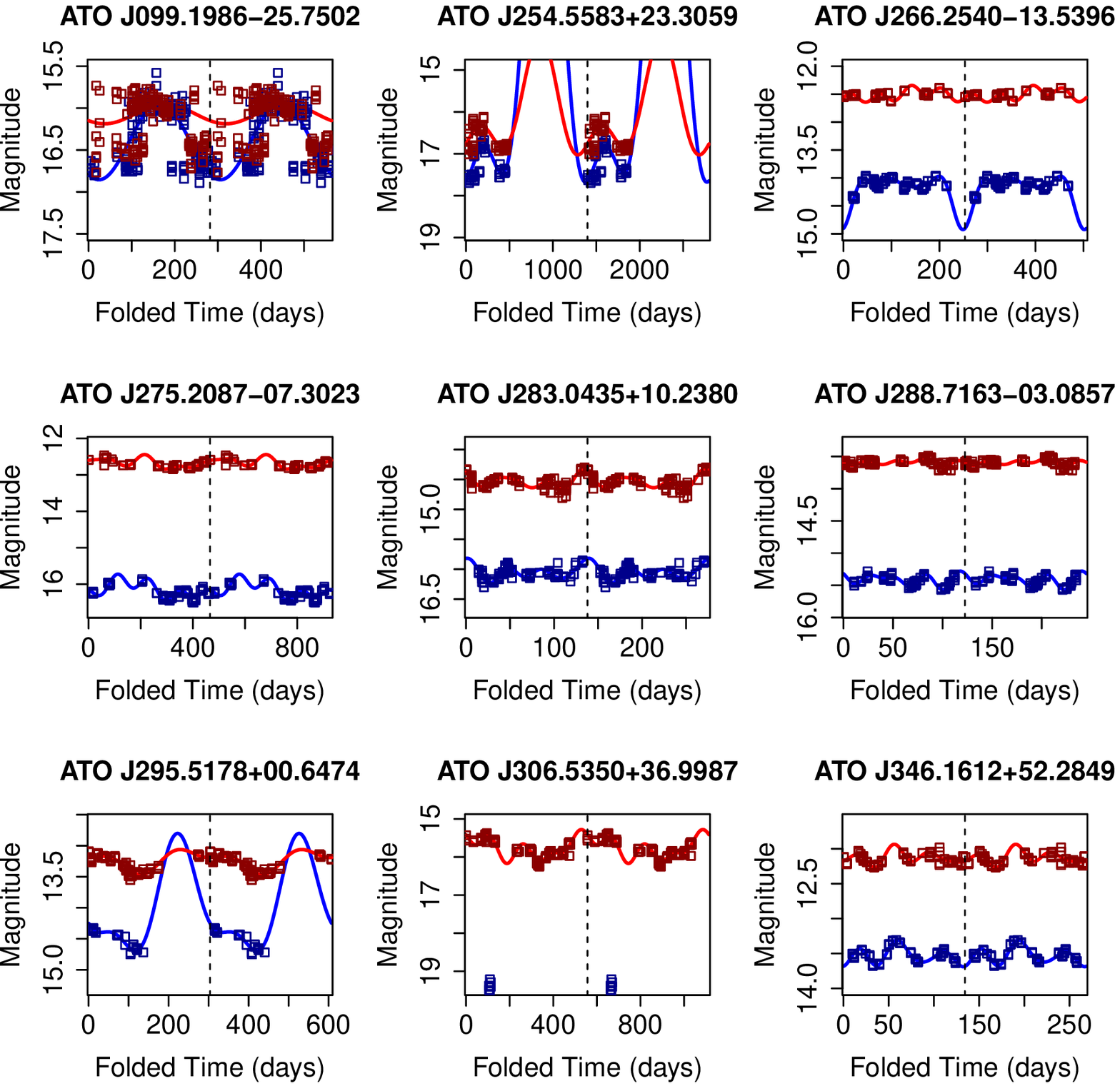}
\caption{\textbf{LPV:}  Examples of our LPV class, which is a catch-all for stars that exhibit variability on a long timescale but do not easily fit into any of our more specific classes. The plots are constructed as in Figure \ref{fig:CBF}, except that the magnitude scales are different for different panels. The offset between the $c$ and $o$ band data still indicates a star's color. Cases where the Fourier fits run away during intervals unconstrained by the data do not compromise our analysis, but sometimes are not plotted to avoid distraction. The objects plotted here are randomly selected from our newly discovered variables classified as LPV.
\label{fig:LPV}}
\end{figure}

\begin{figure}
\includegraphics[scale=0.85]{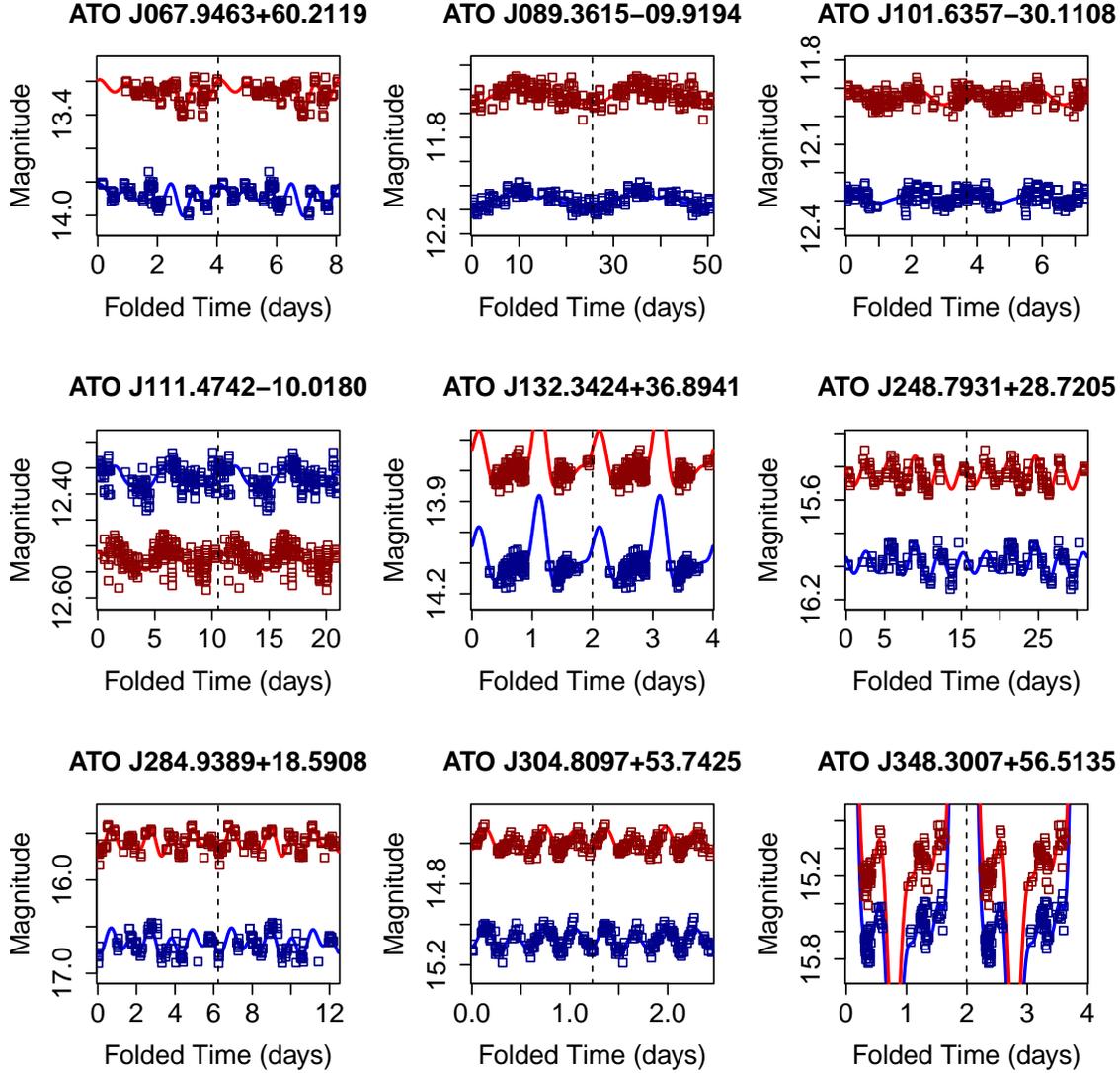}
\caption{\textbf{IRR:} Examples of our IRR class, which contains irregular variables but is also a catch-all for stars that exhibit variability on a short timescale but do not easily fit into any of our more specific classes. The plots are constructed as in Figure \ref{fig:CBF}, except that the magnitude scales are different for different panels. The offset between the $c$ and $o$ band data still indicates a star's color. Cases where the Fourier fits run away during intervals unconstrained by the data do not compromise our analysis. The objects plotted here are randomly selected from our newly discovered variables classified as IRR. Some of them may simply be lower-significance examples of well-known types, while others may be very coherent (hence not really `irregular'), but simply unusual and difficult to classify. Since we do not have a specific class for rotating variables, many of them probably end up as IRR. 
\label{fig:IRR}}
\end{figure}

\begin{figure}
\includegraphics[scale=0.85]{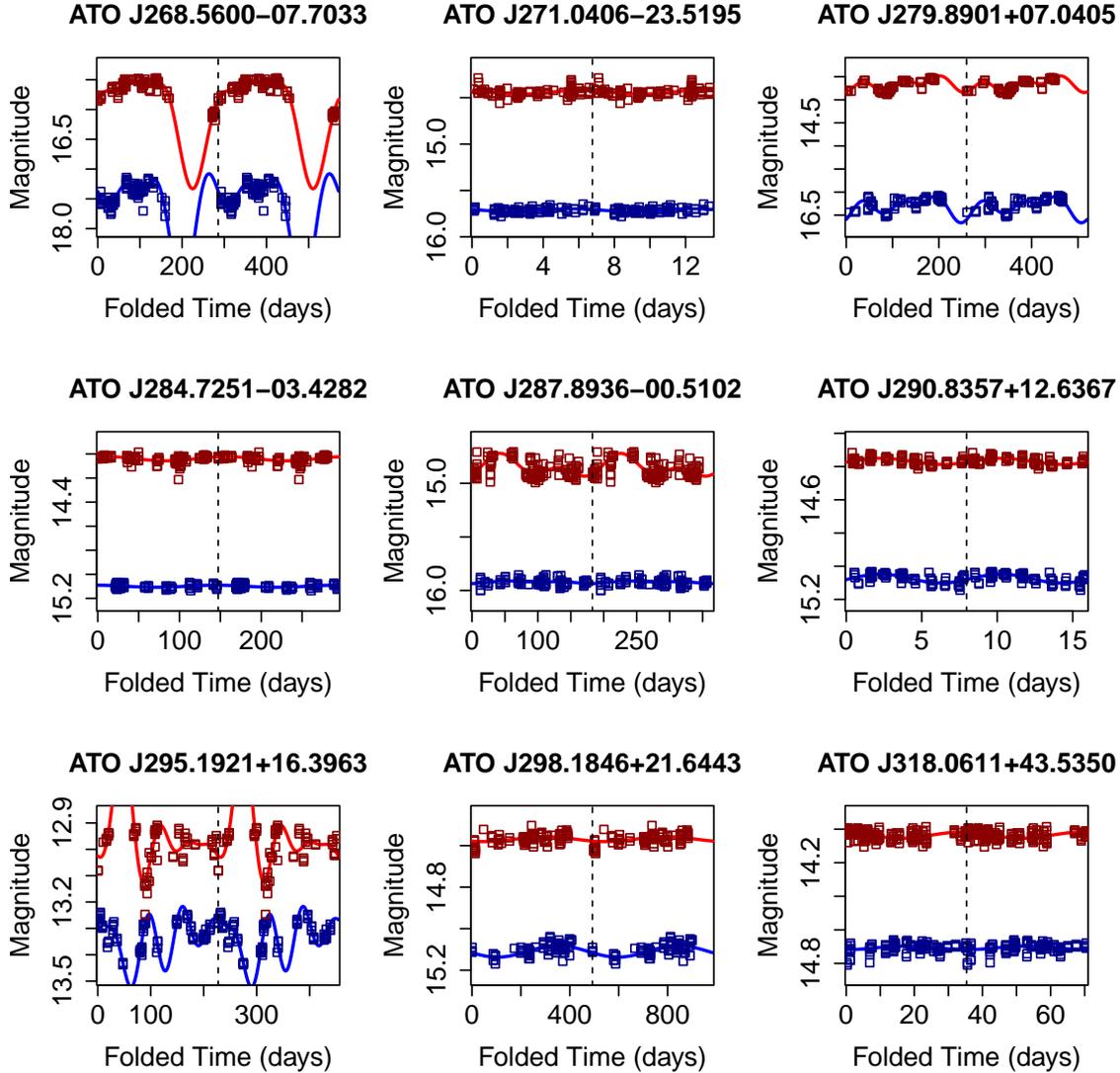}
\caption{\textbf{STOCH:} Examples of our STOCH class, which consists of stars the machine classifier initially designated as `dubious', but which we later reclassified as stochastic variables, based on extremely significant detections in the difference images. The plots are constructed as in Figure \ref{fig:CBF}, so the offset between the $c$ and $o$ band data indicates a star's color. The objects plotted here are randomly selected from our newly discovered variables classified as STOCH. They exhibit a great diversity in amplitudes, with some of them appearing flat in the figure even though the magnitude ranges have been rescaled to mitigate the problem. Closer examination of these objects reveals extremly significant, apparently nonperiodic variations that account for the difference detections. 
\label{fig:STOCH}}
\end{figure}

In this section we provide a brief discussion of each type of variable, and in Figures \ref{fig:CBF} through \ref{fig:STOCH} we present nine examples of each type except `dubious'. These examples are randomly chosen from previously unknown variables in each category, with no attempt to avoid showing failures of our analysis or classification --- hence, readers can use these figures to do their own `quality control' on our classifications. We show only new discoveries because they would be expected to be fainter and more difficult than previously known variables, and hence to provide the most stringent test of our accuracy. The data are phase-folded at the master period output by {\tt fourierperiod}. Two full cycles are shown for each object, with the points overplotted on the best-fit Fourier model. A vertical dotted line indicates the end of the first plotted cycle: hence, its intersection with the x axis gives the period. Data from the ATLAS $c$ band are shown in blue, while the $o$ band data are shown in red. In most cases, a consistent magnitude scale is used for all nine plots so that the diversity in amplitude can be seen at a glance. Since measured magnitudes are shown without re-scaling, the offset between the $c$ and $o$ band data indicates the star's color.

For irregular variables or stars whose period is greater than the temporal span of our data, the master period is not expected to correspond to any true astrophysical frequency. For such systems, the Fourier fit should be interpreted not as a measurement of a true cyclical pattern but merely as a probe of the system's photometric coherence. A few cases exist (especially among the long-period objects) where the Fourier fit runs away to unreasonable values during intervals of time that are unconstrained by the data. Our analysis is designed to avoid any adverse effects from these cases (e.g., in determining the min and max fitted magnitudes, we evaluate the fits only at times corresponding to actual measurements). In some cases where the star is very red and the $c$ band Fourier fit runs away due to the resulting paucity of $c$ band points, we have refrained from plotting the $c$ band fit to avoid distraction.

\textbf{CBF}: Close binary, full period. These stars are contact or near-contact eclipsing binaries for which the Fourier fit has found the correct period and hence fit the primary and secondary eclipses separately. Classification tends to be very definitive in this category, with the rate of serious misclassification being as low as 1\%. Mild errors such as confusion between the CBF and DBF classes; and period errors in which the nominal period is 1.5 times the correct value, may be slightly more common.

\textbf{CBH:} Close binary, half period. These stars are contact or near-contact eclipsing binaries for which the Fourier fit has settled on half the correct period and hence has overlapped the primary and secondary eclipses. Physically, the CBF and CBH stars are expected to differ in that the primary and secondary eclipses are likely to be more similar in depth in the latter class. Like CBF, CBH stars are very rarely misclassified. Even RRc variables, which are notoriously difficult to distinguish from contact binaries because of their symmetrical lightcurves, are well-separated by our Fourier analysis, especially the phase offsets (see Figure \ref{fig:pulsephase}). At the longest periods there may be some contamination from spotted rotators and/or extremely symmetrical Cepheid-type pulsators.

\textbf{DBF:} Distant binary, full period. These stars are detached eclipsing binaries for which the Fourier fit has found the correct period and hence fit the primary and secondary eclipses separately. These stars are challenging because their lightcurves are flat much of the time, causing the PPFAP values to be relatively low, and our maximum of six Fourier terms can be insufficient to fit the narrow eclipses. Hence, a few percent of them may be misclassified, and a larger fraction likely have incorrect periods. Better results could be obtained using the Box Least Squares (BLS) algorithm of \citet{BLS} as was done, e.g., by \citet{ASASSN}. We judged this to be too computationally intensive at present, but will likely use it in DR2.

\textbf{DBH:} Distant binary, half period. These stars are fully detached eclipsing binaries for which the Fourier fit has settled on half the correct period and hence has overlapped the primary and secondary eclipses. Like the DBF class, they are challenging for our analysis, and are more likely to be misclassifed or to have incorrect periods than CBF and CBH. This will be helped in DR2 by our intended use of the BLS algorithm. Nevertheless, most of them are correctly classified even in the current analysis.

\textbf{PULSE:} Pulsating stars showing the classic sawtooth lightcurve, regardless of period. They are expected to include both RR Lyrae and $\delta$ Scuti stars, and some Cepheids. These classes are resolvable based on period, color, amplitude, and the phase offsets of the various Fourier terms. The pulsating stars are in many ways the most interesting class, since they contain the RR Lyrae and Cepheid stars useful for distance determination. Accordingly, we put a great deal of effort into producing a clean set of accurately classified stars. Nearly 6000 were screened by hand to check the machine classification. The final misclassification rate should be as low as 1\%. Most of the misclassifications are likely to be at the longest periods, where there may be some confusion with spotted rotators; and among the faintest objects, where low SNR made the classification difficult.

\textbf{MPULSE:} Stars showing modulated pulsation, such that the Fourier fit has settled on a period double or triple the actual pulsation, in order to render multiple pulses of different amplitudes or shapes. These objects could be multi-modal or Blazhko-effect stars, or stars exhibiting some other kind of variability in addition to their pulsations. In the case of the known high-amplitude $\delta$ Scuti (HADS) star CSS\_J082237.3+030441, we have confirmed multiple pulsation modes using targeted high-precision photometry with the University of Hawaii 2.2 meter telescope on Mauna Kea. 

\textbf{SINE:} Sinusoidal variables. These stars exhibit simple sine-wave variability with little residual noise. Ellipsoidal variables likely dominate this class. There may also be some RR Lyrae stars of type C, especially at faint magnitudes where the lower SNR makes it difficult to detect the non-sinusoidal nature of their lightcurves. Spotted rotators can also show sinusoidal variations: stars whose rotation axis is only modestly inclined to our line of sight may have circumpolar or near-circumpolar spots, which will produce sinusoidal variations due to their changing aspect ratio as long as the inclination is nonzero.

\textbf{MSINE:} Stars showing modulated sinusoids. These are exactly analogous to the MPULSE stars, except that instead of a classic sawtooth pulse lightcurve, the fundamental waveform being modulated is a simple sinusoid. Thus, MSINE stars may show 2, 3, 4, 5, or even 6 cycles through the Fourier fit. Each cycle appears a good approximation to a sine wave, but the amplitude and/or mean magnitude varies from one to the next. Physically, the MSINE stars may include spotted ellipsoidal variables; rotating stars with evolving spots; and sinusoidal pulsators such as RR Lyrae (RRC) stars that have multiple modes or multiple types of variability. Period, color, and amplitude, as well as the exact form of the modulation, will likely elucidate the more detailed classification.

\textbf{NSINE:} Sinusoidal variables with much residual noise, or with evidence of additional variability not captured in the fit. Many spotted rotators with evolving spots likely fall into this class, as well as faint or low-amplitude $\delta$ Scuti stars and ellipsoidal variables.

\textbf{MIRA:} Mira variables. These stars are a subset of the LPV's that have photometric amplitudes exceeding 2.0 mag in either the cyan or orange filter. They generally show coherent periodicity, but the two-year temporal baseline of our data may in many cases be insufficient to solve for the period accurately.

\textbf{SHAV:} These are the slow high-amplitude variables, an extremely rare class with long periods and Mira-like amplitudes, but with color insufficiently red for a true Mira. Only 17 of these were identified in our entire catalog. They include AGN, R Coronae Borealis stars, and at least one apparent nova. As almost all of them are known (one of them is the archetypal variable R Cor Bor itself!), we do not have a figure showing unknown SHAV stars.

\textbf{LPV and IRR:} The acronyms stand for `long period' and `irregular' variables. These classes serve as `catch-all' bins for objects that do not seem to fit into any of our more specific categories. The LPV class contains objects whose variations appear to be dominated by low frequencies, corresponding to $P\gtrsim5$ days, while the IRR class contains objects whose dominant frequencies are higher. Most of the stars classified as LPV or IRR (especially the latter) don't show coherent variations that can be folded cleanly with a single period: hence, both classes are in some sense `irregular', though the characteristic timescales are different. A characteristic timescale (i.e. a dominant frequency) is usually present even though the data cannot be cleanly phased. This timescale likely corresponds to some astrophysical reality such as a rotation, orbital, or pulsation period. Both the LPV and IRR classes contain a significant minority of objects with coherent variations that can be cleanly phased with a single period. That such objects end up in our `catch-all' categories indicates that their periodic waveform, though coherent, is not a good match to any of our more specific classifications. Some of these may simply be faint or noisy examples of variables that should have fallen into one of the specific ATLAS categories, but were not identified by the machine classifier due to the low significance of the signal. However, others are likely periodic variables of well defined astrophysical types that don't fit any of the ATLAS classes --- e.g. spotted rotators such as BY Draconis variables. Among the objects that cannot be cleanly phased to a single period, the LPV class surely includes many semiregular red giant variables, while the IRR class has a large number of cataclysmic binaries. 

While a large majority of objects in both the LPV and IRR classes are expected to be true variables, a larger fraction may be spurious than in the foregoing, more specific categories. Readers interested in studying these objects using our catalog can easily select a purer sample of true variables. A very clean (though greatly reduced) sample would be obtained by simply requiring ddcSTAT=1. Alternatively, a more sophisticated selection could use thresholds on amplitude, PPFAP, the ratio of the raw RMS scatter to the residual RMS from the best fit, Hday, Hlong, or any of a number of other useful statistics described in \S \ref{sec:appendixA}. Examples of database queries relevant for such selections are given in \S \ref{sec:appendixB}.

\clearpage

\textbf{STOCH:} These are the variables that don't fit into any coherent periodic class, not even IRR. They would be classified as `dubious' except that they have ddcSTAT=1, meaning that detections on the difference images demonstrate their genuine variability. Their physical nature is unclear, but many of them do appear to exhibit highly significant stochastic variations with very little coherence on the timescales probed by ATLAS. Some of these may be very high-frequency variables with periods too short to be captured by {\tt lombscar} or {\tt fourierperiod}.

\subsection{The dubious variables}

\begin{figure}
\includegraphics[scale=0.85]{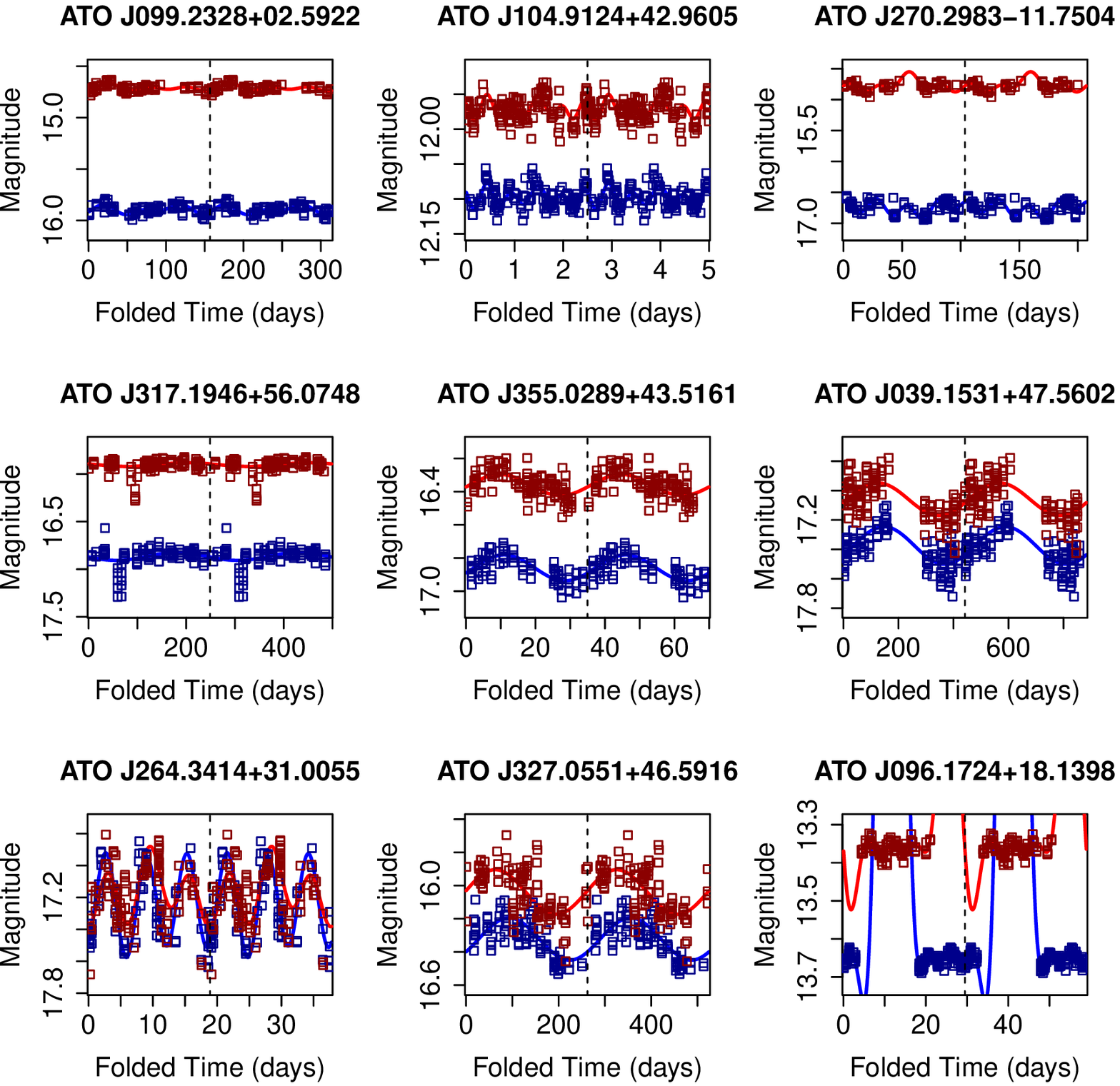}
\caption{\textbf{dubious:} Examples of our `dubious' class, which consists of candidates the machine classifier did not find to be convincingly variable. The plots are constructed as in Figure \ref{fig:CBF}, except that the magnitude scales are different for different panels. The offset between the $c$ and $o$ band data still indicates a star's color. Cases where the Fourier fits run away during intervals unconstrained by the data do not compromise our analysis. Unlike Figures \ref{fig:CBF}--\ref{fig:STOCH}, the stars shown here are not selected at random but are chosen by hand as the best nine variables from a sample of 450. They therefore represent the best 2\% of the `dubious' class, and they do appear to be genuine variables. 
\label{fig:dubious}}
\end{figure}

The majority (more than 90\%) of our candidate variables are designated by the machine classifier as `dubious', indicating that the significance of the variability is so low that we cannot be sure they are real. By-hand examination of randomly chosen samples suggests that at least 2\%, and probably 5--10\%, of these stars are actually real variables that were too faint or low-amplitude to classify definitively. These include faint RR Lyrae stars, eclipsing binaries, spotted rotators, and other classes. Many of these will likely be raised to the status of definitive variables in ATLAS DR2. Figure \ref{fig:dubious} shows the best 2\% from a randomly chosen sample of 450 `dubious' stars screened by hand. We are confident that all of them are true variables.

If our higher estimate of a 10\% variability fraction holds, as many real variables are to be found in our `dubious' category as in all the others put together. Even taking the minimum value of 2\%, we have more than 80,000 true variables classified as `dubious'. This illustrates a common shortcoming of machine classification: in order to obtain a fairly pure sample of true positives, as we have done, one must accept that a considerable number of good objects will be discarded. However, researchers interested in finding `gold in the mine-tailings' can likely use some of the variability features we have calculated for each star to identify many of the objects for which the `dubious' classification was incorrect. Additionally, we have preserved for every star the probabilities output by the machine learning for every class, including the aggregate `HARD' class (see \S \ref{sec:classify}). Presumably most of the `dubious' stars that are real variables will have probabilities significantly below 1.0 for `dubious' and/or `HARD'. It would be easy, for example, to perform a database query (see \S \ref{sec:appendixB}) that would select all the stars that were classified as `dubious' but had a probability less than 0.6 for the `dubious' class --- or, e.g., had a PULSE probability greater than 0.2. Such queries can likely be used not only to find real variables in our `dubious' category, but even real variables of specific types.

\section{Completeness} \label{sec:comp}

The completeness of our catalog can be considered in four stages, with the first being the probability that a variable star with particular properties will be included in the ATLAS lightcurve set; and stages 2, 3, and 4 referring to the probability that ATLAS will flag the star as, respectively, a candidate variable, probable variable, or classified variable assigned to the correct class (see Table \ref{tab:numbers} for an overview of these broad categories).

\begin{figure}
\plottwo{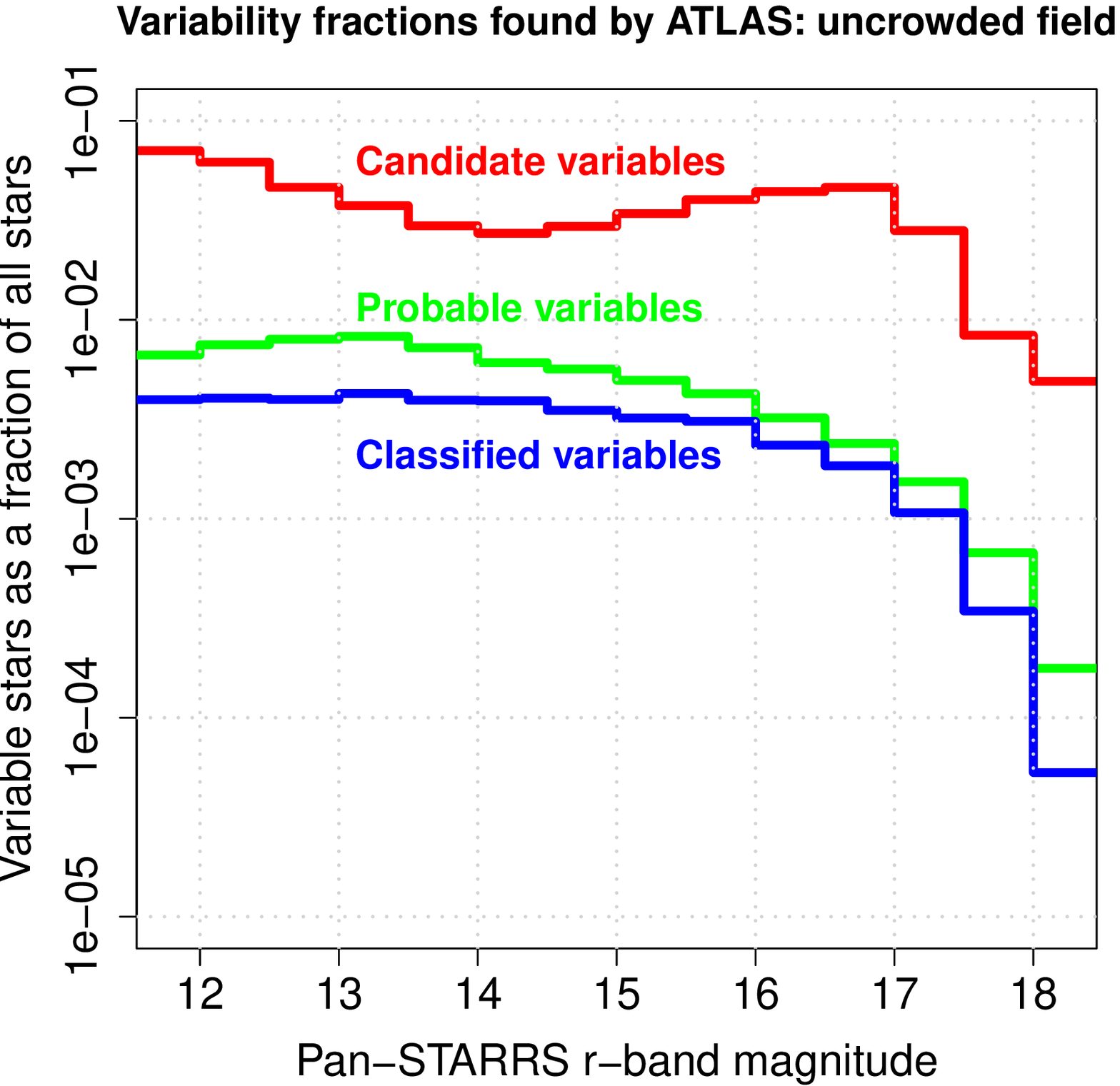}{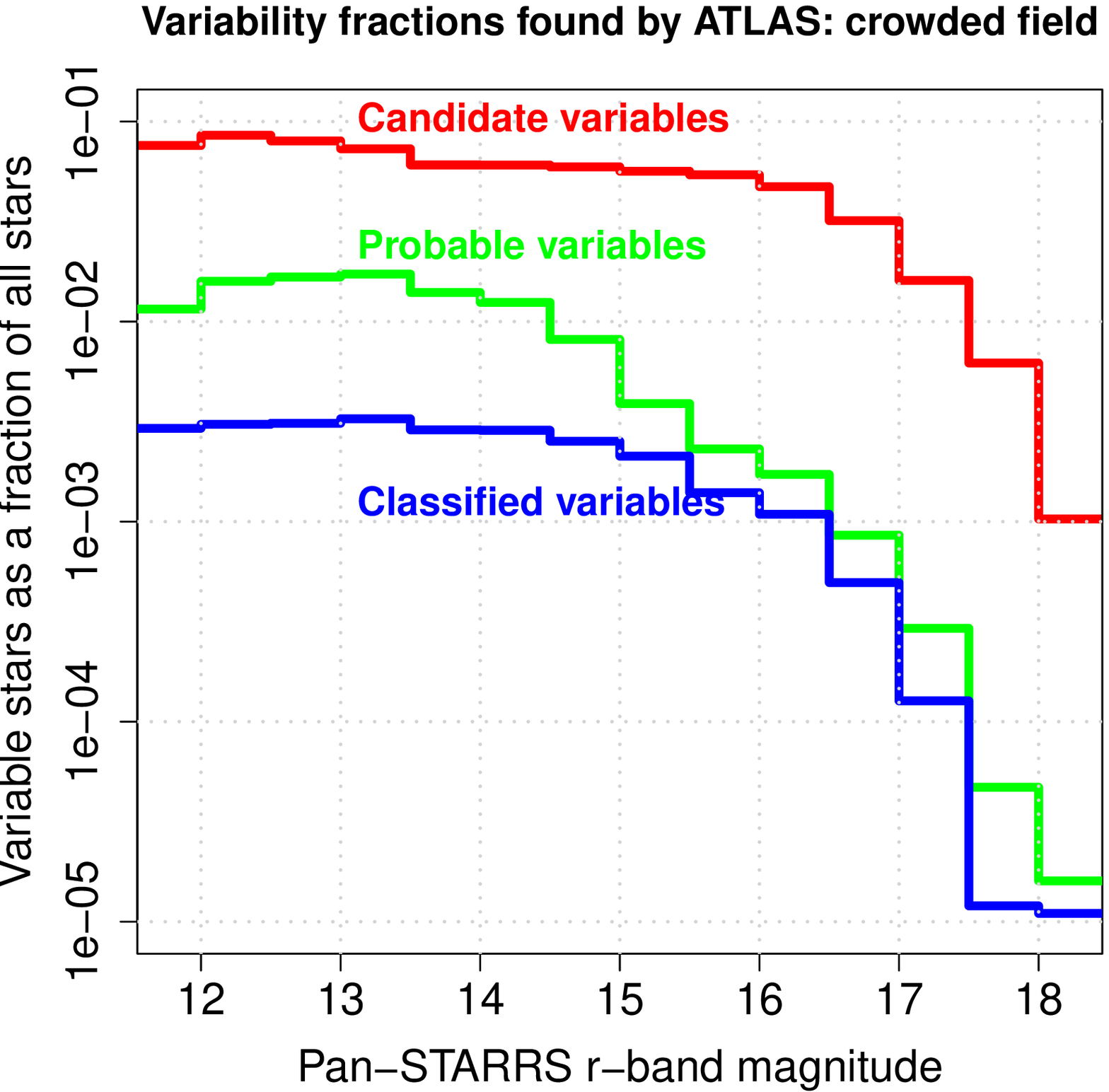}
\caption{Magnitude-dependent fractions of all stars in the object-matching catalog that were identified as variables with different levels of confidence. Bright stars are more likely to be identified as true variables relative to faint stars, because the greatly enhanced photometric precision means that smaller amplitude variations can be detected. Again, as expected, faint variables are particularly difficult to discover and classify in a crowded field.
\label{fig:maghist}}
\end{figure}

\begin{figure}
\plottwo{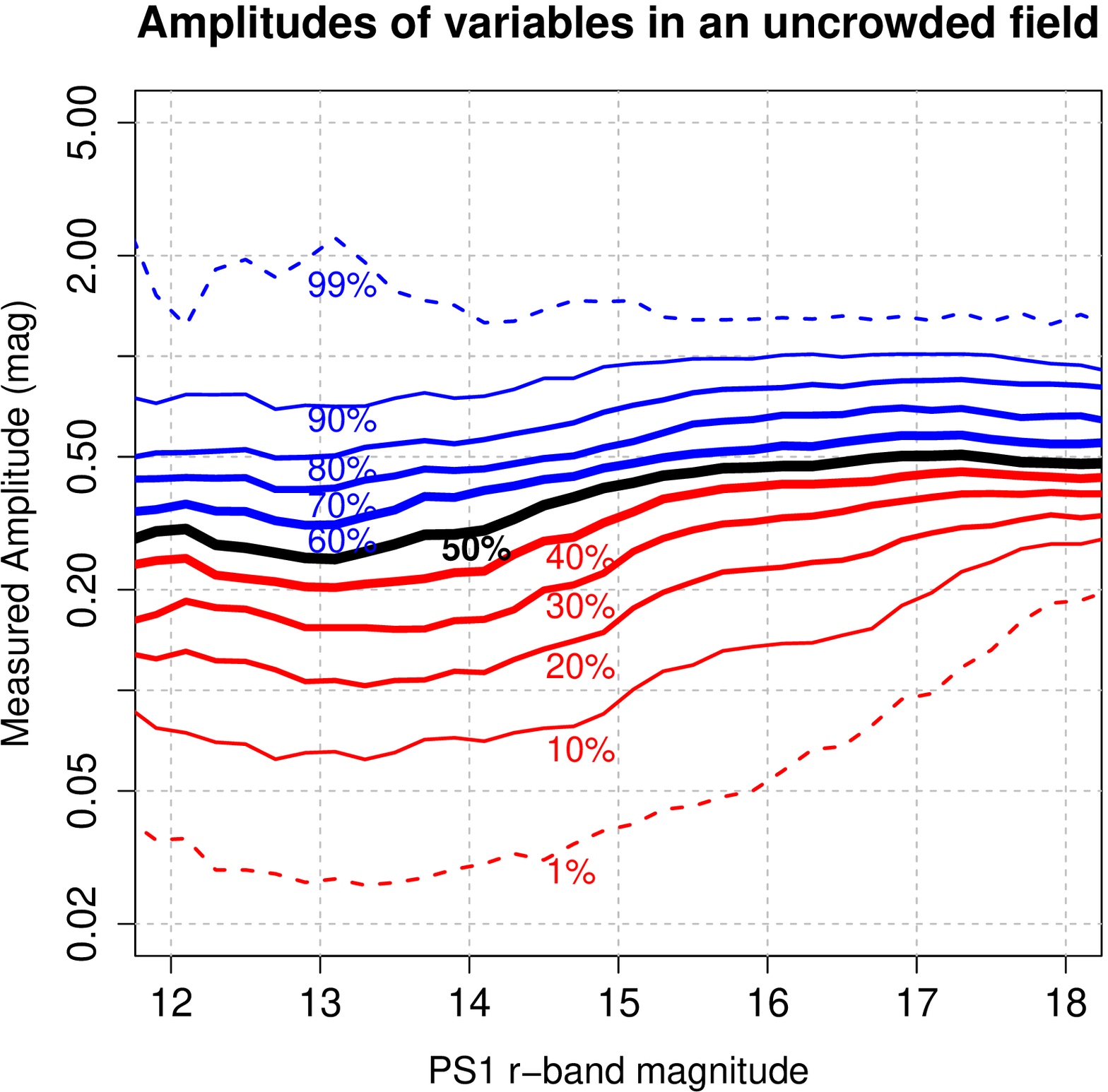}{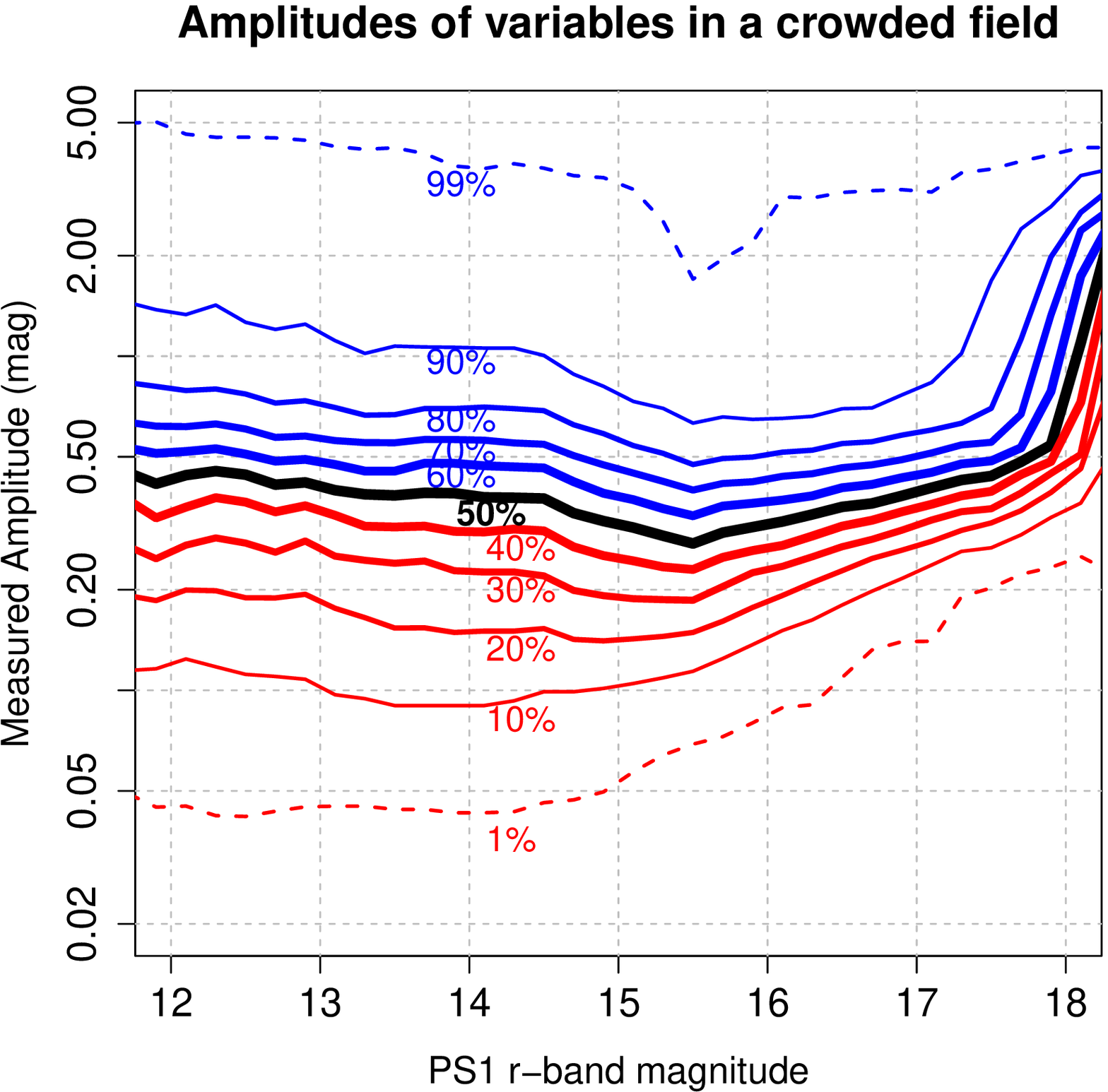}
\caption{Magnitude-dependent percentiles of peak-to-trough amplitudes for classified variables (plus our generic NSINE and LPV categories), in uncrowded and crowded fields. While ATLAS completeness is difficult to disentangle from the underlying distributions of astrophysical amplitudes (which are certainly different for different fields and magnitude regimes), the plots give an indication of the amplitudes to which ATLAS is sensitive.
\label{fig:amplitudes}}
\end{figure}

We have already addressed stage 1 completeness in \S \ref{sec:photdat} and Figure \ref{fig:meashist}. A full analysis of completeness stages 2--4 is far beyond the scope of the current work. It depends on several parameters including stellar magnitude, field-crowding (i.e. Galactic latitude), period, amplitude, and type of variability. A fully satisfactory solution would likely require an extensive simulation involving millions of fake variable stars spanning wide ranges in the parameters listed above. Although ATLAS does not at present have sufficient resources to carry out such an analysis, in this section we attempt to provide some rough indications of our stage 2--4 completeness, illustrated by Figures \ref{fig:maghist} and \ref{fig:amplitudes}. We hope to elucidate, at least, regimes where our stage 4 completeness is probably above 50\%, as compared to others where it is very low. We do not consider here the issue of classified variables with incorrect classifications, because it has already been discussed in \S \ref{sec:classify}.

In Figure \ref{fig:maghist}, we show the fractions of stars in our object-matching catalog that we identified as candidate (stage 2), probable (stage 3), or classified (stage 4) variables (see Table \ref{tab:numbers} and \S \ref{sec:classify} for further definitions of these subsets). The fractions are shown for uncrowded and crowded fields, centered on (but much larger than) the sky areas shown in Figures \ref{fig:imexamp2} and \ref{fig:imexamp}, respectively. Interpretation is made complicated by the reality that different astrophysical populations, with different variability fractions and amplitude distributions, are undoubtedly being probed in the two cases. For example, the crowded field, being near the galactic plane, certainly includes more semi-regular pulsating red giants, which tend to get classified as generic `LPV' stars (see \S \ref{sec:classify}) and hence inflate the `Probable' variables relative to `Classified' stars. Nevertheless it seems clear that crowding has only a modest effect for stars brighter than 15th mag, while by 18th mag it appears to reduce stage 4 completeness by roughly a factor of 10.

In Figure \ref{fig:amplitudes}, we show the percentiles of variability amplitude as as a function of magnitude for the same fields presented by Figure \ref{fig:maghist}. Figure \ref{fig:amplitudes} includes the classified variables plus the LPV and NSINE stars because their amplitudes are likely to be astrophysically meaningful. As with Figure \ref{fig:maghist}, interpretation is made more difficult by the certainty that different populations of stars, with different intrinsic amplitude distributions, are being probed by the two different fields (and even by the same field in different magnitude regimes). Nevertheless, the fact that the median amplitude remains at or below 0.5 mag (peak-to-trough) strongly suggests that ATLAS will identify as probable variables, if not definitively classify, most variable stars with with $r<17.5$ mag and amplitudes of at least 0.5 mag {\em provided they are included in the lightcurve set}. Hence, stage 3 and 4 completeness for variable discovery at an amplitude of 0.5 mag is mostly dependent on mere inclusion in the lightcurve set, i.e. stage 1 completness, which is relatively easy to quantify and which we have already addressed in \S \ref{sec:photdat}. On the other hand, Figure \ref{fig:amplitudes} suggests that our stage 4 completeness at an amplitude of 0.1 mag becomes very low for stars fainter than 16th mag.

We can also conclude that even though we have discovered variables with amplitudes as small as 0.02 mag, our completeness (stage 3 or 4) for them is probably very low at any brightness. We can say this despite the uncertainties of the underlying astrophysical distribution, since it is known \citep[e.g.][]{McQuillan2012} that the occurrence rate of variability in main sequence stars rises from an amplitude of 0.1 mag to 0.01 mag. Thus, the fact that less than 1\% of our stars have measured amplitudes below 0.02 mag suggests that we reliably identify such objects as probable variables only under unusually favorable circumstances --- e.g., an area of the sky with much higher-than-average coverage due to overlapping Dec bands or other effects. On the other hand, we think it likely that a large fraction of all bright ($r<15$ mag) uncrowded stars with $\sim 0.02$ mag variability have been included among our candidate variables, but not confirmed. As such, their ATLAS photometry is publicly available, and interested researchers may be able to identify them with an appropriate query to our database (see \S \ref{sec:atquery}).

\section{Connection of ATLAS Variables with Astrophysics} \label{sec:substructure}

\subsection{Fourier Phase Offsets}

\begin{figure}
\includegraphics{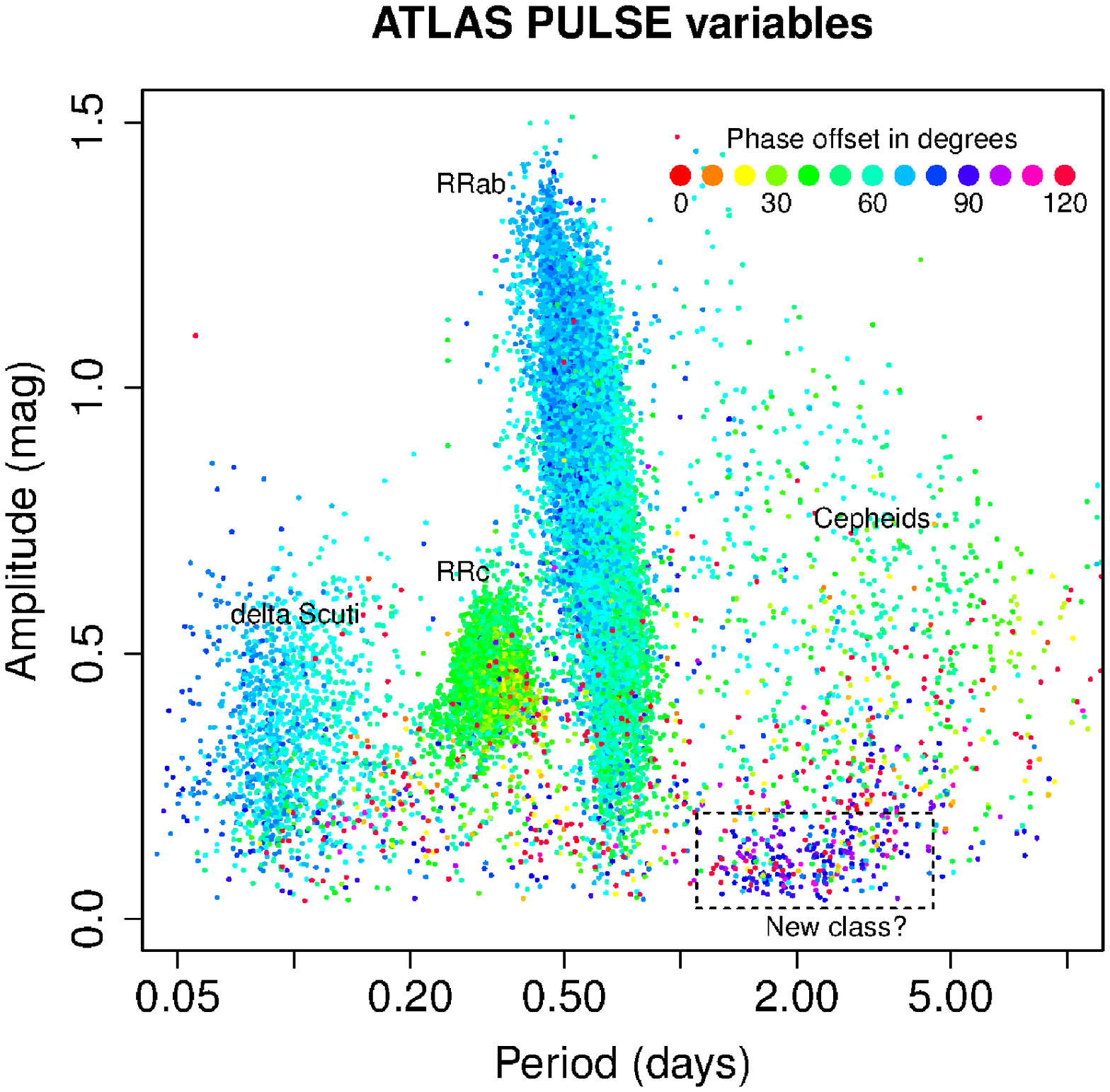}
\caption{
Amplitude vs. period for our PULSE variables. 
Color encodes $\phi_2 - \phi_1$, the phase offset of the first and second Fourier terms. 
Phase offsets $>120^{\circ}$ (very rare in this class) have been scaled 
to $120^{\circ}$ to focus on the interesting regime. The astrophysical sequences of the 
RRab and RRc types are clearly resolved, as are the $\delta$ Scuti variables. 
RRc variables have more symmetrical lightcurves, while $\delta$ Scuti and RRab stars 
have similarly shaped sawtooth lightcurves. The dark blue and purple points in the box 
at lower right are stars with phase shifts centered near 90$^{\circ}$, indicating 
unusual and distinctive lightcurves with narrow, symmetrical maxima --- possibly 
a new class of variables 
(see \S \ref{sec:UCBH}).
\label{fig:pulsephase}}
\end{figure}

Our current analysis using machine learning augmented with screening by hand has barely scratched the surface of the rich data mine comprising the 169 features we calculate for each variable star, much less the photometric data itself. Figure \ref{fig:pulsephase} gives one example of such rich information. The plot shows amplitude vs. period for the variables in the PULSE category, with each star color coded according to the phase offset $\phi_2 - \phi_1$ of its first two Fourier terms. The astrophysical sequences of the RRab and RRc stars are clearly resolved, as are the $\delta$ Scuti variables. RRc variables have smaller phase offsets indicative of their more symmetrical lightcurves, while the similar colors of the $\delta$ Scuti and RRab stars indicate their similar, highly asymmetrical sawtooth lightcurves. 

In the lower right of the plot is a remarkable cluster of variables with small amplitudes ($\le 0.2$ mag, periods from 1--5 days, and phase shifts clustered near $90^{\circ}$. These are lightcurves of a very different type, and may indicate a new class of variables. Whether they are actually pulsators is unclear. These objects are discussed further in \ref{sec:UCBH}.

\begin{figure}
\includegraphics{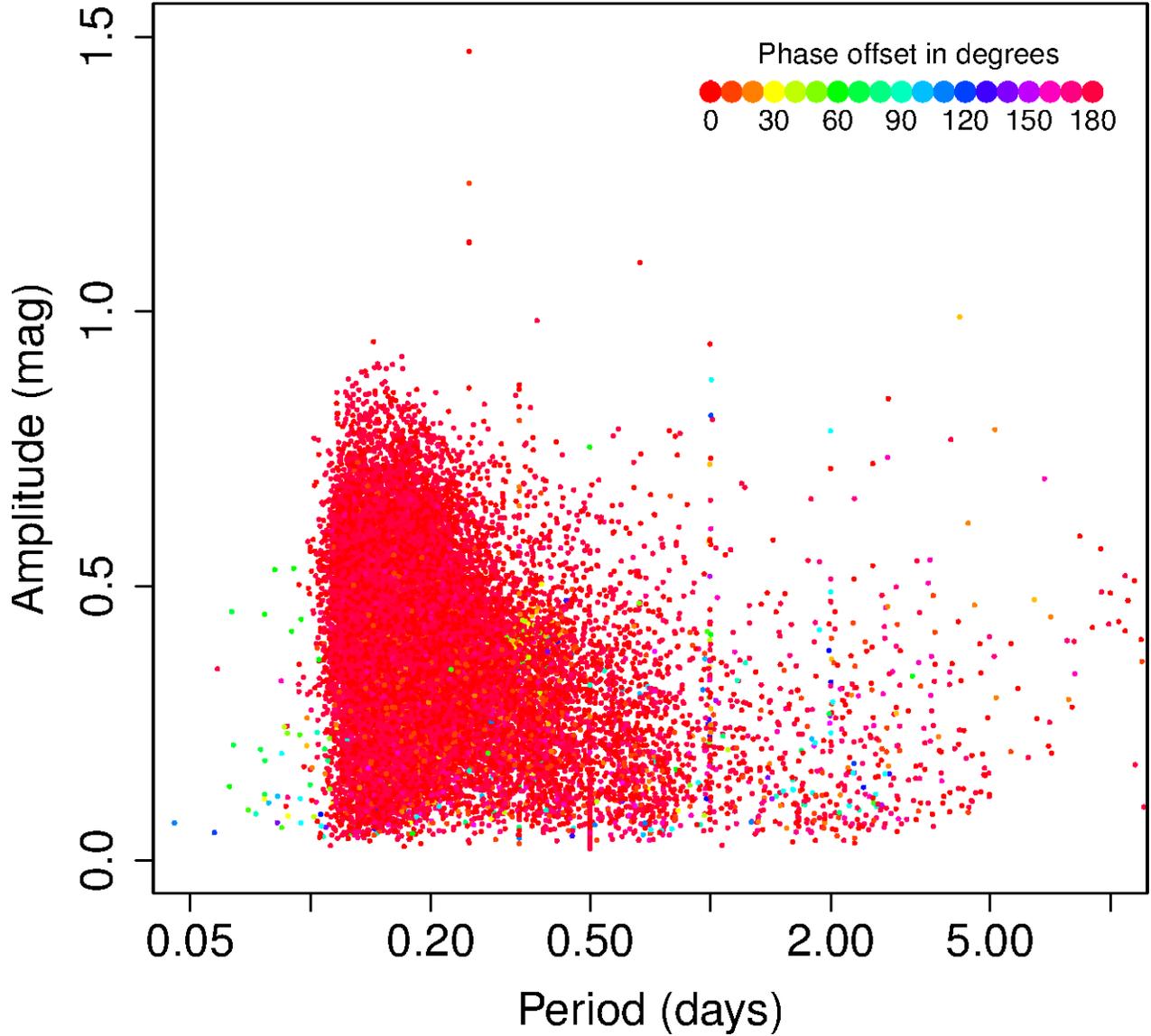}
\caption{Amplitude vs. period for CBH eclipsing binary stars, color-coded by phase offset $\phi_2 - \phi_1$. In striking contrast to the pulsating variables of Figure \ref{fig:pulsephase}, eclipsing binaries in our CBH class almost all have phase offset near 0$^{\circ}$ or 180$^{\circ}$, indicating that the fit has aligned the minima of the first and second Fourier terms to build up a deep, narrow eclipse. Slight concentrations of points at periods of $\frac{1}{3}$, $\frac{1}{2}$, 1, and 2 days are due to aliased long-period variables.
\label{fig:CBHphase}}
\end{figure}

In constrast to the rich diversity of phase offsets among pulsating variables, Figure \ref{fig:CBHphase} shows that eclipsing binaries in our CBH class almost all have phase offsets near $0^{\circ}$ or $180^{\circ}$, indicating that the minima of the first and second Fourier terms are being aligned to produce a relatively deep and narrow eclipse. They are clearly distinguishable from the RRc stars in this analysis, though the two types are commonly confused.

\begin{figure}
\includegraphics{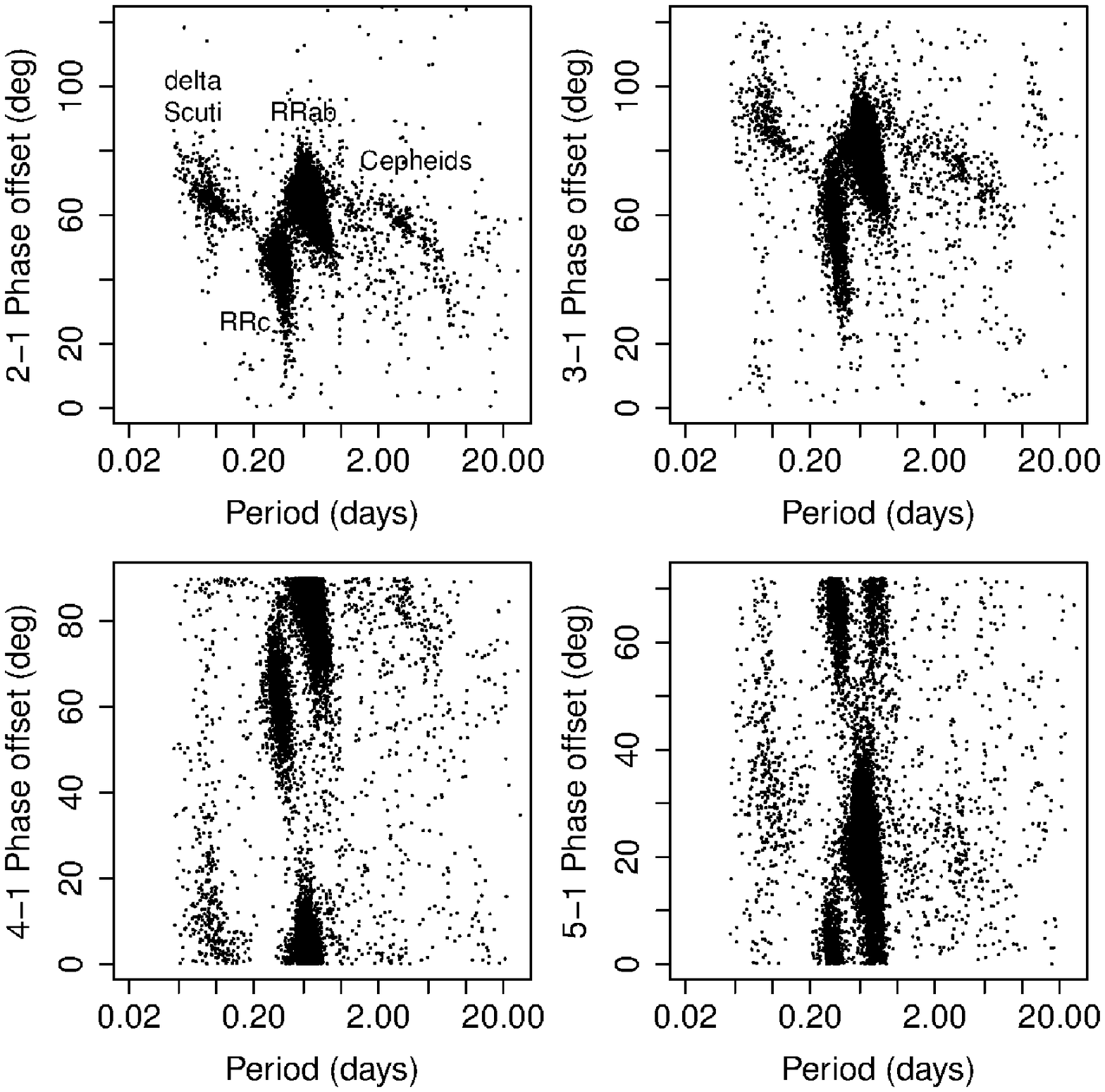}
\caption{Phase offset vs. period for variables classified as pulsators and fit with the maximum of six Fourier terms. The offsets $\phi_2 - \phi_1$, $\phi_3 - \phi_1$, $\phi_4 - \phi_1$, $\phi_5 - \phi_1$ are shown in the respective panels. The astrophysical classes labeled at top left repeat with each different phase offset. Several intriguing sequences can be identified in this figure. For examples of how such groupings can be linked to specific types of Cepheid variables, see \citet{Soszynski2017}.
\label{fig:pulsephase2}}
\end{figure}

Figure \ref{fig:pulsephase2} illustrates even finer substructure by plotting the phase offsets of the 2nd, 3rd, and 4th Fourier term relative to the first term for pulsating variables fit with the maximum number of six Fourier terms. As described in \S \ref{sec:classify}, the maximum possible offset decreases for higher-order Fourier terms, being always equal to 360$^{\circ}/m$ for the mth Fourier term. The differences between RRab and RRc type variables seen in Figure \ref{fig:pulsephase} appear here as well, as do the similarities between the RRab and $\delta$ Scuti stars. Additional substructure in each of these groups also appears, indicating systematic changes in lightcurve shape with period. New, less populated sequences appear at longer periods, which may correspond to Cepheids and W Virginis variables.

\clearpage

\subsection{Astrophysical Nature of SINE and NSINE Classes}

\begin{figure}
\includegraphics{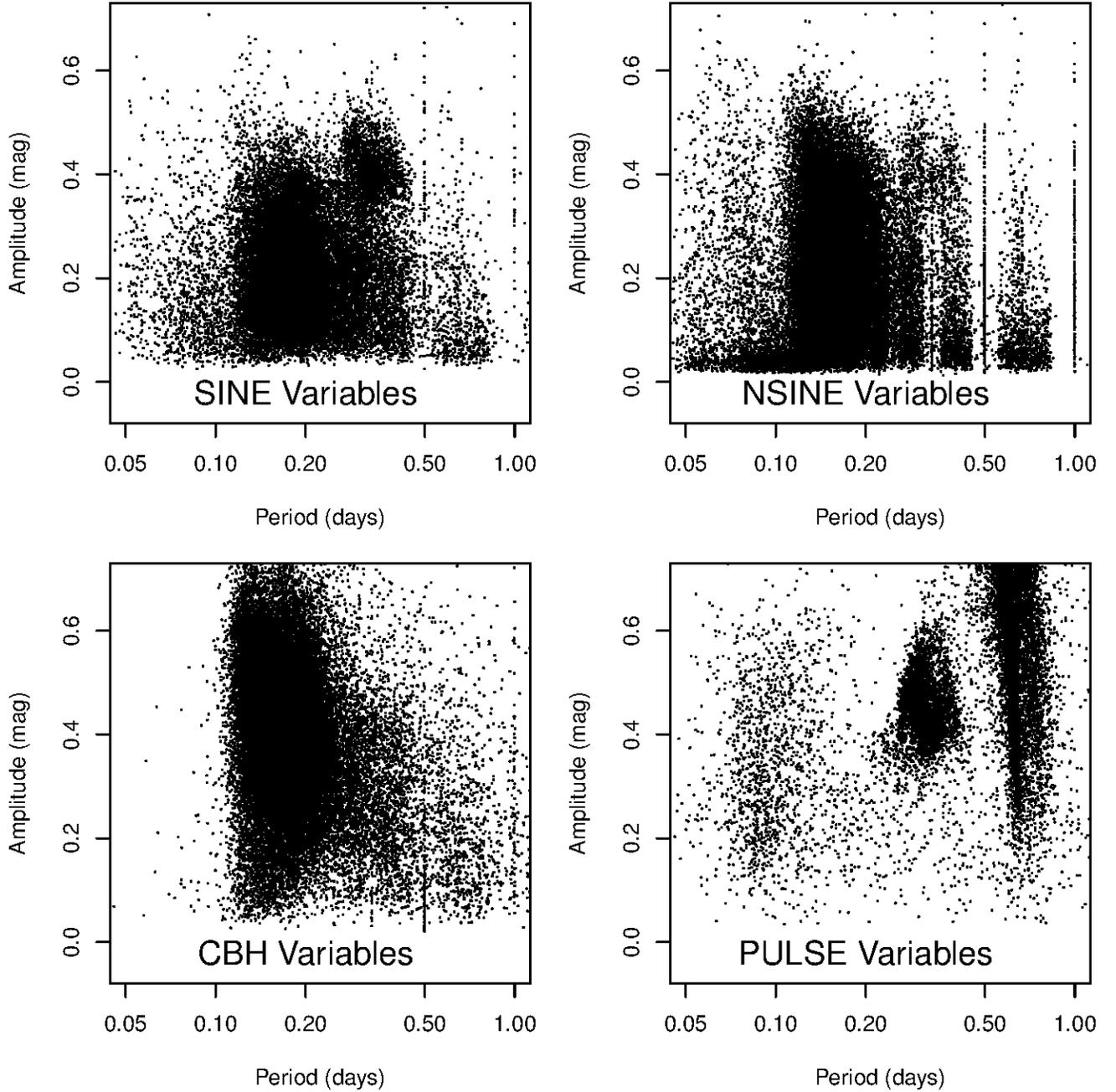}
\caption{Amplitude vs. period plots reveal the astrophysical nature of many of our SINE and NSINE variables. The vertical edge at a $P=0.11$ (corresponding to a true orbital period of 0.22 days) is present in the CBH, SINE, and NSINE plots, indicating that many stars in the latter two classes are close binaries --- i.e., ellipsoidal variables or misclassified low-amplitude eclipsers. From shortest period to longest, the three clumps in the PULSE plot are the $\delta$ Scuti, RRc, and RRab variables. Comparison with the SINE diagram indicates that some RRc but very few RRab are classified as SINE. Most of the objects shortward of the close binary period limit in both SINE and NSINE are $\delta$ Scuti stars. As in Figure \ref{fig:CBHphase}, aliased long-period variables produce mild concentrations at periods of $\frac{1}{3}$, $\frac{1}{2}$, and 1.0 days.
\label{fig:peramp}}
\end{figure}

Even without phase offset information, which is not applicable to stars fit with a pure sinusoid, we can use features of the amplitude vs. period distributions to probe the astrophysical nature of SINE and NSINE stars. Figure \ref{fig:peramp} shows the amplitude and period for the SINE and NSINE stars compared with CBH and PULSE. 

Several interesting facts emerge. A considerable number of RRc variables, but very few RRab, have been classified as SINE. This is not surpising since the lightcurves of RRc stars are much more sinusoidal than the strong sawtooth curves of the RRab --- a fact that is also indicated by Figure \ref{fig:pulsephase}. Interestingly, the number of RRc stars classified as NSINE is much smaller even though the NSINE class is far more populous than SINE.

A clear vertical edge is seen at a period of about 0.11 days in the CBH, SINE, and NSINE plots. This corresponds to the well known 0.22 day orbital period cutoff for contact binaries \citep{Drake2014b,Soszynski2015}, and indicates that many of the SINE and NSINE stars are close binary systems. The SINE and especially NSINE classes extend to far lower amplitudes than CBH. Although some of the high amplitude SINE and NSINE variables are certainly misclassified eclipsing binaries, the low-amplitude majority are mostly ellipsoidal variables in systems astrophysically similar to the CBH stars but insufficiently inclined to our line of sight to exhibit eclipses.  

Comparison between PULSE, SINE, and NSINE distributions in Figure \ref{fig:peramp} strongly suggests that the shortest-period $\delta$ Scutis in our sample have been designated SINE or NSINE. The likely explanation is that their sawtooth waveforms contained frequencies outside the range of our Fourier analysis.

The lowest-amplitude SINE and especially NSINE stars at periods longer than 0.3 days are likely to be spotted rotators rather than ellipsoidal variables. Additionally, many SINE and NSINE stars have periods longer than the maximum covered by Figure \ref{fig:peramp}. We believe most of these longer-period stars are spotted rotators, although some ellipsoidal variables will exist among them, especially in the SINE class.

\subsection{The Color-Dependent Short-Period Limit of Eclipsing Binaries} \label{sec:colorperiod}

\begin{figure}
\includegraphics{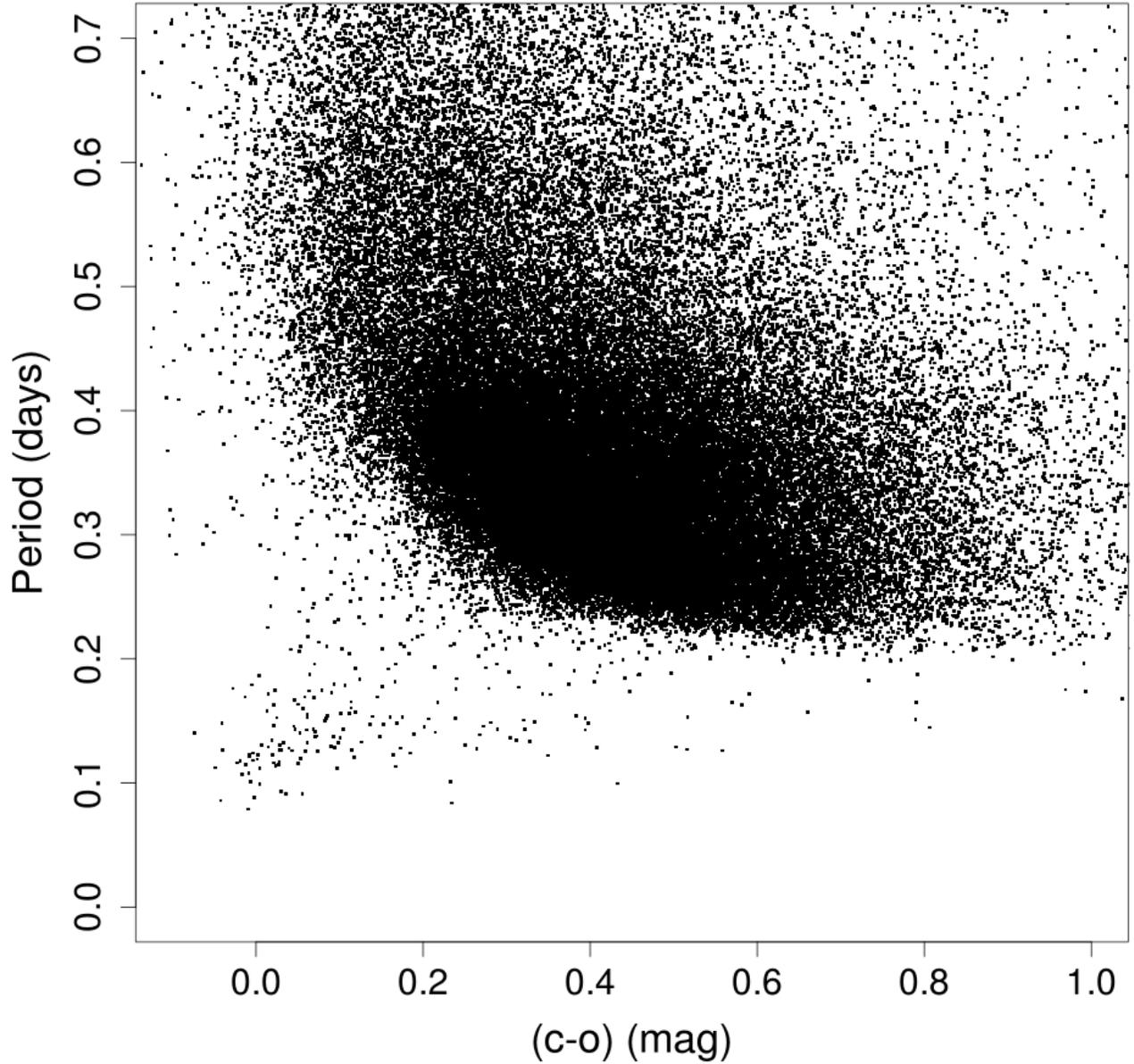}
\caption{The color-dependent short-period limit of eclipsing binaries. As predicted in the text, the lower envelope of the period distribution goes to shorter periods for redder stars, which have lower masses and higher mean densities on the main sequence. The envelope turns sharply when it encounters the well-known period limit at 0.22 days, but interestingly it continues to descend at a much slower rate. The sparse distribution of blue objects at very short periods may be subdwarf binaries or misclassified $\delta$ Scuti stars.
\label{fig:colorperiod}}
\end{figure}

A simplified and approximate analysis using Kepler's Third Law suggests the period of a contact binary (at least, one with equal-mass stars) should scale as $\rho^{-1/2}$, where $\rho$ is the mean density of the stars. Since more massive stars have lower mean densities when on the main sequence, it follows that contact binaries composed of more massive main-sequence stars should have longer periods. Since evolved stars have lower densities than main sequence stars of equal mass, and contact binaries have shorter periods than detached binaries of equal mass, it follows that for stars of a given mass, the shortest-period binary star will be a contact binary composed of main-sequence stars --- and that the period of this shortest-period binary star will increase with the masses of the components. Since more massive main sequence stars have bluer colors, we predict that a plot of period vs. color for eclipsing binary systems will show a lower envelope in period that increases toward the blue. We test this prediction with Figure \ref{fig:colorperiod}.

Although the train of reasoning above ignores many details (such as the distinction between contact and over-contact systems, and that fact that mutual interactions tend to cause contact binaries {\em not} to be normal main-sequence stars), Figure \ref{fig:colorperiod} confirms its basic prediction: from the bluest objects plotted redward to a $c-o$ color of about 0.3 mag, the lower envelope of the period distribution descends sharply. This suggests that, indeed, more massive contact binaries have longer periods.

The trend we have predicted experiences a sharp change at a $c-o$ color of 0.3 mag, and the envelope does not continue steeply descending at redder colors. This is because it has run into the short-period limit at of $P=0.22$ days for contact binaries, which is well known although its astrophysical causes are not understood in detail \citep{Drake2014b,Soszynski2015}. It seems likely that it has to do with a limiting mass near 0.6 $M_{\sun}$, below which main-sequence binaries are not able to evolve into a state of contact, perhaps due to the greatly reduced efficiency of wind-driven angular momentum loss in such low mass stars (\citet{Soszynski2015} and references therein). In this context, it is interesting that the envelope in Figure \ref{fig:colorperiod} {\em does} continue to descend, albeit much more slowly, from a $c-o$ color of 0.3 mag out to the reddest objects plotted. This may be a clue to the detailed astrophysics of the 0.22 day period limit.

\subsection{Galactic Distributions of Variable Classes}

\begin{figure}
\includegraphics[scale=0.85]{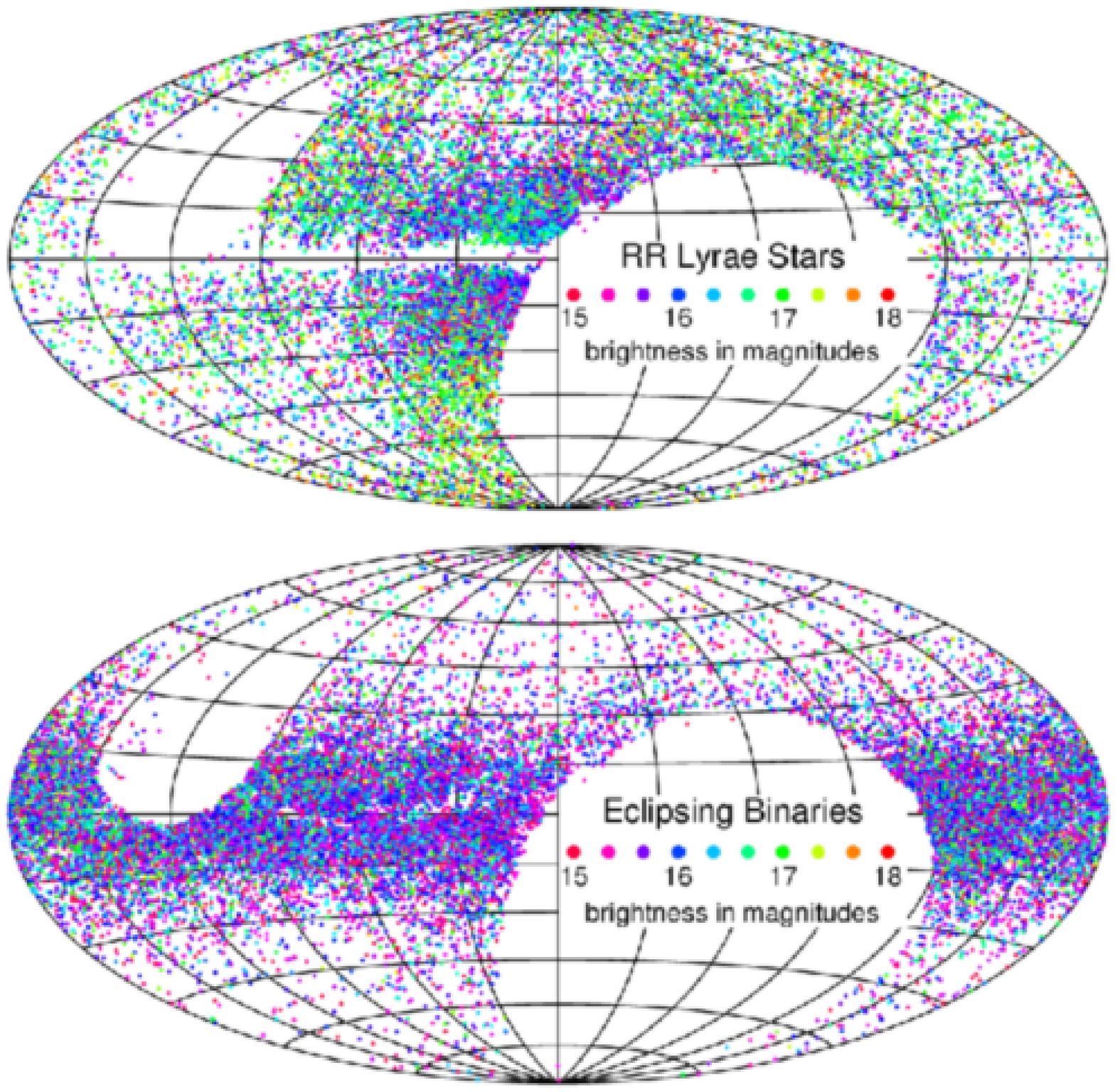}
\caption{Distribution of RR Lyrae stars and eclipsing binaries in Galactic coordinates, color coded to give the `o' band magnitudes. Empty areas are at Dec$<-30^{\circ}$ and $>+60^{\circ}$. The RR Lyrae stars populate the bulge and halo --- while the eclipsing binaries, being a younger population of mostly main-sequence objects, are found in the Galactic disk. The fainter RR Lyrae stars begin to reveal substructure in the inner halo, though the current ATLAS catalog doesn't extend to faint enough magnitudes to probe tidal streams in the outer halo as was done, e.g., by \citet{Drake2013a,Drake2013b,Hernitscheck2016}; and \citet{Cohen2017} using much larger telescopes.
\label{fig:galdist01}}
\end{figure}

It is very interesting to plot the on-sky distribution of our different classes of variables in Galactic coordinates. We show such plots for RR Lyrae stars and for all eclipsing binaries (CBF, CBH, DBF, and DBH) in Figure \ref{fig:galdist01}. The eclipsing binaries are strongly clustered to the Galactic plane, clearly indicating a disk population. By contrast, the RR Lyrae stars are distributed widely across the sky with very little preference for the Galactic plane, but with a strong concentration toward the Galactic center --- the signature of an old, halo population. Although our sensitivity is not sufficient to probe star streams in the outer halo as was done by \citet{Drake2013a,Drake2013b,Hernitscheck2016} and \citet{Cohen2017} using much larger telescopes, the faintest RR Lyrae stars plotted in Figure \ref{fig:galdist01} do indicate significant substructure in the halo at distances less than 30 kpc. 

\begin{figure}
\plottwo{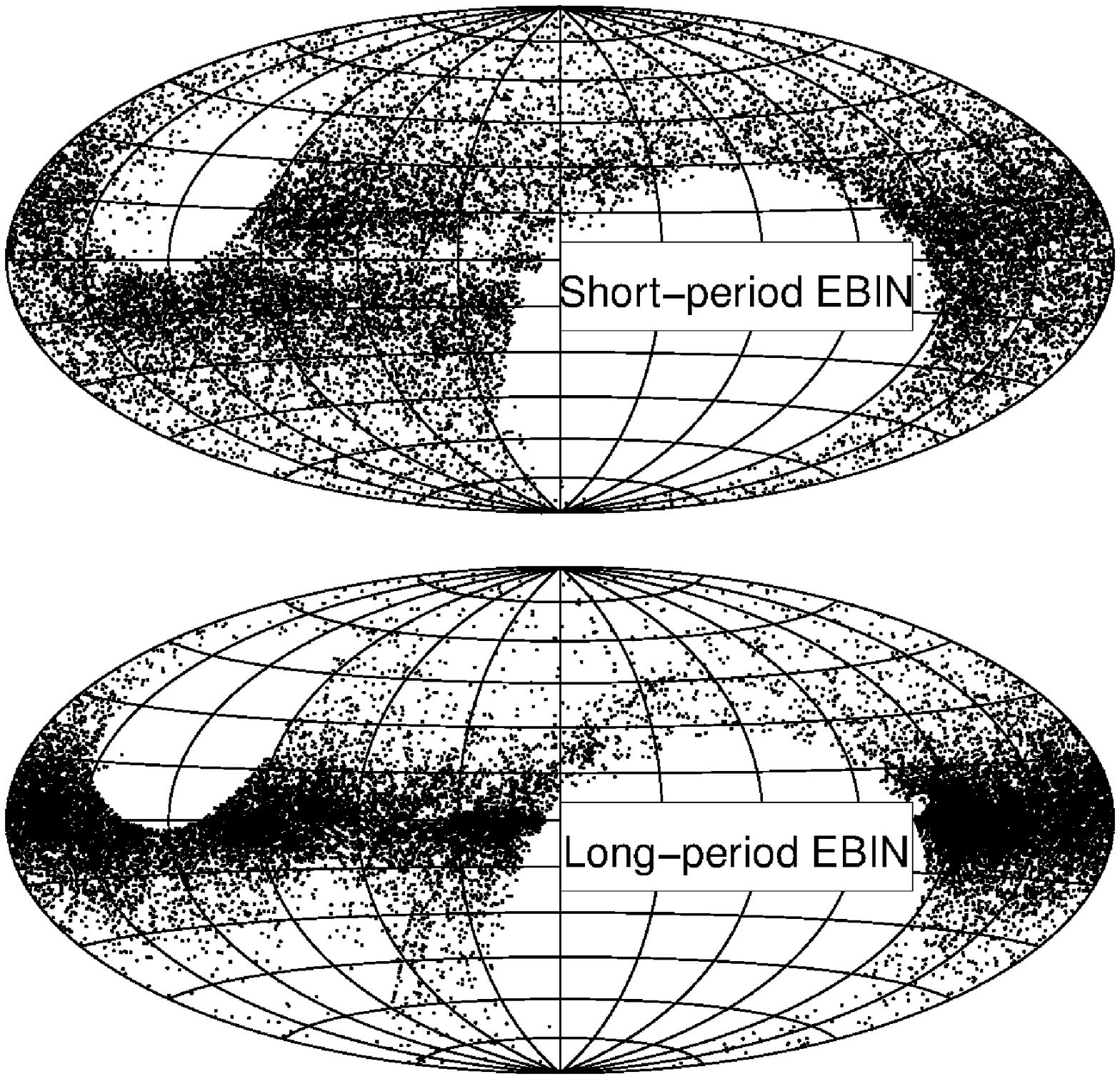}{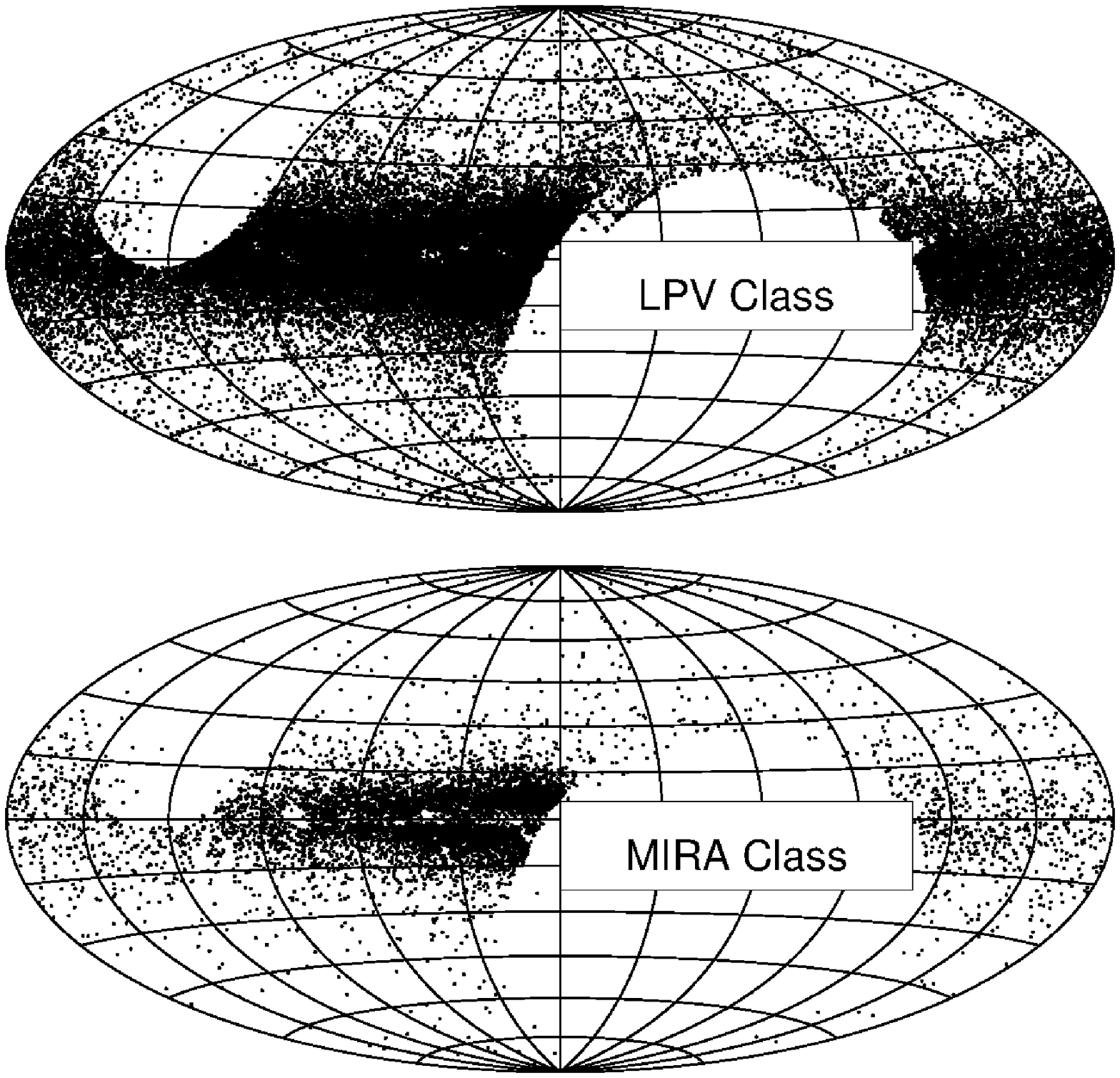}
\caption{\textbf{Left:} Distribution in Galactic coordinates of eclipsing binaries (CBF, CBH, DBF, and DBH classes) with orbital periods shorter than 0.3 days (top) and longer than 1.04 days (bottom). The period thresholds were chosen to yield equal-sized samples from both extremes of the period distribution. \textbf{Right:} Distribution in Galactic coordinates of our LPV (top) and MIRA classes (bottom). The short-period binaries are least luminous and can only be seen at a distances less than a few times the thickness of the Galactic disk --- hence, they appear only mildly concentrated toward the Galactic equator. The very luminous Mira stars, by contrast, are visible all the way across the Galaxy, and therefore they appear strongly concentrated in the direction of the bulge.
\label{fig:ShortLong}}
\end{figure}

While Figure \ref{fig:galdist01} probed disk vs. halo populations, it is interesting also to attempt to divide stars into different luminosity classes based on the characteristics of their variability. From the discussion in \S \ref{sec:colorperiod}, we would predict that the shortest-period eclipsing binaries would be low-mass stars with very low intrinsic luminosities, while longer-period eclipsing binaries should generally be more luminous. Many of the LPVs will be evolved stars and hence should be even more luminous than the longer-period eclipsing binaries, which we expect will mostly still be on or near the main sequence. Finally, the Mira variables, being asymptotic giant branch (AGB) stars, should have the greatest mean luminosity of any class. Figure \ref{fig:ShortLong} shows the on-sky distribution of each of these four types of objects in Galactic coordinates.

We would expect that the lowest-luminosity objects will be invisble to ATLAS at large distances, and hence all of them will be quite local. If the maximum distance at which we can detect them is only a few times the thickness of the Galactic disk, then the on-sky distribution will show only mild clustering toward the Galactic plane. By contrast, more luminous objects that can be seen at distances equal to many times the thickness of the disk should show a stronger clustering in the Galactic plane. The enormous concentration of most types of stars in the Galactic bulge should dictate that the on-sky distribution of any type of star that is luminous enough to be seen at the distance of the bulge will be strongly concentrated in the direction of the Galactic center. Figure \ref{fig:ShortLong} bears all of these expectations out beautifully. While all four types of objects are clearly disk rather than halo populations, their differing luminosities result in very different distributions on the sky, with the Mira variables, being visible at enormous distances, concentrated in the direction of the Galactic center.

\section{Interesting and Mysterious Subtypes} \label{sec:mystery}

\subsection{`Upside-down CBH' Variables} \label{sec:UCBH}

\begin{figure}
\includegraphics{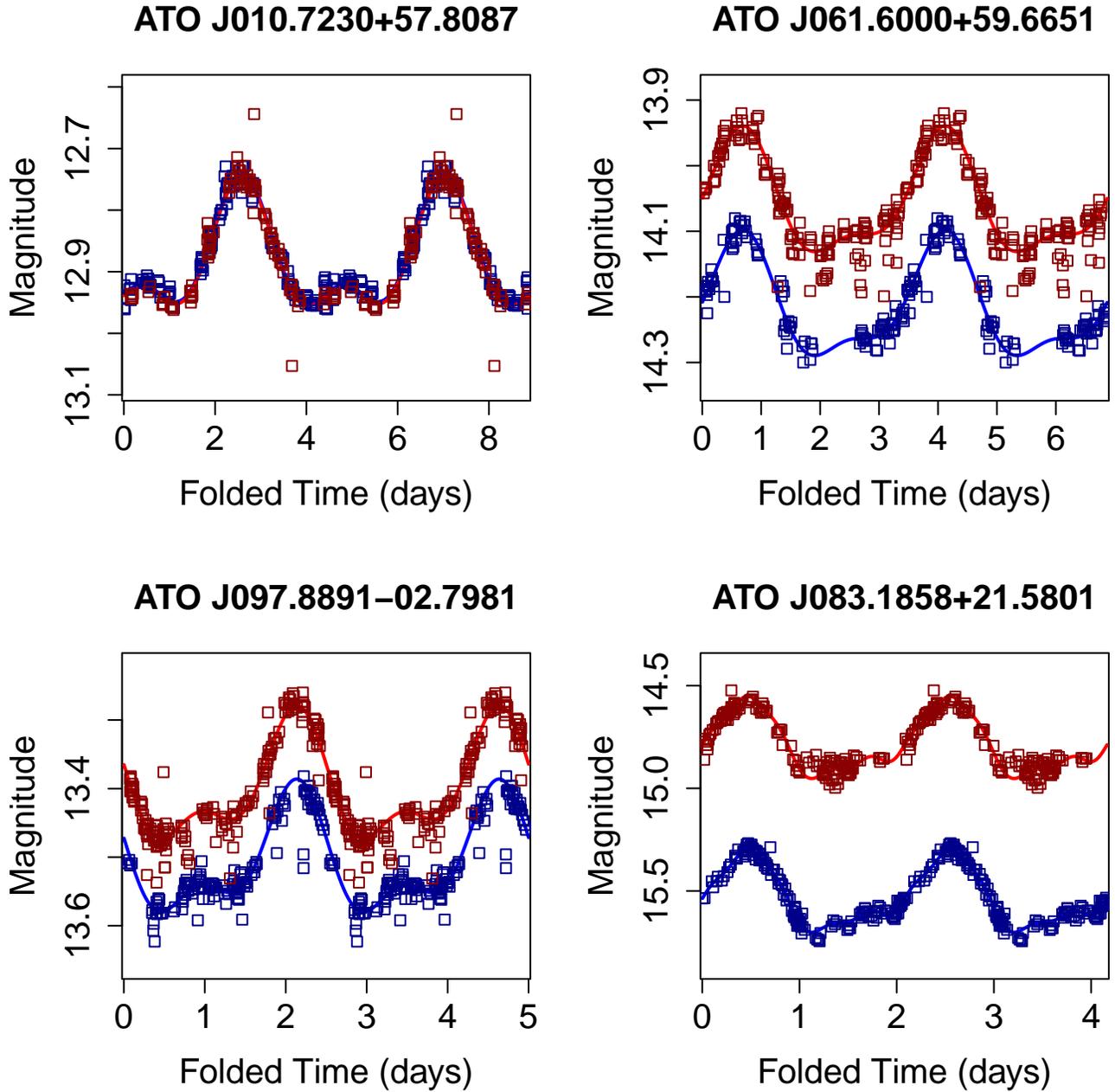}
\caption{Plots of 4 representative `upside-down CBH' variables. These stars are the possible new class of variables labeled at the lower right in Figure \ref{fig:pulsephase}: they are symmetrical, low amplitude variables with periods from 1 to 5 days. Their astrophysical nature is unclear.
\label{fig:UCBH}}
\end{figure}

These objects correspond to the possible new class of variables labeled in Figure \ref{fig:pulsephase}. We first noticed them long before constructing the figure, when we were screening lightcurves manually in order to construct the training set for machine learning. A distinctive lightcurve shape, not matching any known type of variable, was seen repeatedly in the course of our screening. When we made Figure \ref{fig:pulsephase}, we were able to confirm the connection between the unusual lightcurves and the unusual cluster of points. As implied by their location in that figure, these stars exhibit low-amplitude variations with periods ranging from about 1 to 5 days --- and consistent with the clustering of their phase offsets near 90$^{\circ}$, they have narrow, symmetrical maxima very similar to the minima of CBH eclipsing binaries. Figure \ref{fig:UCBH} shows four of the most representative lightcurves. In the course of by-hand screening, we have identified a total of about 70 such objects, but there are probably many more in our catalog.

For lack of a better term, we refer to these objects as `upside-down CBH' variables, since their lightcurves do indeed look almost exactly like a CBH turned upside down. This inversion explains why they have $\phi_2 - \phi_1$ phase offsets near 90$^{\circ}$, in contrast to 0$^{\circ}$ or 180$^{\circ}$ for real CBH systems. For CBH stars, the minima of the first two Fourier terms align to produce a deep, narrow eclipse, while for the upside down objects, the maxima align to produce a tall, narrow peak. The upside-down CBH variables are certaintly not any type of eclipsing binary. Our machine classification designates most of them as PULSE. It seems conceivable that they do represent a new type of pulsating variable, although other possibilities exist, as we discuss below.

Almost all of these variables are ATLAS discoveries. In the few cases that do appear in the VSX, classifications are not consistent: examples include `EW' (contact binary), `ROT' (spotted rotator), and `R' (binary star with strong reflection effects). Given the shapes of our lightcurves, the `EW' and `R' classifications can be ruled out immediately (the latter because it should produce a pure sinusoid). Since a conspiracy of spots could produce almost any type of waveform, a spotted rotator cannot be ruled out in any particular case. However, the fact that the objects appear to constitute a well-defined class with consistent waveforms does seem to rule out ordinary spotted rotators.

Another possible explanation is that they are binary systems in which a compact object is accreting material from a giant companion through an optically thick accretion disk. The stream of accreting material could then be making a bright spot where it impacts the accretion disk (presumably setting up a standing shock), and the periodic appearance and disappearence of this spot as the stars orbit one another would explain the variability we see. However, the luminosity of the standing shock would have to be remarkably consistent over time to explain the very coherent lightcurves we have observed for these objects.

\subsection{Eclipsing Binaries Showing the O'Connell Effect} 

\begin{figure}
\includegraphics{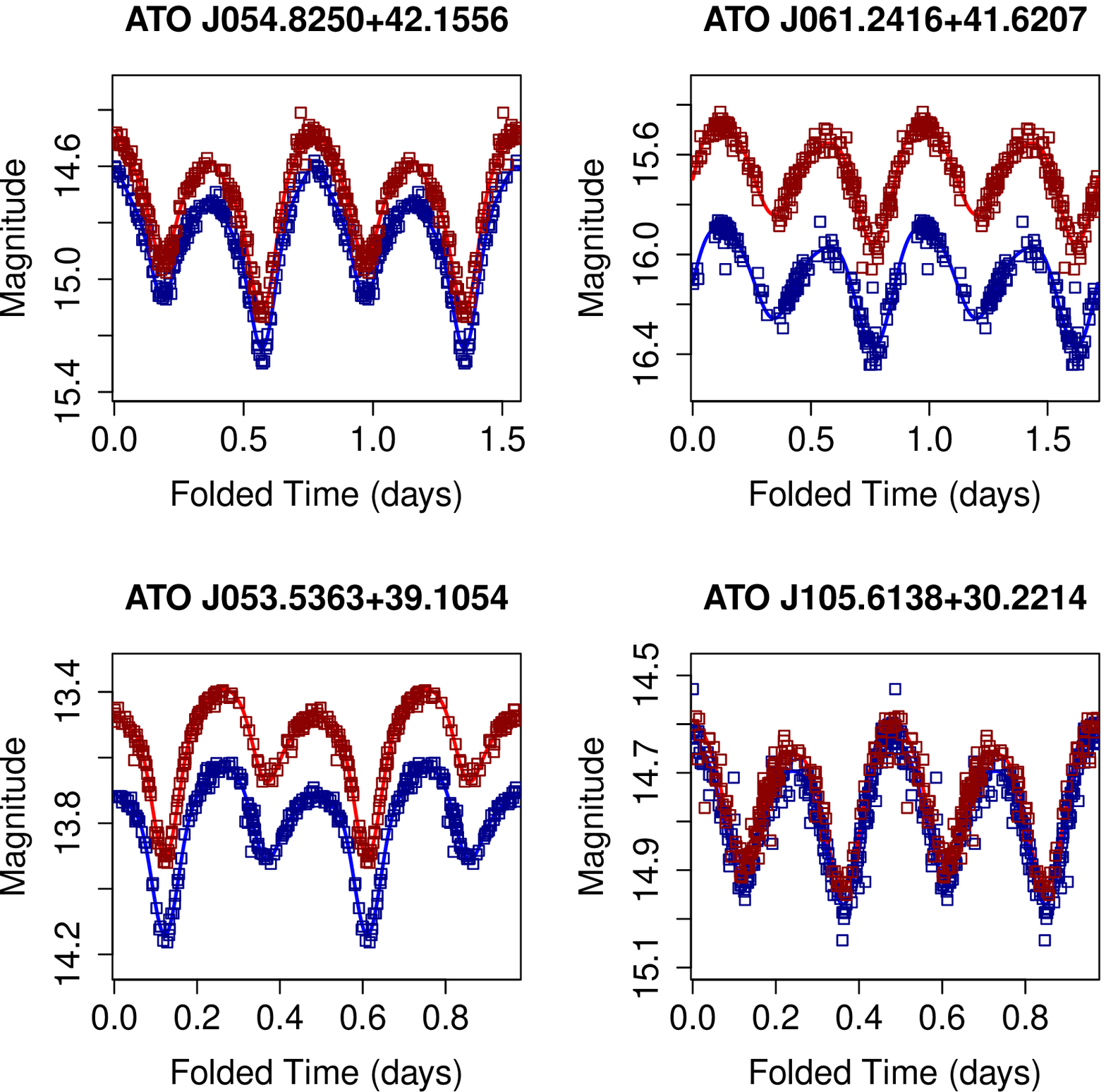}
\caption{Plots of 4 representative eclipsing binaries showing the O'Connell effect, in which the two maxima have significantly different brightness. The astrophysical cause is unknown. Starspots are often suggested, but the long-term stability of the effect appears inconsistent with the generally shorter lifetimes of starspots.
\label{fig:MOC}}
\end{figure}

Figure \ref{fig:MOC} shows example eclipsing binaries showing the O'Connell effect \citep{oconnell}, in which the brightness of the two maxima are different. The astrophysical cause is unknown \citep{Wilsey2009}. There are several hundred such systems in our data.

\citet{Drake2014a} concluded based on lightcurves from the Catalina Sky Survey that the O'Connell effect is probably not caused by starspots, because it remains coherent over periods of years. The ATLAS data support this assessment in a majority of cases. \citet{Wilsey2009} note cases where the O'Connell effect is definitely inconstant, and suggest that in these cases it is due to spots --- however, they argue that more than one astrophysical cause may be at work. It seems possible that the large number of O'Connell stars in the ATLAS data set may enable the deduction of the true astrophysical explanation(s) through a statistical analysis.

\subsection{Two-Cycle Modulated Sine Waves} 

\begin{figure}
\includegraphics{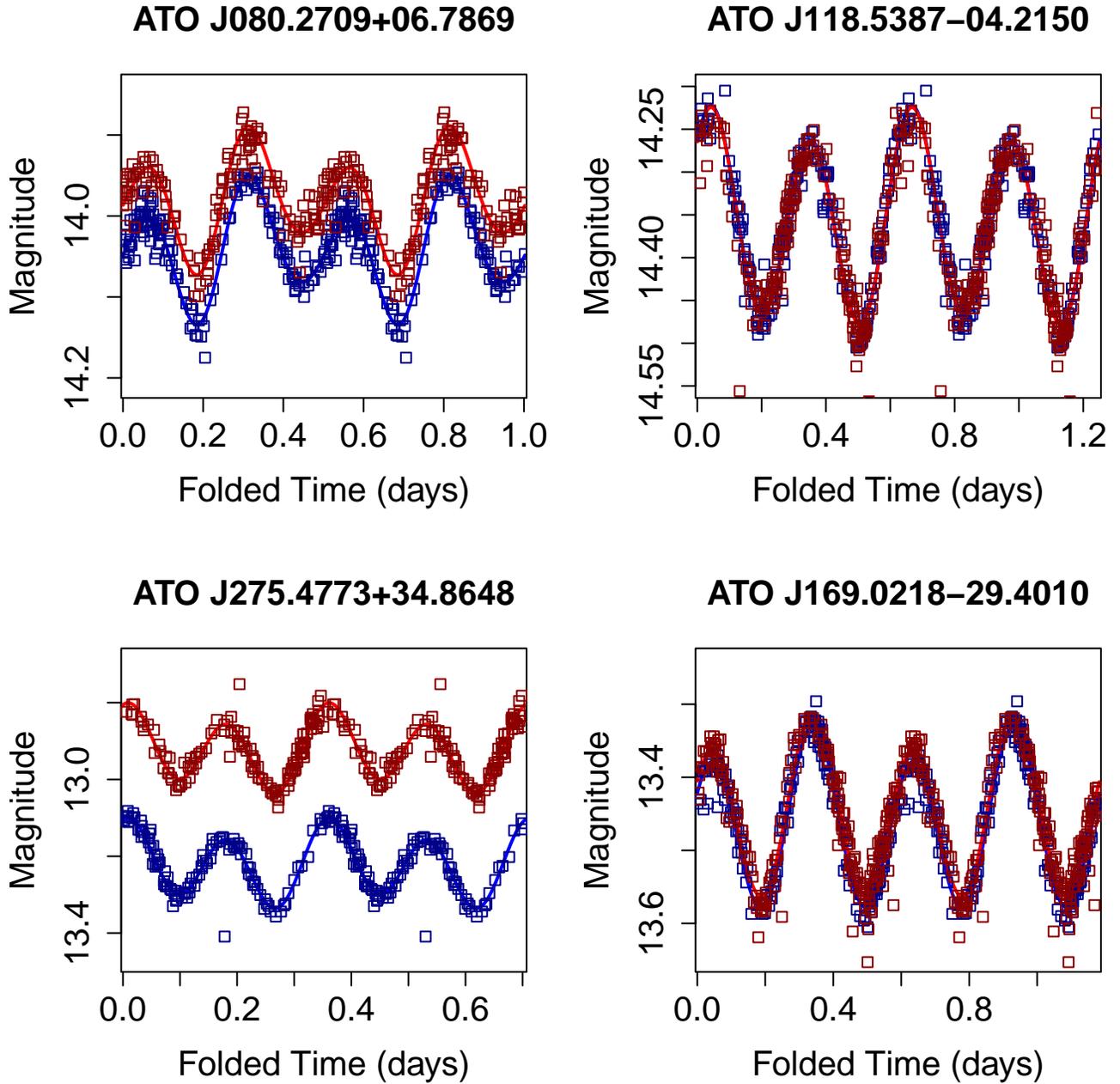}
\caption{Plots of 4 representative variables showing a 2-cycle modulated sine wave. These may be ellipsoidal variables whose maxima differ in brightness due to the same cause that produces the O'Connell effect in eclipsing binaries -- in which case, studying them in more detail may elucidate the nature of the O'Connell effect.
\label{fig:2SINE}}
\end{figure}

Figure \ref{fig:2SINE} shows representative lightcurves with 2-cycle modulated sine waves. There are hundreds of these in our data. It seems most likely that they are ellipsoidal variables (or eclipsing binaries with grazing eclipses) whose maxima differ in brightness due to the same astrophysical cause that produces the O'Connell effect. If so, their statistics may also provide a clue to the physical nature of the effect. At present, however, we cannot rule out the alternative hypothesis that they are multi-mode pulsators with very symmetrical lightcurves.

\subsection{Eclipsing Binaries that are Apparently Extreme Examples of the O'Connell Effect} 

\begin{figure}
\includegraphics{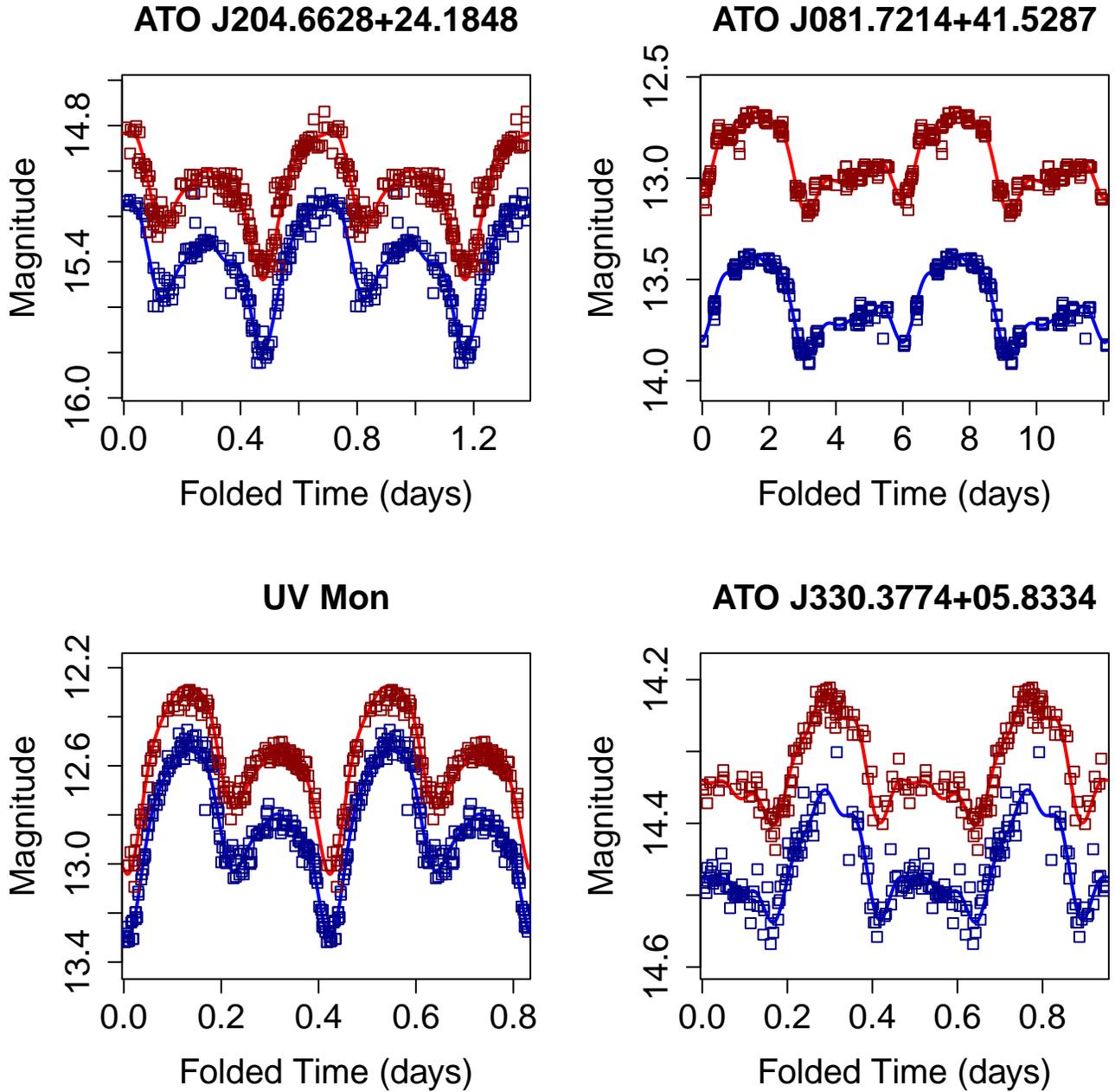}
\caption{Plots of 4 representative variables that appear to be eclipsing binaries showing the O'Connell Effect in an extreme form. UV Mon has been analyzed in detail by \citet{Wilsey2009} as an eclipsing binary exhibiting the O'Connell effect. \citet{Soszynski2016} present the lightcurve of a star they classify as an eclipsing binary (OGLE-BLG-ECL-334012), which is almost identical to that of ATO J330.3774+05.8334.
\label{fig:EOC}}
\end{figure}

Figure \ref{fig:EOC} shows apparent eclipsing binaries that exhibit the O'Connell effect in such an extreme form that it seems to call into question the nature of the systems: are they really eclipsing binaries, or are they peculiar multi-mode pulsators or some other exotic type? There are are at least a few dozen like this in our data. 

UV Mon, the only known variable among these four examples, is classifed as an eclipsing binary by VSX and was analyzed in detail by \citet{Wilsey2009}. The two variables on the right in Figure \ref{fig:EOC} are even more extreme than UV Mon. Are even they real eclipsing binaries? \citet{Soszynski2016} provide at least tentative support for a `yes' answer: they show the lightcurve of a star (OGLE-BLG-ECL-334012), which they classify as a peculiar eclipsing binary and which closely resembles those on the right hand side of Figure \ref{fig:EOC}. In particular, the lightcurve of OGLE-BLG-ECL-334012 is almost identical to ATO J330.3774+05.8334 (though the ATLAS star is two magnitudes brighter and has a longer period).

Thus, previous classifications seems to support the idea that objects like those in Figure \ref{fig:EOC} are bona fide eclipsing binaries. If so, the extreme O'Connell signature that they exhibit should place strong constraints on the astrophysical cause of the effect. For example, it seems improbable that starspots could have sufficient contrast with the photosphere to cause such a strong effect --- a conclusion that \citet{Wilsey2009} come to even in the milder case of UV Mon.

\subsection{Notched Stars} \label{sec:notched}

\begin{figure}
\includegraphics{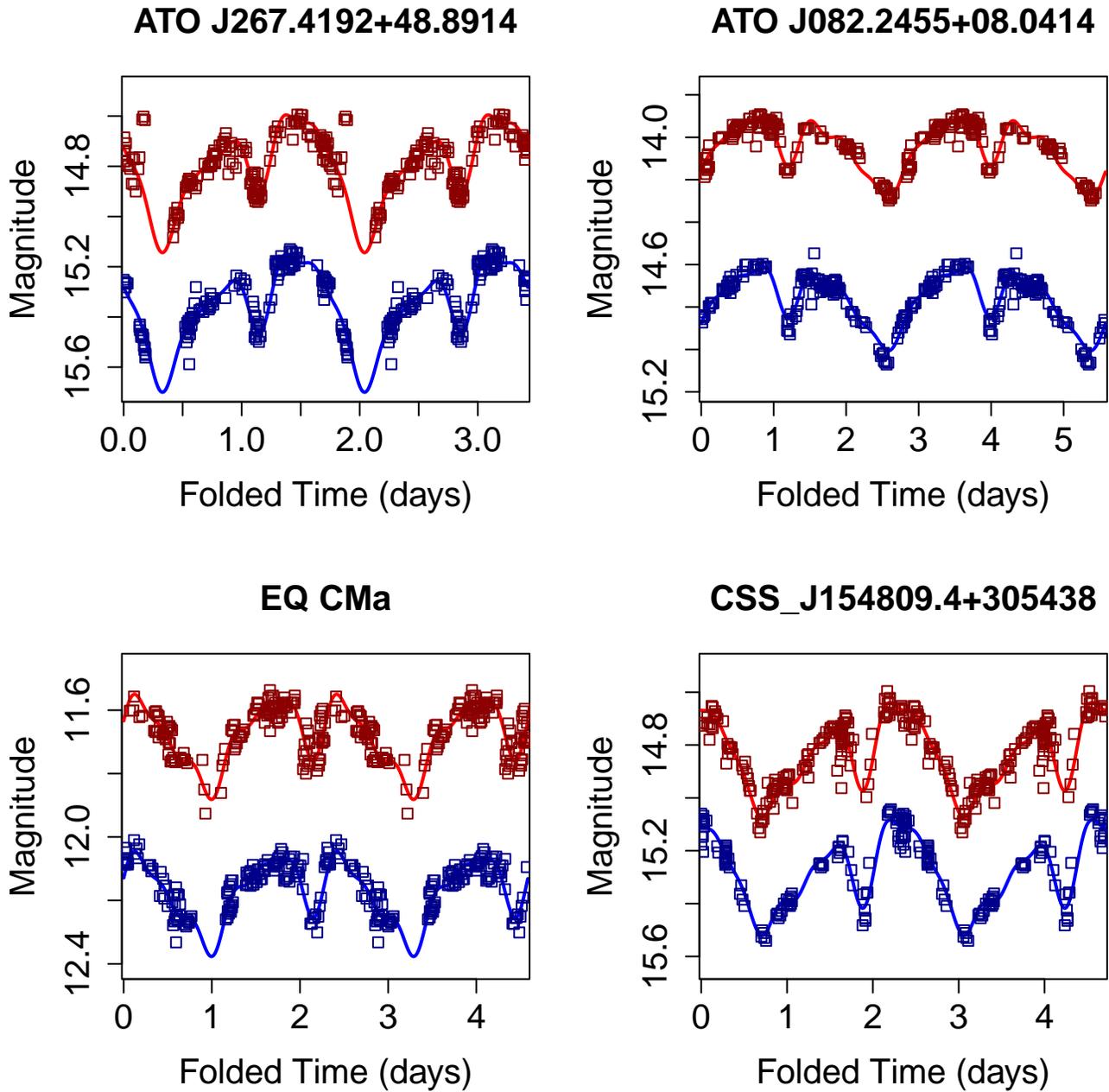}
\caption{Plots of 4 representative variables that appear to show pulse-like variability interrupted by narrow eclipses. We refer to these as notched stars in \S \ref{sec:notched}. They cannot actually be pulsating variables in eclipsing binary systems, because the pulsation and orbital period would have to be perfectly in phase, which is not plausible. EQ CMa is classified in VSX as an RS CVn star (a spotted rotator), but its lightcurve is far from typical for that class.
\label{fig:PUEC}}
\end{figure}

Figure \ref{fig:PUEC} shows examples of a rare type of variable star (only a few tens appear to exist in our data) that we initially interpreted as pulsating stars in eclipsing systems. This is probably not the case, however, since the pulsations would then have to be improbably synchronized with the orbital period. We refer to them provisionally as `notched stars', since their eclipses appear as narrow notches imposed on some other type of variability.

Of the two known objects in Figure \ref{fig:PUEC}, VSX classifies CSS\_J154809.4+305438 simply as an Algol-type eclipsing binary, while EQ CMa is classified as an RS Canum Venaticorum (RS CVn) variable, a spotted rotator sometimes also showing eclipses. This may be the correct diagnosis, but we have examined the ATLAS lightcurves of dozens of known variables classified as RS CVn, and found that most resemble our MSINE or NSINE classes and look nothing like the stars in Figure \ref{fig:PUEC}. If indeed the out-of-eclipse modulation of our notched stars is caused by starspots, the contrast and size of the spots must be quite extreme since they create a modulation comparable in amplitude to the eclipses. This suggests an astrophysical link between the notched stars and the extreme O'Connell stars: perhaps a thorough understanding of one class would also explain the other.

\section{Conclusion}

We analyze 142 million stars with a full magnitude range of about 11--19 mag in the ATLAS bands, and with at least 100 photometric measurements each during the first two years of ATLAS operations. Among these stars, we identify 4.7 million candidate variables. For each of these candidates, we calculate 169 variability features. We use 70 of these features as input for a machine classifier that sorts the stars into broad, astrophysically useful classes based on lightcurve morphology.

We briefly explore just a small part of the rich astrophysical potential of these data. We find them to be useful for the detailed categorization of pulsating variables and eclipsing binaries, and for probing the astrophysical nature and Galactic distribution of many variable types. The enormous statistical power of the large, homogeneously analyzed data set may enable the elucidation of longstanding astrophysical mysteries such as the cause of the O'Connell effect \citep{oconnell} in eclipsing binaries. Our data may also result in the discovery of new classes of variables (see \S \ref{sec:UCBH}).

For all 4.7 million candidate variables, we publicly release all of the ATLAS photometry, plus the vector of variability features and the probabilities from our machine classifier. This is the largest catalog of candidate variables yet released with photometry. We hope that many in the astronomical community will query the data through STScI (see \S \ref{sec:appendixB}) and exploit its potential for exciting discoveries and productive synergies with other projects.

\section{Acknowledgements}

We acknowledge useful discussions with Mark Huber, Dan Huber, Ben Shappee, and Jennifer van Saders.
Support for the ATLAS survey was provided by NASA grant NN12AR55G under the guidance of Lindley Johnson and Kelly Fast.
This publication makes extensive use of data products from the Pan-STARRS1 Surveys and the PS1 public science archive, which have been made possible through contributions by the Institute for Astronomy, the University of Hawaii, the Pan-STARRS Project Office, the Max-Planck Society and its participating institutes, the Max Planck Institute for Astronomy, Heidelberg and the Max Planck Institute for Extraterrestrial Physics, Garching, The Johns Hopkins University, Durham University, the University of Edinburgh, the Queen's University Belfast, the Harvard-Smithsonian Center for Astrophysics, the Las Cumbres Observatory Global Telescope Network Incorporated, the National Central University of Taiwan, the Space Telescope Science Institute, the National Aeronautics and Space Administration under Grant No. NNX08AR22G issued through the Planetary Science Division of the NASA Science Mission Directorate, the National Science Foundation Grant No. AST-1238877, the University of Maryland, Eotvos Lorand University (ELTE), the Los Alamos National Laboratory, and the Gordon and Betty Moore Foundation.
This publication makes use of the SIMBAD online database,
operated at CDS, Strasbourg, France; the VizieR online database \citep[see][]{vizier}; and the 
International Variable Star Index (VSX) database, operated at AAVSO, Cambridge, MA, USA.
%We have also made extensive use of information and code from \citet{nrc}. 

%We have used digitized images from the Palomar Sky Survey 
%(available from \url{http://stdatu.stsci.edu/cgi-bin/dss\_form}),
% which were produced at the Space 
%Telescope Science Institute under U.S. Government grant NAG W-2166. 
%The images of these surveys are based on photographic data obtained 
%using the Oschin Schmidt Telescope on Palomar Mountain and the UK Schmidt Telescope.

\section{Appendix A: Description of Variable Star Features} \label{sec:appendixA}

Here we provide a brief description of all the variability features included in the analysis vector we have calculated for each star, as given in the `object' table hosted by STScI (described in \S \ref{sec:appendixB} below). As described in the text, there are 185 statistics calculated by ATLAS. The database gives 10 additional columns giving Pan-STARRS photometry, and two additional star identification columns, for a total of 197 columns. All of them are described in Table \ref{tab:classvector}.

Note well that the median magnitudes for each star (vf\_c\_med and vf\_o\_med) are given rather late in the table, in columns 98 and 99 --- an idiosyncrasy of the current catalog that we will correct in DR2.

In Table \ref{tab:classvector}, two-character prefixes encode which stage of the ATLAS analysis produced the feature: `fp' means {\tt fourierperiod}; `vf' means {\tt varfeat}; `df' means the feature came from our statistical analysis of detections in the difference images; `ps' means from the proximity statistics (i.e., angular distance to nearest neighboring star); and `ls' means {\tt lombscar}. Features used in our machine-classification are indicated with table note marks: $^a$ for quantities used without alteration, and $^b$ for the sine and cosine amplitudes of the Fourier terms, which were converted into amplitude and phase using Equations \ref{eq:onetermsincos} and \ref{eq:onetermamp} before being used in the machine classification.

\clearpage

\begin{deluxetable}{lll}
\tabletypesize{\small}
\tablewidth{0pt}
\tablecaption{Variable Star Features \label{tab:classvector}}
\tablehead{ \colhead{Column \#} & \colhead{Column name} & \colhead{Description} }
\startdata
1 & ATO\_ID & official ATLAS name (add ATO prefix: e.g. ATO J054.8250+42.1556)\\
2 & ra & J2000.0 RA in degrees \\
3 & dec & J2000.0 Dec in degrees \\
4 & fp\_c\_pts & Number of $c$-band points {\tt fourierperiod} identified as good\\
5 & fp\_o\_pts & Number of $o$-band points {\tt fourierperiod} identified as good\\
6 & fp\_LSperiod & Original period from {\tt fourierperiod's} Lomb Scargle periodogram \\
7 & fp\_origLogFAP\tablenotemark{a} & PPFAP for {\tt fourierperiod's} Lomb Scargle periodogram \\
8 & fp\_origRMS & RMS scatter of median-subtracted input magnitudes that \\
& & {\tt fourierperiod} identified as good\\
9 & fp\_magrms\_c\tablenotemark{a} & RMS scatter of median-subtracted $c$-band magnitudes that\\
& & {\tt fourierperiod} identified as good\\
10 & fp\_magrms\_o\tablenotemark{a} & RMS scatter of median-subtracted $o$-band magnitudes that\\
& & {\tt fourierperiod} identified as good\\
11 & fp\_lngfitper\tablenotemark{a} & Final master period from long-period Fourier fit (days)\\
12 & fp\_lngfitrms & RMS scatter from final long-period fit (magnitudes)\\
13 & fp\_lngfitchi & $\chi^2/N$ for long-period Fourier fit\\
14 & fp\_lngfournum\tablenotemark{a} & Number of Fourier terms used in long-period fit\\
15 & fp\_lngmin\_c\tablenotemark{a} & Minimum brightness reached by long-period fit to the $c$-band\\
& & photometry at any time corresponding to an actual measurement (magnitudes) \\
16 & fp\_lngmax\_c\tablenotemark{a} & Maximum brightness reached by long-period fit to the $c$-band\\
& & photometry at any time corresponding to an actual measurement (magnitudes) \\
17 & fp\_lngmin\_o\tablenotemark{a} & Minimum brightness reached by long-period fit to the $o$-band\\
& & photometry at any time corresponding to an actual measurement (magnitudes) \\
18 & fp\_lngmax\_o\tablenotemark{a} & Maximum brightness reached by long-period fit to the $o$-band\\
& & photometry at any time corresponding to an actual measurement (magnitudes) \\
19 & fp\_lngfitrms\_c\tablenotemark{a} & RMS scatter of residuals from long-period fit to the $c$-band data (magnitudes)\\
20 & fp\_lngfitrms\_o\tablenotemark{a} & RMS scatter of residuals from long-period fit to the $o$-band data (magnitudes)\\
21 & fp\_lngfitchi\_c & $\chi^2/N$ for long-period Fourier fit to the $c$-band data \\
22 & fp\_lngfitchi\_o & $\chi^2/N$ for long-period Fourier fit to the $o$-band data \\
23 & fp\_lngconst\_c & Constant term in the long-period fit to the $c$-band data (magnitudes) \\
24 & fp\_lngconst\_o & Constant term in the long-period fit to the $o$-band data (magnitudes) \\
25 & fp\_sin1\_c\tablenotemark{b} & Sine coefficient of first Fourier term in the long-period fit to\\
& & the $c$-band data (magnitudes)\\
26 & fp\_cos1\_c\tablenotemark{b}  & Cosine coefficient of first Fourier term in the long-period fit to\\
& & the $c$-band data (magnitudes)\\
27 & fp\_sin1\_o\tablenotemark{b} & Sine coefficient of first Fourier term in the long-period fit to\\
& & the $o$-band data (magnitudes)\\
28 & fp\_cos1\_o\tablenotemark{b} & Cosine coefficient of first Fourier term in the long-period fit to\\
& & the $o$-band data (magnitudes)\\
29 & fp\_PPFAPlong1 & PPFAP of residuals after subtraction of the best long-period\\
& & Fourier fit with one term \\
30 & fp\_sin2\_c\tablenotemark{b} & Sine coefficient of second Fourier term in the long-period fit to\\
& & the $c$-band data (magnitudes)\\
31 & fp\_cos2\_c\tablenotemark{b} & Cosine coefficient of second Fourier term in the long-period fit to\\
& & the $c$-band data (magnitudes)\\
32 & fp\_sin2\_o\tablenotemark{b} & Sine coefficient of second Fourier term in the long-period fit to\\
& & the $o$-band data (magnitudes)\\
33 & fp\_cos2\_o\tablenotemark{b} & Cosine coefficient of second Fourier term in the long-period fit to\\
& & the $o$-band data (magnitudes)\\
34 & fp\_PPFAPlong2 & PPFAP of residuals after subtraction of the best long-period\\
& & Fourier fit with two terms \\
35 & fp\_sin3\_c\tablenotemark{b} & Obvious by analogy to columns 25 and 30 \\
36 & fp\_cos3\_c\tablenotemark{b} & Obvious by analogy to columns 26 and 31 \\
37 & fp\_sin3\_o\tablenotemark{b} & Obvious by analogy to columns 27 and 32 \\
38 & fp\_cos3\_o\tablenotemark{b} & Obvious by analogy to columns 28 and 33 \\
39 & fp\_PPFAPlong3 & Obvious by analogy to columns 29 and 34 \\
40 & fp\_sin4\_c\tablenotemark{b} & Obvious by analogy to columns 25 and 30 \\
41 & fp\_cos4\_c\tablenotemark{b} & Obvious by analogy to columns 26 and 31 \\
42 & fp\_sin4\_o\tablenotemark{b} & Obvious by analogy to columns 27 and 32 \\
43 & fp\_cos4\_o\tablenotemark{b} & Obvious by analogy to columns 28 and 33 \\
44 & fp\_PPFAPlong4 & Obvious by analogy to columns 29 and 34 \\
45 & fp\_hifreq\_c & A measure of the relative power in the high-frequency vs low-frequency\\
& & terms in the long-period Fourier fit to the $c$-band\\
46 & fp\_hifreq\_o & A measure of the relative power in the high-frequency vs low-frequency\\
& & terms in the long-period Fourier fit to the $o$-band\\
47 & fp\_timerev\_c & A measure of the degree of invariance of the long-period Fourier fit\\
& & to the $c$-band data with respect to reversal (i.e. mirroring) of the \\
& & time-axis about the time of minimum light (large value = invariant)\\
48 & fp\_timerev\_o & A measure of the degree of invariance of the long-period Fourier fit\\
& & to the $o$-band data with respect to reversal (i.e. mirroring) of the \\
& & time-axis about the time of minimum light (large value = invariant)\\
49 & fp\_phase180\_c & A measure of the degree of invariance of the long-period Fourier fit\\
& & to the $c$-band data with respect to a 180$^{\circ}$ phase-shift \\
& & (large value = invariant)\\
50 & fp\_phase180\_o & A measure of the degree of invariance of the long-period Fourier fit\\
& & to the $o$-band data with respect to a 180$^{\circ}$ phase-shift \\
& & (large value = invariant)\\
51 & fp\_powerterm\_c\tablenotemark{a} & Highest-amplitude Fourier term in the long-period fit to the\\
& & $c$-band data \\
52 & fp\_powerterm\_o\tablenotemark{a} & Highest-amplitude Fourier term in the long-period fit to the\\
& & $o$-band data \\
53 & fp\_domper\_c & Period corresponding to fp\_powerterm\_c (days)\\
54 & fp\_domper\_o & Period corresponding to fp\_powerterm\_o (days)\\
55 & fp\_shortfit\tablenotemark{a} & Was a short-period fit performed? (0 = no)\\
56 & fp\_period\tablenotemark{a} & Final master period from short-period Fourier fit (days)\\
57 & fp\_fitrms & RMS scatter from final short-period fit (magnitudes)\\
58 & fp\_fitchi & $\chi^2/N$ for short-period Fourier fit\\
59 & fp\_fournum\tablenotemark{a} & Number of Fourier terms used in short-period fit\\
60 & fp\_alias & Diurnal alias $j$ of final period relative to fp\_LSperiod (see Equation \ref{eq:alias})\\
61 & fp\_multfac & Multiplication factor $f$ of final period relative to fp\_LSperiod (see Equation \ref{eq:alias})\\
62 & fp\_phaseoff & Offset of final period relative to fp\_LSperiod, in cycles over the\\
& & full temporal span of our data\\
63 & fp\_min\_c\tablenotemark{a} & Minimum brightness reached by short-period fit to the $c$-band\\
& & photometry at any time corresponding to an actual measurement (magnitudes) \\
64 & fp\_max\_c\tablenotemark{a} & Maximum brightness reached by short-period fit to the $c$-band\\
& & photometry at any time corresponding to an actual measurement (magnitudes) \\
65 & fp\_min\_o\tablenotemark{a} & Minimum brightness reached by short-period fit to the $o$-band\\
& & photometry at any time corresponding to an actual measurement (magnitudes) \\
66 & fp\_max\_o\tablenotemark{a} & Maximum brightness reached by short-period fit to the $o$-band\\
& & photometry at any time corresponding to an actual measurement (magnitudes) \\
67 & fp\_fitrms\_c\tablenotemark{a} & RMS scatter of residuals from short-period fit to the $c$-band data (magnitudes)\\
68 & fp\_fitrms\_o\tablenotemark{a} & RMS scatter of residuals from short-period fit to the $o$-band data (magnitudes)\\
69 & fp\_fitchi\_c & $\chi^2/N$ for short-period Fourier fit to the $c$-band data \\
70 & fp\_fitchi\_o & $\chi^2/N$ for short-period Fourier fit to the $o$-band data \\
71 & fp\_const\_c\tablenotemark{a} & Constant term in the short-period fit to the $c$-band data (magnitudes) \\
72 & fp\_const\_o\tablenotemark{a} &  Constant term in the short-period fit to the $o$-band data (magnitudes) \\
73 & fp\_sin1\tablenotemark{b} & Sine coefficient of first Fourier term in the short-period fit (magnitudes)\\
74 & fp\_cos1\tablenotemark{b} & Cosine coefficient of first Fourier term in the short-period fit (magnitudes)\\75 & fp\_PPFAPshort1 & PPFAP of residuals after subtraction of the best short-period\\
& & Fourier fit with one term \\
76 & fp\_sin2\tablenotemark{b} & Sine coefficient of second Fourier term in the short-period fit (magnitudes)\\
77 & fp\_cos2\tablenotemark{b} & Cosine coefficient of second Fourier term in the short-period fit (magnitudes)\\
78 & fp\_PPFAPshort2 & PPFAP of residuals after subtraction of the best short-period\\
& & Fourier fit with two terms \\
79 & fp\_sin3\tablenotemark{b} & Obvious by analogy to columns 73 and 76\\
80 & fp\_cos3\tablenotemark{b} & Obvious by analogy to columns 74 and 77\\
81 & fp\_PPFAPshort3 & Obvious by analogy to columns 75 and 78\\
82 & fp\_sin4\tablenotemark{b} & Obvious by analogy to columns 73 and 76\\
83 & fp\_cos4\tablenotemark{b} & Obvious by analogy to columns 74 and 77\\
84 & fp\_PPFAPshort4 & Obvious by analogy to columns 75 and 78\\
85 & fp\_sin5\tablenotemark{b} & Obvious by analogy to columns 73 and 76\\
86 & fp\_cos5\tablenotemark{b} & Obvious by analogy to columns 74 and 77\\
87 & fp\_PPFAPshort5 & Obvious by analogy to columns 75 and 78\\
88 & fp\_sin6\tablenotemark{b} & Obvious by analogy to columns 73 and 76\\
89 & fp\_cos6\tablenotemark{b} & Obvious by analogy to columns 74 and 77\\
90 & fp\_PPFAPshort6 & Obvious by analogy to columns 75 and 78\\
91 & fp\_hifreq & A measure of the relative power in the high-frequency vs low-frequency\\
& & terms in the short-period Fourier fit\\
92 & fp\_timerev\tablenotemark{a} & A measure of the degree of invariance of the short-period Fourier fit\\
& & with respect to reversal (i.e. mirroring) of the time-axis about the\\
& & time of minimum light (large value = invariant)\\
93 & fp\_phase180\tablenotemark{a} & A measure of the degree of invariance of the short-period Fourier fit\\
& & with respect to a 180$^{\circ}$ phase-shift (large value = invariant)\\
94 & fp\_powerterm\tablenotemark{a} & Highest-amplitude Fourier term in the short-period fit\\
95 & fp\_domperiod & Period corresponding to fp\_powerterm (days)\\
96 & vf\_Nc & Number of $c$ band observations\\
97 & vf\_No & Number of $o$ band observations\\
98 & vf\_c\_med\tablenotemark{a} & Weighted median $c$ magnitude\\
99 & vf\_o\_med\tablenotemark{a} & Weighted median $o$ magnitude\\
100 & vf\_percentile5\tablenotemark{a} & 5th percentile of median-subtracted magnitudes\\
101 & vf\_percentile10\tablenotemark{a} & 10th percentile of median-subtracted magnitudes\\
102 & vf\_percentile25\tablenotemark{a} & 25th percentile of median-subtracted magnitudes\\
103 & vf\_percentile75\tablenotemark{a} & 75th percentile of median-subtracted magnitudes\\
104 & vf\_percentile90\tablenotemark{a} & 90th percentile of median-subtracted magnitudes\\
105 & vf\_percentile95\tablenotemark{a} & 95th percentile of median-subtracted magnitudes\\
106 & vf\_Hday\tablenotemark{a} & A statistic probing the significance of intra-night variations\\
107 & vf\_Hlong\tablenotemark{a} & A statistic probing the significance of inter-night (long-term) variations\\
108 & vf\_wsd & Weighted standard deviation \citep{Sokolovsky2017}\\
109 & vf\_iqr & Interquartile range \citep{Sokolovsky2017}\\
110 & vf\_chin\tablenotemark{a} & Reduced $\chi^2$ = $\chi^2/(N-1)$ \citep{Sokolovsky2017}\\
111 & vf\_roms\tablenotemark{a} & Robust median statistic \citep{Sokolovsky2017}\\
112 & vf\_nxs & Normalized excess variance \citep{Sokolovsky2017}\\
113 & vf\_nppa & Normalized peak-to-peak amplitude \citep{Sokolovsky2017}\\
114 & vf\_inu\tablenotemark{a} & Inverse von Neumann ratio \citep{Sokolovsky2017}\\
115 & vf\_WS\_I\tablenotemark{a} & Welch-Stetson I \citep{Sokolovsky2017}\\
116 & vf\_S\_J\tablenotemark{a} & Stetson J \citep{Sokolovsky2017}\\
117 & vf\_S\_K & Stetson K \citep{Sokolovsky2017}\\
118 & df\_numdet & Number of detections at this location in the difference images\\
119 & df\_medmag & Median magnitude of detections in the difference images (negative-going\\
& & detections are included by calculating magnitudes from the absolute\\
& & value of the flux)\\
120 & df\_meanmag &  Mean magnitude of detections in the difference images (negative-going\\
& & detections are included by calculating magnitudes from the absolute\\
& & value of the flux)\\
121 & df\_medsig & Median SNR of detections in the difference images\\
122 & df\_meansig & Mean SNR of detections in the difference images\\
123 & df\_r2sig & SNR of second most significant difference image detection\\
124 & df\_r1sig & SNR of most significant difference image detection\\
125 & df\_medchin & Median $\chi^2/N$ of PSF fits on the difference images\\
126 & df\_numbright & Number of positive-going detections on the difference images\\
127 & df\_medPvar & Median value of Pvr (probability of being a variable star) from {\tt vartest}\\
& & (max=999) \\
128 & df\_meanPvar & Mean value of Pvr (max=999)\\
129 & df\_r2Pvar & Second highest value of Pvr\\
130 & df\_r1Pvar & Highest value of Pvr\\
131 & df\_medPscar &  Median value Psc (probability of being a star subtraction residual)\\
& & from {\tt vartest} (max=999) \\
132 & df\_meanPscar & Mean value Psc (probability of being a star subtraction residual)\\
& & from {\tt vartest} (max=999) \\
133 & ps\_dist & Angular distance to the nearest star in our Pan-STARRS reference catalog\\
& & (arcsec)\\
134 & ps\_dist0 & Angular distance to the nearest star in our Pan-STARRS reference catalog\\
& & that is at least equally bright(arcsec) \\
135 & ps\_dist2 & Angular distance to the nearest star in our Pan-STARRS reference catalog\\
& & that is at least two magnitudes brighter (arcsec)\\
136 & ps\_dist4 & Angular distance to the nearest star in our Pan-STARRS reference catalog\\
& & that is at least four magnitudes brighter (arcsec)\\
137 & ls\_Npt & Number of photometric measurements input to {\tt lombscar}\\
138 & ls\_Nuse & Number of photometric measurements {\tt lombscar} identified as good\\
139 & ls\_c\_med & Median $c$-band magnitude calculated by {\tt lombscar}\\
140 & ls\_o\_med & Median $o$-band magnitude calculated by {\tt lombscar}\\
141 & ls\_Pday & Period output by {\tt lombscar} (days)\\
142 & ls\_PPFAP & PPFAP from Lomb-Scargle periodogram in {\tt lombscar}\\
143 & ls\_Chin & $\chi^2/N$ for Fourier+polynomial fit performed by {\tt lombscar}\\
144 & ls\_Cchin & $\chi^2/N$ for constant-magnitude fit performed by {\tt lombscar}\\
145 & ls\_Pchin & $\chi^2/N$ for polynomial-only fit performed by {\tt lombscar}\\
146 & ls\_Xchin & $\chi^2/N$ for polynomial-only fit performed by {\tt lombscar}, without\\
& & outlier trimming\\
147 & ls\_Fraclo & Fraction of points with magnitudes more than 5$\sigma$ below the median\\
148 & ls\_Frachi & Fraction of points with magnitudes more than 5$\sigma$ above the median\\
149 & ls\_txclo & Fraction of low outliers with time diff less than 0.06 day\\
150 & ls\_txchi & Fraction of high outliers with time diff less than 0.06 day\\
151 & ls\_Chin\_minus\_1 & $\chi^2/N$ for {\tt lombscar} Fourier fit to $j=-1$ alias\\
152 & ls\_Chin\_minus\_h & $\chi^2/N$ for {\tt lombscar} Fourier fit to $j=-0.5$ alias\\
153 & ls\_Chin\_plus\_h & $\chi^2/N$ for {\tt lombscar} Fourier fit to $j=+0.5$ alias\\
154 & ls\_Chin\_plus\_1 & $\chi^2/N$ for {\tt lombscar} Fourier fit to $j=+1$ alias\\
155 & ls\_Ply1 & Linear coefficient of polynomial fit by {\tt lombscar} (mag/year)\\
156 & ls\_Ply2 & Quadratic coefficient of polynomial fit by {\tt lombscar} (mag/year$^2$)\\
157 & ls\_Phgap & biggest time gap with no points in folded lightcurve (fraction of ls\_Pday)\\
158 & ls\_D & Period doubling (1 if {\tt lombscar} output period has been doubled relative to\\
& & the highest peak in the Lomb-Scargle periodogram, or 0 if not)\\
159 & ls\_RMS & RMS of residuals from {\tt lombscar} Fourier fit\\
160 & ls\_F0 & Amplitude of {\tt lombscar} constant Fourier term divided by RMS\\
161 & ls\_F1cos & Amplitude of {\tt lombscar} cos1 Fourier term divided by RMS\\
162 & ls\_F1sin & Amplitude of {\tt lombscar} sin1 Fourier term divided by RMS\\
163 & ls\_F2cos & Amplitude of {\tt lombscar} cos2 Fourier term divided by RMS\\
164 & ls\_F2sin & Amplitude of {\tt lombscar} sin2 Fourier term divided by RMS\\
165 & ls\_F3cos & Amplitude of {\tt lombscar} cos3 Fourier term divided by RMS\\
166 & ls\_F3sin & Amplitude of {\tt lombscar} sin3 Fourier term divided by RMS\\
167 & ls\_F4cos & Amplitude of {\tt lombscar} cos4 Fourier term divided by RMS\\
168 & ls\_F4sin & Amplitude of {\tt lombscar} sin4 Fourier term divided by RMS\\
169 & CLASS & Final ATLAS variable classification\\
170 & ddcSTAT & Difference image statistic (1=probably variable independent of any other\\
& & information)\\
171 & proxSTAT & Proximity statistic (1=variability detection probably {\em not} caused by blending)\\
172 & prob\_CBF & Machine classifier probability that this star is in the CBH category\\
173 & prob\_CBH & Machine classifier probability that this star is in the CBF category\\
174 & prob\_DBF & Machine classifier probability that this star is in the DBF category\\
175 & prob\_DBH & Machine classifier probability that this star is in the DBH category\\
176 & prob\_HARD & Machine classifier probability that this star is IRR, LPV, or `dubious'\\
177 & prob\_MIRA & Machine classifier probability that this star is in the MIRA category\\
178 & prob\_MPULSE & Machine classifier probability that this star is in the MPULSE category\\
179 & prob\_MSINE & Machine classifier probability that this star is in the MSINE category\\
180 & prob\_NSINE & Machine classifier probability that this star is in the NSINE category\\
181 & prob\_PULSE & Machine classifier probability that this star is in the PULSE category\\
182 & prob\_SINE & Machine classifier probability that this star is in the SINE category\\
183 & prob\_IRR & Machine classifier probability that this star is in the IRR category\\
184 & prob\_LPV & Machine classifier probability that this star is in the LPV category\\
185 & prob\_dubious & Machine classifier probability that this star is in the `dubious' category\\
186 & gmag & Pan-STARRS1 DR1 $g$-band magnitude\\
187 & gerr & Uncertainty on $g$-band magnitude\\
188 & rmag & Pan-STARRS1 DR1 $r$-band magnitude\\
189 & rerr & Uncertainty on $r$-band magnitude\\
190 & imag & Pan-STARRS1 DR1 $i$-band magnitude\\
191 & ierr & Uncertainty on $i$-band magnitude\\
192 & zmag & Pan-STARRS1 DR1 $z$-band magnitude\\
193 & zerr & Uncertainty on $z$-band magnitude\\
194 & ymag & Pan-STARRS1 DR1 $Y$-band magnitude\\
195 & yerr & Uncertainty on $Y$-band magnitude\\
196 & starID & Old version of ATLAS star ID (historical interest only)\\
197 & objid & Object ID, useful for linkage with the `detection' database\\
\enddata
\tablenotetext{a}{Used for machine classification.}
\tablenotetext{b}{Used for machine classification after conversion of sine and cosine coefficients to an overall amplitude and phase (see Equations \ref{eq:onetermsincos} and \ref{eq:onetermamp}).}
\end{deluxetable}

\clearpage

\section{Appendix B: Sample Database Queries} \label{sec:appendixB}

In this appendix we supply instructions for querying the databases described above. Most importantly, we give example queries for the ATLAS variable star databases presented herein. First, however, we show how to query the Pan-STARRS1 DR1 database to produce a catalog like the one we used for matching ATLAS photometric dections to unique stars (see \S \ref{sec:photdat}).

To query either the Pan-STARRS or ATLAS databases, begin by going to the website:

{\tt http://mastweb.stsci.edu/ps1casjobs/}

Create an account, and login.

\subsection{Pan-STARRS1 DR1 queries used to construct our object-matching catalog} \label{pscat}

Having logged in to the website given above, click on the `Query' tab. Select PanSTARRS\_DR1 from the `Context' dropdown menu. Type distinctive names for `Table' and `Task Name'. The name you type under `Table' is very important, since it will be the filename of the catalog produced by your query. Be sure to avoid attempting two different queries with the same table name. 

Paste a query based on the example below into the big empty box filling most of the page. Click the `Syntax' button at upper right to check for errors. If the syntax is OK, click the `Submit' button. If you think your query might run very fast, you can use the `Quick' button instead of `Submit', but the PanSTARRS\_DR1 database is so huge that almost nothing is quick.

When the query is finished, which could take several hours, click on the `MyDB' tab to see the table that has been produced. Click on the filename and then click the `Download' button that will appear near the middle of a row of buttons at top right. When the download is finished, click the `Output' button to finally access the file that has been produced.

Here is an example query that select objects with ra between 120 and 135 from the PanSTARRS\_DR1 database:

\begin{verbatim}
SELECT objectThin.objID, raMean, decMean,
gMeanKronMag,gMeanKronMagErr,
rMeanKronMag,rMeanKronMagErr,
iMeanKronMag,iMeanKronMagErr,
zMeanKronMag,zMeanKronMagErr,
yMeanKronMag,yMeanKronMagErr 
FROM ObjectThin
JOIN MeanObject ON objectThin.uniquePspsOBid = meanObject.uniquePspsOBid
JOIN stackObjectThin ON objectThin.objID = stackObjectThin.objID
WHERE
bestdetection = 1 AND
primarydetection = 1 AND
((gKronMag < 19 AND gKronMag > 0) OR
(rKronMag < 19 and rKronMag > 0 )  OR
(iKronMag < 19 and iKronMag > 0)  OR
(zKronMag < 19 and zKronMag > 0))
AND raMean >= 120. AND raMean < 135.
\end{verbatim}

\subsection{How to Query the ATLAS variable star databases} \label{sec:atquery}

Just as for Pan-STARRS, begin by logging it at

{\tt http://mastweb.stsci.edu/ps1casjobs/}

and use the `Context' dropdown menu to select HLSP\_ATLAS\_VAR.

There are three databases: object (the catalog of candidate variable stars, with 4.7 million entries); detection (all ATLAS photometric measurements of all the candidate variables, nearly a billion entries); and observation (a catalog of all ATLAS images used for DR1). To get examples of the columns present in each of the three catalogs, try the following three queries using the `Quick' button:

\begin{verbatim}
select top 10 * from object

select top 10 * from detection

select top 10 * from observation
\end{verbatim}

The column names in the detection and observation databases should be mostly self-explanatory. The detection database gives the time (mjd = Modified Julian Day), celestial coordinates (ra and dec), magnitude and magnitude uncertainty (m and dm) and filter ($o$ or $c$) for each ATLAS measurement. 

\textbf{Note well} that the Modified Julian Day (MJD) values in the detection and observation databases do {\em not} have a light-travel-time correction applied: that is, they are {\em not} the Heliocentric or Barycentric MJD. For precision timing or accurate phasing of short period variables, you {\em must} apply a light-travel-time correction to produce HMJD or BMJD. Formulae for this correction are available online and in the Astronomical Almanac.

The object database has a huge number of columns, which are described in \S \ref{sec:appendixA} above. We remind the reader that \textbf{the median magnitudes you will most likely want for generic queries (vf\_c\_med and vf\_o\_med) are given oddly late in the table, in columns 98 and 99}.

Here are some more sophisticated example queries:

\noindent{Find very short-period objects ($P < 0.05$ days) with very significant intra-night variations:}
\begin{verbatim}
select fp_LSperiod,fp_Period, objid from object where fp_period < 0.05 
and vf_hday > 20 and fp_shortfit =1
\end{verbatim}

\noindent{Get detections for one of the objects identified in the short-period query:}
\begin{verbatim}
select * from detection where objid = 98841395045605667
\end{verbatim}

Select candidate long-period Cepheids in four stages:

\noindent{1: PULSE variables with only a long-period fit:}
\begin{verbatim}
select * from object where fp_shortfit =0 and CLASS='PULSE'
\end{verbatim}

\noindent{2: PULSE variables with a short-period fit leading to a master period longer than 6.2 days:}
\begin{verbatim}
select * from object where fp_shortfit =1 and fp_period > 6.2 
and fp_origLogFAP > 20 and CLASS='PULSE'
\end{verbatim}

\noindent{3: Variables with a short-period fit that are not classified as PULSE but nevertheless have highly significant variations that might be consistent with a Cepheid. We demand highly significant variability by requiring that the PPFAP from {\tt fourierperiod's} original periodogram be more than 20, and we demand a coherent periodic fit by requiring that the ratio of raw to residual RMS be at least 4 and that $\chi^2/N$ for the short-period fit must be less than 5.}
\begin{verbatim}
select * from object where fp_shortfit =1 and fp_period > 6.2 
and fp_origLogFAP > 20 
and (fp_origRMS/fp_fitrms) > 4 and fp_fitchi < 5 and CLASS!='PULSE'
\end{verbatim}

\noindent{4: Variables with a long-period fit that are not classified as PULSE but nevertheless have highly significant variations that might be consistent with a Cepheid. Besides using constraints analogous to the short-period query, we excluded MIRA stars and other extremely long-period variables by requiring $P < 50$ days.}
\begin{verbatim}
select * from object where fp_shortfit =0 and fp_lngfitper < 50 
and fp_origLogFAP > 20 and (fp_origRMS/fp_lngfitrms) > 4 and fp_lngfitchi < 5 
and CLASS!='PULSE'
\end{verbatim}

The total number of Cepheid candidates matching any of the four queries is only about 600 out of the 4.7 million stars in the database.

Then here are some additional, more general queries:

\noindent{Find RR Lyrae stars --- extract all PULSE variables with periods between 0.3 and 0.9 days:}
\begin{verbatim}
select ATO_ID, class, fp_period, objid from object
where class='PULSE' and fp_period between 0.3 and 0.9
order by fp_period
\end{verbatim}

\noindent{Find irregular stars with huge amplitudes greater than 1.5 magnitudes, for which ddcSTAT=1, indicating that the statistics from the difference images show the variability is not spurious:}
\begin{verbatim}
select ATO_ID, class, (fp_min_c-fp_max_c) as delta1, (fp_min_o-fp_max_o) 
as delta2 from object
where class='IRR' and (((fp_min_c-fp_max_c)>1.5) or ((fp_min_o-fp_max_o)>1.5)) 
and ddcSTAT=1
order by delta1, delta2
\end{verbatim}

\noindent{Find eclipsing binaries with orbital periods below the well-known cutoff at 0.22 days:}
\begin{verbatim}
select ATO_ID, class, fp_period from object
where (class in ('CBH','DBH') and fp_period < 0.11) 
or (class in ('CBF','DBF') and fp_period < 0.22)
order by class, fp_period
\end{verbatim}

\noindent{Find SINE stars that might misclassified RRc based on their periods and amplitudes:}
\begin{verbatim}
select ATO_ID, class, fp_period, fp_min_c - fp_max_c as delta from object
where class = 'SINE' and (fp_period between 0.3 and 0.48
and ((fp_min_c - fp_max_c) between 0.36 and 0.6))
order by fp_period, delta
\end{verbatim}

\noindent{Find LPV stars that are strong, high-amplitude variables: }
\begin{verbatim}
select ATO_ID, class, fp_origLogFAP, (fp_min_c - fp_max_c) as delta1, 
(fp_min_o - fp_max_o) as delta2 from object
where class='LPV' and fp_origLogFAP>20
and (((fp_min_c - fp_max_c)>0.5) or ((fp_min_o - fp_max_o))>0.5)
order by fp_origLogFAP, delta1, delta2
\end{verbatim}

\noindent{Find LPV stars that are probably coherent and regular, regardless of their amplitudes:}
\begin{verbatim}
select ATO_ID, class, fp_origLogFAP, fp_lngfitper, 
(fp_lngmin_c - fp_lngmax_c)  as delta1, 
(fp_lngmin_o - fp_lngmax_o) as delta2 from object
where class='LPV' and fp_origLogFAP>20 and fp_lngfitchi < 5
and fp_powerterm_c=1 and fp_powerterm_o=1
\end{verbatim}

\end{document}